\documentclass[twocolumn,aps,pra,superscriptaddress,amsmath,amssym,floatfix]{revtex4-1}

\usepackage{graphicx}
\usepackage{bm}
\usepackage{color}
\usepackage{amssymb}
\usepackage{enumerate}
\usepackage{comment}

\def\la{\langle}
\def\ra{\rangle}

\def\be{\begin{equation}}
\def\ee{\end{equation}}

\begin{document}

\newcommand{\bigjprob}{{\mathcal{P}}}
\newcommand{\bigprob}{_{\bm{q}_F}{\mathcal{P}}_{\bm{q}_I}}
\newcommand{\cum}[1]{\llangle #1 \rrangle}       					
\newcommand{\op}[1]{\hat{\bm #1}}                					
\newcommand{\vop}[1]{\vec{\bm #1}}
\newcommand{\opt}[1]{\hat{\tilde{\bm #1}}}
\newcommand{\vopt}[1]{\vec{\tilde{\bm #1}}}
\newcommand{\td}[1]{\tilde{ #1}}
\newcommand{\mean}[1]{\la#1\ra}                  					
\newcommand{\cmean}[2]{ { }_{#1}\mean{#2}}       				
\newcommand{\pssmean}[1]{ { }_{\bm{q}_F}\mean{#1}_{\bm{q}_I}}
\newcommand{\ket}[1]{\vert#1\ra}                 					
\newcommand{\bra}[1]{\la#1\vert}                 					
\newcommand{\ipr}[2]{\left\la#1\mid#2\right\ra}            				
\newcommand{\opr}[2]{\ket{#1}\bra{#2}}           					
\newcommand{\pr}[1]{\opr{#1}{#1}}                					
\newcommand{\Tr}[1]{\text{Tr}(#1)}               					
\newcommand{\Trd}[1]{\text{Tr}_d(#1)}            					
\newcommand{\Trs}[1]{\text{Tr}_s(#1)}            					
\newcommand{\intd}[1]{\int \! \mathrm{d}#1 \,}
\newcommand{\dd}{\mathrm{d}}
\newcommand{\fullint}{\iint \! \mathcal{D}\mathcal{D} \,}
\newcommand{\drv}[1]{\frac{\delta}{\delta #1}}
\newcommand{\partl}[3]{ \frac{\partial^{#3}#1}{ \partial #2^{#3}} }		
\newcommand{\smpartl}[4]{ \left( \frac{\partial^{#3} #1}{ \partial #2^{#3}} \right)_{#4}}
\newcommand{\smpartlmix}[4]{\left( \frac{\partial^2 #1}{\partial #2 \partial #3 } \right)_{#4}}
\newcommand{\limit}[2]{\underset{#1 \rightarrow #2}{\text{lim}} \;}
\newcommand{\funcd}[2]{\frac{\delta #1}{\delta #2}}
\newcommand{\funcdiva}[3]{\frac{\delta #1[#2]}{\delta #2 (#3)}}
\newcommand{\funcdivb}[4]{\frac{\delta #1 (#2(#3))}{\delta #2 (#4)}}
\newcommand{\funcdivc}[3]{\frac{\delta #1}{\delta #2(#3)}}
\definecolor{dgreen}{RGB}{30,130,30}

\title{Prediction and Characterization of Multiple Extremal Paths in Continuously Monitored Qubits}

\author{Philippe Lewalle} 
\email{plewalle@pas.rochester.edu}
\affiliation{Department of Physics and Astronomy, University of Rochester, Rochester, NY 14627, USA}
\affiliation{Center for Coherence and Quantum Optics, University of Rochester, Rochester, NY 14627, USA}
\author{Areeya Chantasri} 
\affiliation{Department of Physics and Astronomy, University of Rochester, Rochester, NY 14627, USA}
\affiliation{Center for Coherence and Quantum Optics, University of Rochester, Rochester, NY 14627, USA}
\author{Andrew N. Jordan}
\affiliation{Department of Physics and Astronomy, University of Rochester, Rochester, NY 14627, USA}
\affiliation{Center for Coherence and Quantum Optics, University of Rochester, Rochester, NY 14627, USA}
\affiliation{Institute for Quantum Studies, Chapman University, Orange, CA 92866, USA}

\date{\today}

\begin{abstract}
We examine optimal paths between initial and final states for diffusive quantum trajectories in continuously monitored pure-state qubits, obtained as extrema of a stochastic path integral. We demonstrate the possibility of ``multipaths'' in the dynamics of continuously-monitored qubit systems, wherein multiple optimal paths travel between the same pre- and post-selected states over the same time interval. Optimal paths are expressed as solutions to a Hamiltonian dynamical system. The onset of multipaths may be determined by analyzing the evolution of a Lagrangian manifold in this phase space, and is mathematically analogous to the formation of caustics in ray optics or semiclassical physics. Additionally, we develop methods for finding optimal traversal times between states, or optimal final states given an initial state and evolution time; both give insight into the measurement dynamics of continuously-monitored quantum states. We apply our methods in two systems: a qubit with two non-commuting observables measured simultaneously, and a qubit measured in one observable while subject to Rabi drive. In the two-observable case we find multipaths due to caustics, bounded by a diverging Van-Vleck determinant, and their onset time. We also find multipaths generated by paths with different ``winding numbers'' around the Bloch sphere in both systems.
\end{abstract}


\maketitle
\section{Introduction}
Continuous quantum measurement, and the accompanying theory of diffusive quantum trajectories, has become a standard research tool in quantum optics, and related fields, over the past three decades. Continuous monitoring of a quantum system introduces a noisy backaction, but provides the experimenter with a corresponding stochastic readout containing information about the system. There has been considerable research interest in this area, using technologies based on quantum electronics, cavity quantum electrodynamics (QED), and circuit QED \cite{BookWiseman,BookBarchielli,BookCarmichael,Jacobs2006,Murch2013,Jordan2013rev,Korotkov1999,Korotkov2001,Korotkov2011,Korotkov2016}. Experimental implementation of a weak, continuous measurement of a qubit is typically done, in circuit QED, with a superconducting transmon qubit dispersively coupled to a cavity, such that a homodyne or heterodyne measurement of the quadrature of the cavity field weakly measures the qubit state \cite{Murch2015teach,Gambetta2008,Jordan2015flor}.
Continuous monitoring is important for quantum feedback control applications \cite{BookWiseman,Wiseman1994,Doherty2000,Ibarcq2013,Sayrin2011,Vijay2012}, and has been shown to be useful for entanglement generation \cite{Ruskov2003-2,Trauzettel2006,Williams2008,Riste2013deterministic,Roch2014,Chantasri2016}, among other tasks of great interest for the areas of quantum computing and quantum information.  

\par {\color{black} Among the more recent theoretical work concerning diffusive quantum trajectories is the development of a stochastic path integral (SPI) formalism; extremizing the path integral allows us to compute optimal paths (OPs) \cite{Chantasri2013, Chantasri2015, Langouche1978, Dykman1994, BookKamenev}. OPs can, in principle, be most-likely paths (MLPs), least-likely paths (LLPs), or saddle-paths (SPs); in practice most OPs are MLPs, which is the case of physical interest, as discussed below.
The SPI/OP formalism is framed in terms of an initial state represented by $\mathbf{q}_i$ and a final state represented by $\mathbf{q}_f$, and gives us the optimal route(s) between these coordinates in the elapsed time $T$. Noisy quantum trajectories pre- and post-selected over corresponding boundary conditions will tend to cluster around the MLP. The MLPs for continuously monitored qubits have been shown to be in good agreement with experiment \cite{Weber2014,Chantasri2016,Mahdi2016}. After the SPI optimization procedure, OPs are smooth curves, mathematically obtained as solutions to a Hamiltonian dynamical system, generated by a ``stochastic Hamiltonian'' $H(\mathbf{q},\mathbf{p})$.} The coordinates $\mathbf{q}$ represent the quantum state (these are coordinates on the Bloch sphere if we monitor a qubit), and the conjugate variables $\mathbf{p}$ (generalized ``momenta'') can be understood as Lagrange multipliers imposing the state update in the optimization process. The OP formalism is our primary analysis tool below, and we will see that the relative computational simplicity of a Hamiltonian dynamical system, compared to stochastic differential equations, leads to many attractive features in this approach.

\par The momenta $\mathbf{p}$ are never directly assigned or measured in an experiment, but are used to advance our understanding of the physics of continuously-measured quantum systems. The SPI formalism naturally frames OPs as solutions to a boundary value problem, with boundary conditions $\mathbf{q}_i$, $\mathbf{q}_f$ and $T$, which a finite number of OPs may satisfy. 
Another approach is to set up an initial value problem, using $\mathbf{q}_i$ and $\mathbf{p}_i$, which specifies exactly one OP at any later time. Thus we may use our momenta to understand the number of paths between different states, leading to our main result: 
We predict the existence of ``multipath'' solutions, a dynamical instability in which more than one MLP meeting the boundary conditions $\mathbf{q}_i$, $\mathbf{q}_f$, $T$, appears in the quantum trajectories of qubit systems. We are not aware of any previous work studying this type of instability in continuously monitored quantum systems, although similar studies have been carried out for classical statistical systems \cite{Dykman1994-2,BookKamenev}. Finding multipaths in the OP picture is directly analogous to finding caustics in classical (ray) optics, and is more generally the type of problem dealt with in catastrophe theory \cite{BookArnoldClassical, BookArnoldCatastrophe,BookKravtsov}.
Our demonstration of multipaths' existence in physical qubit systems suggests the possibility of previously unexplored challenges for, and applications of, feedback control and error correction protocols in quantum information processing. Experimental observation of multipaths has recently been achieved in a driven fluorescing qubit, in collaboration with the Murch group at Washington University, St. Louis \cite{Mahdi2016}.
\par {\color{black} We also present results concerning the relative probabilities of different OPs generated by different initial momenta. By relaxing the constraints on either $\mathbf{q}_f$ or $T$, we are able to use the choice of initial momentum to identify the time which maximizes the probability of arriving at a chosen final state, (the time where more stochastic trajectories are passing through that final state than any other). We can also find the optimal final state at a given time; the momenta control the weighting between different OPs and the probability density for evolution to different quantum states.}
{\color{black} By combining our understanding of OPs corresponding to statistically dominant behaviors with the presence of multipaths}, this work opens the door to further explorations of chaos and/or the long-term predictability of continuously-measured quantum systems.
\par This paper is laid out according to the following scheme: In section II we introduce the mathematical tools we need, including the derivation of the stochastic Hamiltonian from the SPI, the Lagrange manifold we use to find multipath solutions, and specific results for one-dimensional Hamiltonians. We then apply these tools to two idealized, physical, systems: In section III we consider the dynamics of a qubit being simultaneously monitored along the $x$ and $z$ axes \cite{Jordan2005,Leigh2016}, and in section IV we consider a qubit being monitored along the $z$ axis while subject to a Rabi drive \cite{Chantasri2013,Weber2014}. In both cases we make a simplifying assumption of perfect measurement efficiency, such that we can restrict our analysis to pure states on a great circle of the Bloch sphere, and the phase-space of the corresponding stochastic Hamiltonian is two-dimensional. We give our conclusions in section V. 

\section{General Comments on Optimal Paths and the Role of Momenta}
We will begin by reviewing the main points of the SPI formalism we use to compute OPs, described more fully in Refs. \cite{Chantasri2013,Chantasri2015}. We will use discretized time steps, indexed by $k$. The probability density to take a given step in the quantum state space is $P(\mathbf{q}_{k+1},\mathbf{r}_k|\mathbf{q}_k) = P(\mathbf{q}_{k+1}|\mathbf{q}_k,\mathbf{r}_k) P(\mathbf{r}_k|\mathbf{q}_k)$, where the joint probability density to move between some pre- and post-selected state $\mathbf{q}_i$ and $\mathbf{q}_f$ along a path $\lbrace \mathbf{q}_0, \mathbf{q}_1,...,\mathbf{q}_n \rbrace$ is 
\be \mathcal{P} = \delta^d(\mathbf{q}_0 - \mathbf{q}_i) \delta^d(\mathbf{q}_n - \mathbf{q}_f) \prod_{k=0}^{n-1} P(\mathbf{q}_{k+1},\mathbf{r}_k|\mathbf{q}_k). \ee 
The dimensionality of the coordinate $\mathbf{q}$ parameterizing the quantum state is $d$, which is \emph{not} necessarily the same as the dimensionality of $\mathbf{r}$ which represents the number of independent measurement readouts. The readout $\mathbf{r}$ changes stochastically according to its distribution $P(\mathbf{r}_k|\mathbf{q}_k) \approx \exp[ \mathcal{G}_k [\mathbf{q}_k,\mathbf{r}_k] dt + \mathcal{O}(dt^2)]$ where $dt$ is the elapsed time between $k$ and $k+1$. Given the stochastic $\mathbf{r}_k$, the rest of the evolution can be expressed deterministically as $P(\mathbf{q}_{k+1}| \mathbf{q}_k,\mathbf{r}_k) = \delta^d(\mathbf{q}_{k+1} - \mathbf{q}_k - \mathcal{F}_k[\mathbf{q}_k,\mathbf{r}_k] dt)$ where $\mathcal{F}_k$ describes the quantum state evolution. We discuss how the functions $\mathcal{F}$ and $\mathcal{G}$ can be determined from either a stochastic master equation (SME) approach \cite{BookBarchielli,BookWiseman,Jacobs2006,BookCarmichael}, or a quantum Bayesian state update formalism \cite{Korotkov2011,Korotkov1999,Korotkov2001,Korotkov2016} in appendix \emph{A}. By substituting the above probability density relations into the expression for $\mathcal{P}$ and using the Fourier representation of all the $\delta$-functions, we may express the probability density $\mathcal{P}$ in the form of a functional integral,
\be
\mathcal{P} \propto \int \mathcal{D}[\mathbf{p}] e^S,
\ee
where $\int \mathcal{D}[\mathbf{p}] \propto \int... \int d\mathbf{p}_{-1} ... d\mathbf{p}_{n}$.
The stochastic action is 
\be
S = B + \sum_{k=0}^{n-1} (-\mathbf{p}_k \cdot(\mathbf{q}_{k+1} - \mathbf{q}_k - \mathcal{F}_k dt) + \mathcal{G}_k dt),
\ee
with a boundary term $B = - \mathbf{p}_{-1}\cdot(\mathbf{q}_0 - \mathbf{q}_i) - \mathbf{p}_n\cdot(\mathbf{q}_n - \mathbf{q}_f)$.
In association with a classical action, and in the continuum limit, this can equivalently be expressed as
\be
S = \int_0^T dt\left( -\mathbf{\dot{q}} \cdot \mathbf{p} + H(\mathbf{q},\mathbf{p},\mathbf{r})\right),
\ee 
with the stochastic Hamiltonian \be\label{spif_h} 
H= \mathbf{p} \cdot \mathcal{F}[\mathbf{q},\mathbf{r}] + \mathcal{G}[\mathbf{q},\mathbf{r}], 
\ee
and implicit boundary conditions $\mathbf{q}(t=0)  = \mathbf{q}_0= \mathbf{q}_i$ and $\mathbf{q}(t=T) = \mathbf{q}_T= \mathbf{q}_f$. {\color{black} In our subsequent notation we will use $\mathbf{q}_f$ when we constrain the final state, and $\mathbf{q}_T$ when we refer to a final state obtained by fixing $T$. A complete post-selection consists of choosing both $q_f$ and $T$, but certain problems we solve below will require us to fix only one at a time.}
\par A dynamical system of equations for the OPs are derived through a least action principle, optimizing the readout(s) $\mathbf{r}$ such that $\delta S = 0$. The action is approximated as Gaussian in the readout variables (see appendix \emph{A}), in which case we may equivalently integrate out the measurement results $\mathbf{r}$. The probability density associated with an OP goes as $\mathcal{P} \sim e^S$, hence extremizing the action extremizes the probabilities, giving us OPs, which may be MLPs, LLPs, or SPs. (Note that by neglecting pre-factors in $\mathcal{P}\sim e^S$, we are effectively writing this path probability in a small-noise limit.) When there is only one OP for some boundary conditions, it will always be a MLP, not a LLP or SP. OPs derived from $\delta S = 0$ satisfy Hamilton's equations $-\partial_\mathbf{q} H = \dot{\mathbf{p}}$ and $\partial_\mathbf{p} H = \dot{\mathbf{q}}$, and the constraint $\partial_\mathbf{r} H = 0$ \cite{Chantasri2013}. Notice that this implies that these OPs satisfy $\dot{\mathbf{q}} = \mathcal{F}$, meaning that they are themselves possible quantum trajectories.

\subsection*{Multipaths and the Lagrange Manifold}
To formally discuss multipaths, we must begin by noting that their existence actually depends on the momenta $\mathbf{p}$ being un-observable. It is impossible to get paths which cross in a complete Hamiltonian phase-space defined by $(\mathbf{q},\mathbf{p})$ \cite{BookStrogatz}, but we are physically only directly interested in the projection of OPs into the $\mathbf{q}$-space, where they can cross. When we apply the OP formalism to a continuously monitored qubit where the coordinate-space is the Bloch sphere, trajectories in the 6-dimensional phase space \emph{cannot} cross, but their projection down into the 3-dimensional Bloch sphere \emph{can}. It is these crossings which we are searching for when we investigate multipaths.
\par We introduce a mathematical object we call the Lagrange Manifold, which is a conceptually elegant way of understanding how multipaths appear in Hamiltonian systems \cite{AlonsoChat2016, ChapterAlonso,BookArnoldClassical,BookMaslov,Littlejohn1992,Dykman1994-2}. For an $N$-dimensional Hamiltonian ($2N$-dimensional phase space), we consider a specific $N$-dimensional manifold defined by \emph{all} possible $\mathbf{p}_i$ for a fixed $\mathbf{q}_i$. For $N=1$, this means that we initialize our manifold as a vertical line in the phase portrait ($p$ vs. $q$) at $t=0$; evolving the system forward allows that manifold to deform as all of the $p_i$ forming the curve generate different paths. If we can draw a vertical line of all $p$ and some $q_T$ which intersects the manifold more than once (the manifold, drawn as a function of $q_T$, fails the vertical line test), then at least two values of $p_i$ generated paths originating at $q_i$ which will arrive at the same $q_T$. Multipaths, in short, occur when the manifold described above cannot be projected \emph{injectively} onto $q$-space. We may use numerically-generated plots of the manifold to predict the onset of multipaths. Explicit examples are shown in sections III and IV (\emph{e.g.} Fig.~\ref{fig-LMonset}).
\par We may examine the point(s) at which the manifold folds over itself more closely. In a one dimensional system (two-dimensional phase space), we consider the Jacobian $J$ transforming the initial momenta $p_i$ into a final coordinate $q_f$ (or $q_T$):
\be \label{jac1d}
J = \partl{q_f}{p_i}{}.
\ee
$J$ is necessarily zero at the point where new multipaths are forming, \emph{i.e.}~where the mapping from $p_i$ to $q_f$ is not invertible. We note that $J^{-1}$ is related to the slope of the Lagrange Manifold in phase-space. Consider the inverse of $|J|$,
\be \label{vvd1}
V = \left|\partl{p_i}{q_f}{} \right| =  \left|\frac{\partial^2 S}{\partial q_f \partial q_i}\right|.
\ee
The quantity $V$ is a one-dimensional version of the Van-Vleck determinant found in classical and semi-classical physics \cite{VanVleck1928,BookMaslov,Littlejohn1992,ChaosBook37}, and necessarily diverges when the slope of the Lagrange manifold is infinite. This makes $V$ a useful quantity for finding the onset of multipaths, as has been noted in the literature \cite{ChaosBook37,BookArnoldClassical,Littlejohn1992}. Generalizing to higher dimensional phase-spaces, we have a matrix
\be
\mathtt{J}_{jk} = \partl{q_{f,j}}{p_{i,k}}{} ,
\ee
($i$ and $f$ still denote initial and final, whereas $j$ and $k$ index spatial coordinates), and define the Van-Vleck determinant (VVD)
\be \label{vvd2}
\mathtt{V} = \det(\mathtt{J}^{-1}) = \det \left( \frac{\partial^2 S}{\partial \mathbf{q}_f \partial \mathbf{q}_i} \right),
\ee
which diverges where new multipaths are forming. Points (or curves, in higher-dimensional systems) along which $V$ diverges define the boundary of a caustic region. These regions are so named in reference to their manifestation in optics, where many rays of light cluster or focus \cite{BookKravtsov}. Given an initial state and manifold, there are by definition several OPs which lead to any final position within a caustic region, meaning we have multipaths there.

\subsection*{One-Dimensional Systems}
\par We here impose some simplifications which reflect the physical systems we will analyze in sections III and IV. Let us suppose that we have a one-dimensional Hamiltonian system (two-dimensional phase space), defined by $H(q,p)$. We assume that our $H$ has no explicit time dependence, such that the ``stochastic energy'' $E = H$ is a conserved quantity; then the initial values of $q$ and $p$ ($q_i$ and $p_i$) determine $E$ of an OP for all time, and either $p_i$ or $E$ can be regarded as a degree of freedom. In such a case it is always possible to solve for a function $p(q,E)$; these curves can be plotted directly for many $E$ to construct the phase portrait. The stochastic action, which is related to the probability density for different paths, reads
\be \label{2dps_action}
S = ET - \int_0^T \dot{q} p(q,E) dt = ET - \int_{q_i}^{q_f \: \text{or} \: q_T} p(q,E) dq.
\ee
\par We reduce the range of possibilities implied by \eqref{spif_h} which we explore in one dimension, by imposing further assumptions. To begin, suppose (i) that for one-dimensional $\mathbf{q}\rightarrow q$ we may decompose $\mathcal{F}$ into a sum of terms $f_i$ which each depend on only one readout channel $r_i$, such that
\be \label{hgen0}
H = p\left( \sum_i f_i(q,r_i) \right) + g(q,\mathbf{r}),
\ee
and that (ii) each $f_i$ is linear in its $r_i$, such that 
\be
f_i(q,r_i) = \alpha_i(q) + r_i \beta_i(q).
\ee
We have let $\mathcal{G}[\mathbf{q},\mathbf{r}] \rightarrow g(q,\mathbf{r})$ for our one-dimensional problem. We impose no particular constraints on the functions $\alpha$ and $\beta$, other than that they be continuously differentiable everywhere except a finite number of singular points in the physically appropriate domain for $q$.
We use the same measurement model as in Ref. \cite{Chantasri2013} (see Appendix \emph{A} as well), wherein we may write
\be\label{assum_iii}
g(q,\mathbf{r}) = -\sum_i \frac{r_i^2-2 r_i \gamma_i(q)+1}{2\tau_i},
\ee
which we interpret below as a ``cost function''. Here $\tau_i$ is the characteristic measurement time in the $i^{th}$ measurement channel, and $\gamma_i$ is some function of $q$ ($\gamma$ will be sinusoidal for the physical systems we analyze later). We take the form \eqref{assum_iii} to be a third assumption (iii).
The optimal readout(s) $r_i^\star$ is (are) obtained by solving $\partial_{r_i} H = 0$, which means $r_i^\star = \gamma_i + p \tau_i \beta_i.$
We see that assumption (ii) forces the readout to be linear in $p$. Substituting this result back into \eqref{hgen0} leads to a Hamiltonian which is necessarily quadratic in $p$, specifically
\be \label{hgen2}
H = p^2 a(q) + p b(q) + c(q),
\ee
for $a(q) = \sum_i \tau_i \beta_i^2/2$, $b(q) = \sum_i(\alpha_i - \gamma_i \beta_i)$, and $c(q) = \sum_i (\gamma_i^2-1)/2\tau_i$. Note that this makes $a(q) \geq 0$ for all $q$.
We could reach \eqref{hgen2} directly by integrating out the readout(s) $\mathbf{r}$ in the SPI, because $e^g$ is Gaussian in $\mathbf{r}$.
\par A special case occurs when the condition 
\be \label{cond_iv}
\frac{\gamma_i^2-1}{2\tau_i} = - \frac{\tau_i \beta_i^2}{2} \:\text{ for all } \: i 
\ee
is met, \emph{i.e.} $c = - a$. Then \eqref{hgen2} reduces to
\be \label{hgen3}
H = (p^2-1)a(q) + p b(q).
\ee
It is straightforward to show that both of the systems treated in sections III and IV satisfy \eqref{cond_iv} and \eqref{hgen3}, which we subsequently refer to as assumption (iv). For qubit systems where $q$ is an angle $\theta$ on the Bloch sphere (this means we only consider pure states), the geometry of the unit circle forces $b$ and $c$ to be sines or cosines when the measurements are orthogonal or along a convenient axis, and then the identity $\sin^2\theta + \cos^2 \theta = 1$ leads to \eqref{cond_iv} being satisfied. 
\par Even without assumption (iv) however, \eqref{hgen2} will always give
\be \begin{split} \label{dotq_form}
\dot{q} = \partl{H}{p}{} &= 2 p a(q)+ b(q) \\
&= \pm \sqrt{4a^2 + 4 a E + b^2},
\end{split}\ee
so that at a given $q$, the velocities $\dot{q}$ of OPs scale linearly with $p$. In the second line of \eqref{dotq_form} we have substituted in the solutions of \eqref{hgen3} with $H=E$ (since $E$ is a conserved quantity)
\be \label{pqE}
p_\pm(q,E) = - \frac{b}{2a} \pm \sqrt{1 + \frac{E}{a} + \frac{b^2}{4a^2}}.
\ee 
Qualitatively, \eqref{dotq_form} shows that for non-singular $a$ and $b$, {\color{black} and a $q$ representing an angular variable on the Bloch sphere or similar,} we have clockwise-rotating paths far into the upper part of the phase portrait (large positive $p$), and counterclockwise-rotating paths far into the lower part of the phase portrait (large $|p|$ and negative $p$). The ends of our Lagrange manifold will be pulled in opposite directions, which in turn implies that any folding in the manifold will generate even numbers of points with diverging VVD, and odd numbers of OPs meeting boundary conditions to form multipaths in caustic regions {\color{black} (at least two of which will be MLPs)}. Velocities increase monotonically as a function of $E$ (and $E$ is the only quantity inside the square root which may be negative). It follows that in the systems we are studying, outside of either periodic islands or otherwise asymptotically bounded regions of phase-space, velocities do not change direction and the Lagrange manifold cannot fold over itself.
The function $q(t)$ cannot be inverted uniquely into $t(q)$ for positions and times where multipaths due to a true caustic exist (\emph{i.e.} where the manifold has folded into a caustic bounded by a diverging Van-Vleck determinant).
\par We also examine the integrand of the action $\dot{S}$. $S$ determines the relative probability density for OPs, since any additive constants which do not depend on $p$ or $q$ fall off when we take ratios of probability densities $\mathcal{P}_1/\mathcal{P}_2 = e^{S_1 - S_2}$ between some paths 1 and 2. $\dot{S}$ therefore contains information about the rate at which the relative probability density changes between paths.  For optimal paths where $\dot{q} = \mathcal{F}$ we have
\be
\dot{S} =-p \dot{q} +H =-p \dot{q} + p \mathcal{F} + g = g(q,\mathbf{r}),
\ee or, with the optimal readouts substituted in or integrated out,
\be \label{dots_form}
\dot{S} = - a p^2 + c \;\; \text{(i-iii)},\quad \dot{S} = -(1+p^2)a \;\; \text{(i-iv)}.
\ee
We have noted that $a(q) \geq 0 \: \forall \: q$, which means that the system under assumptions (i-iv) necessarily has $\dot{S} \leq 0$ everywhere in phase space. Paths which traverse regions of phase-space with more negative values of $\dot{S}$ incur a higher loss to their probability density per unit time, and hence we may regard $\dot{S}$ as a cost-function for the probability. Note that $\dot{S}\rightarrow -\infty$ as $p\rightarrow \pm\infty$ except at values of $q$ where $a(q) = 0$; we conclude that paths which travel infinitely fast also have a vanishingly small probability density to actually appear. 
{\color{black} Generally, regions of the OP phase-portrait where $\dot{S}$ is very negative describe behaviors which relatively few stochastic trajectories exhibit, as compared to behaviors described by regions where $\dot{S}$ is closer to zero. Thus, the combination of the phase-portrait and $\dot{S}$-portrait summarize both the possible behaviors of the underlying trajectories, and the relative frequency with which those behaviors occur.}

\subsection*{Extremal-Probabilities within Optimal Paths}
When we continuously monitor an ensemble of qubits prepared in the same state, not all final states have equal probability density at any given later time. In the OP formalism, fixing $q_i$, $q_f$, and $T$ completely constrains us to a finite number of optimal solutions. If we constrain either the final state or elapsed time, however, we still have an infinite number of possible solutions available, defined by at least some subset of possible values of $p_i$. 
We consider the OPs in two different cases: (A) We can fix $q_i$ and $q_f$ but allow the evolution time $T$ to vary, and (B) we can fix $q_i$ and $T$ while putting no constraints on the final $q_T$. Finding an optimal solution under case (A) is equivalent to asking what the most-probable evolution time between the specified $q_i$ and $q_f$ is, whereas in case (B) we are asking what the most-likely final state is after some time. Note that by interpreting all of these results with an open final boundary condition in terms of a probability density, we assume a small noise approximation where $\mathcal{P} \sim e^S$.

\subsubsection*{Case A: Optimal Traversal Times}
\par We search for the most-probable path with $q_i$ and $q_f$ fixed, but variable evolution time $T$. Assuming that $q(t)$ is invertible (valid in regions without caustics, or for times short enough that one has not formed yet), we may write
\be \label{timedef}
T = \int_0^T dt = \int_{q_i}^{q_f} \frac{dt}{dq} dq = \int_{q_i}^{q_f} \partl{p}{E}{} dq,
\ee
which can then be substituted into the action \eqref{2dps_action} to give
\be
S = - \int_{q_i}^{q_f} p(q,E)dq + E \int_{q_i}^{q_f} \partl{p}{E}{} dq.
\ee
The path with maximum probability density is the one which maximizes the action (because $\mathcal{P} \sim e^S$). Consider
\be \label{zeroE}
\partl{S}{E}{} = E \int_{q_i}^{q_f} \partl{p}{E}{2} dq= E \partl{T}{E}{} \bigg|_{q_i}^{q_f} = 0.
\ee
We infer from this that the optimal solution between $q_i$ and $q_f$ must satisfy either $E = 0$ or $\partial_E T|_{q_i}^{q_f}=0$. 
\par It turns out that only the condition $E=0$ is useful. We note that the form \eqref{hgen3} (\emph{i.e.} assumptions (i-iv)) implies that $p(q,E)$ takes the form \eqref{pqE},
and recall that $a(q) \geq 0 \: \forall \: q$.
We may consider the consequences this has for the traversal time \eqref{timedef} between $q_i$ and $q_f$ by substituting in \eqref{pqE}, to get
\be \label{time2}
T = \pm \int_{q_i}^{q_f} \frac{dq}{\sqrt{4 a^2 + 4 a E + b^2}},
\ee
where the $\pm$ is chosen to give a positive time for the desired boundary conditions.
This leads to
\be\label{dTdE}
\partl{T}{E}{}\bigg|_{q_i}^{q_f} = \mp \int_{q_i}^{q_f} \frac{2 a dq}{(4 a^2 + 4 a E + b^2)^\frac{3}{2}}.
\ee
The choices of $E$ are limited to values such that $T$ is always real, \emph{i.e.} such that $4a^2 + 4 a E + b^2$ is a positive number (we work out a specific example of this following from \eqref{pqEsqrt}). But then with $a(q) \geq 0 \: \forall \: q$, the integrand of \eqref{dTdE} is necessarily also always positive for all $q$. If the integrand can never change sign, then the only way to obtain $\partial_E T|_{q_i}^{q_f} = 0$ is by the trivial choice $q_i = q_f$, and for a physically meaningful optimization of the traversal time, we must take $E=0$ to be our solution. Effectively, we have shown that we may relax the boundary condition in $T$, and then optimize over that degree of freedom, thereby finding that the $E=0$ path moves between boundary conditions $q_i$ and $q_f$ in an optimal traversal time {\color{black} \footnote{ We consider the sub-ensemble of trajectories that pass through $q_f$, and with an emphasis on \emph{when} they pass through that desired $q_f$; more trajectories are at $q_f$ at the time the $E=0$ path reaches $q_f$ than at any other time. }}; the result is valid anywhere outside of a caustic in phase space. An example is discussed in section IV (see Fig.~\ref{fig-act}).

\subsubsection*{Case B: Optimal Final States}
\par We now fix the evolution time, and ask which $q_T$ is the most probable given $q_i$ and $T$. As above, we optimize the action, but this time we do it with respect to $q_T$ rather than $E$. It is useful to recast the stochastic action $S(q_T,p_T)$ as a generating function $S_G(q_T,q_i)$, (closely related to Hamilton's principle function \cite{ChaosBook37}), which is valid when the initial and final coordinates can be computed from each other via canonical transformation. Under these conditions, we have \cite{BookArnoldClassical,BookTabor,Clineder,Littlejohn1992}:
\be\label{sg}
\partl{S_G}{q_i}{} = -p_i \quad\text{and}\quad \partl{S_G}{q_T}{} = p_T.
\ee 
But then the condition which extremizes the action of an OP over $q_T$ is simply that $p_T = 0$ at the desired time $T$. 

\subsection*{Discussion: Role of Momenta}
The momenta in the SPI/OP formalism, while unphysical and unobservable, are key to understanding very real physics in diffusive quantum trajectories in two ways. First, a range of initial momenta can be used to define a Lagrange manifold, which may fold into catastrophes / caustics corresponding to the presence of multipaths in the quantum trajectories. Second, we have defined a function $\dot{S}$ which describes the ``probability cost'' of traveling through different points of the OP phase space, and depends sensitively on our momenta. Thus, our momenta also play a role in understanding the likelihood of different measurement outcomes in the physical system. 
We will see below that in combination with the optimizations described in \eqref{zeroE} {\color{black} ($E=0$ gives optimal $T$ between $q_i$ and $q_f$)} and \eqref{sg} {\color{black} ($p_T = 0$ corresponds to an extremal probability in $q_T$)}, we have developed a powerful and computationally simple tool for understanding dominant long-term behaviors in continuously monitored qubit systems. 

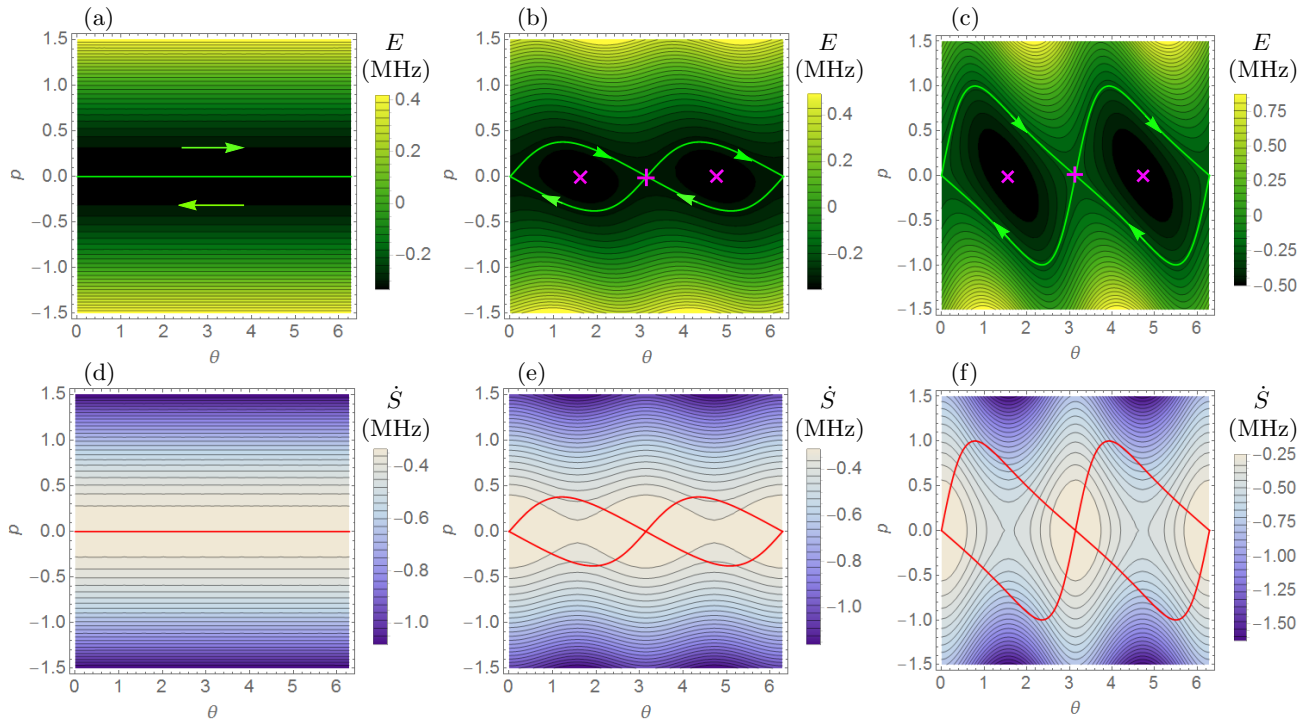
\begin{figure*}
\begin{tabular}{ccc}
\begin{picture}(100,100) \put(-100,-45){\includegraphics[width=0.32\textwidth]{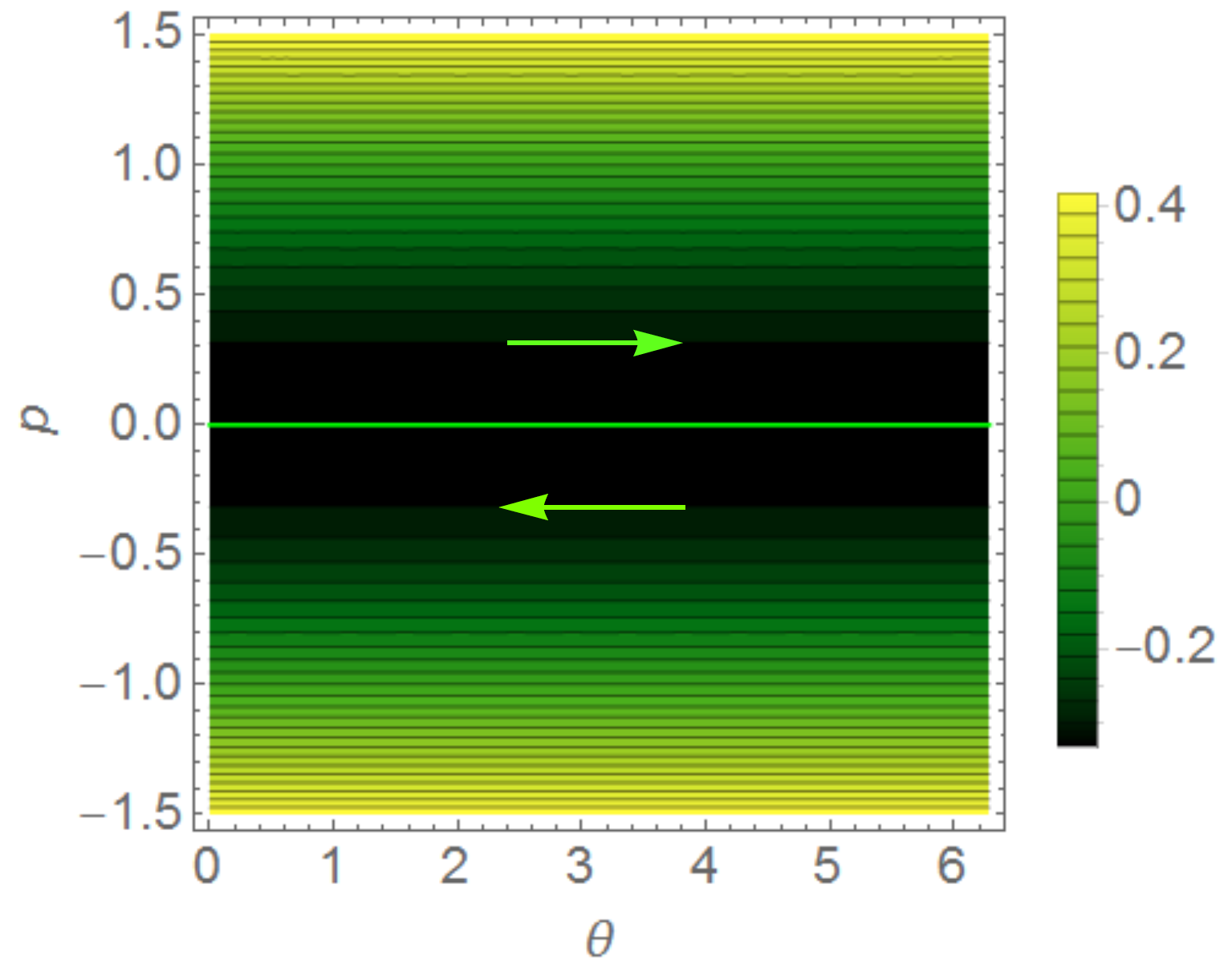}} \put(35,70){\begin{minipage}{0.05\textwidth} $E$ \\ (MHz) \end{minipage}} \put(-70,85){(a)}\end{picture} & 
\begin{picture}(100,100) \put(-40,-45){\includegraphics[width=0.32\textwidth]{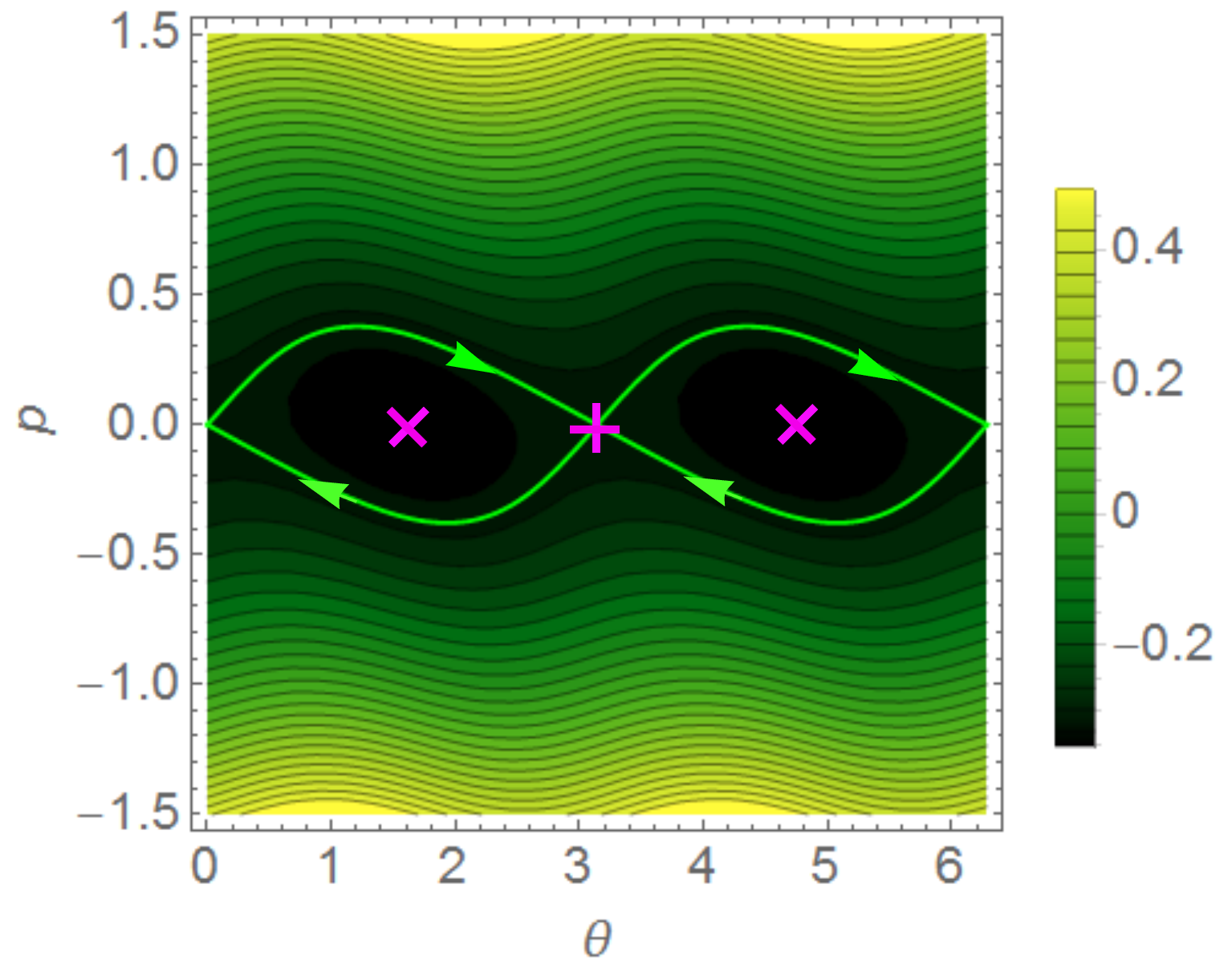}} \put(95,70){\begin{minipage}{0.05\textwidth} $E$ \\ (MHz) \end{minipage}} \put(-10,85){(b)}\end{picture} &  
\begin{picture}(100,100) \put(20,-43){\includegraphics[width=0.32\textwidth]{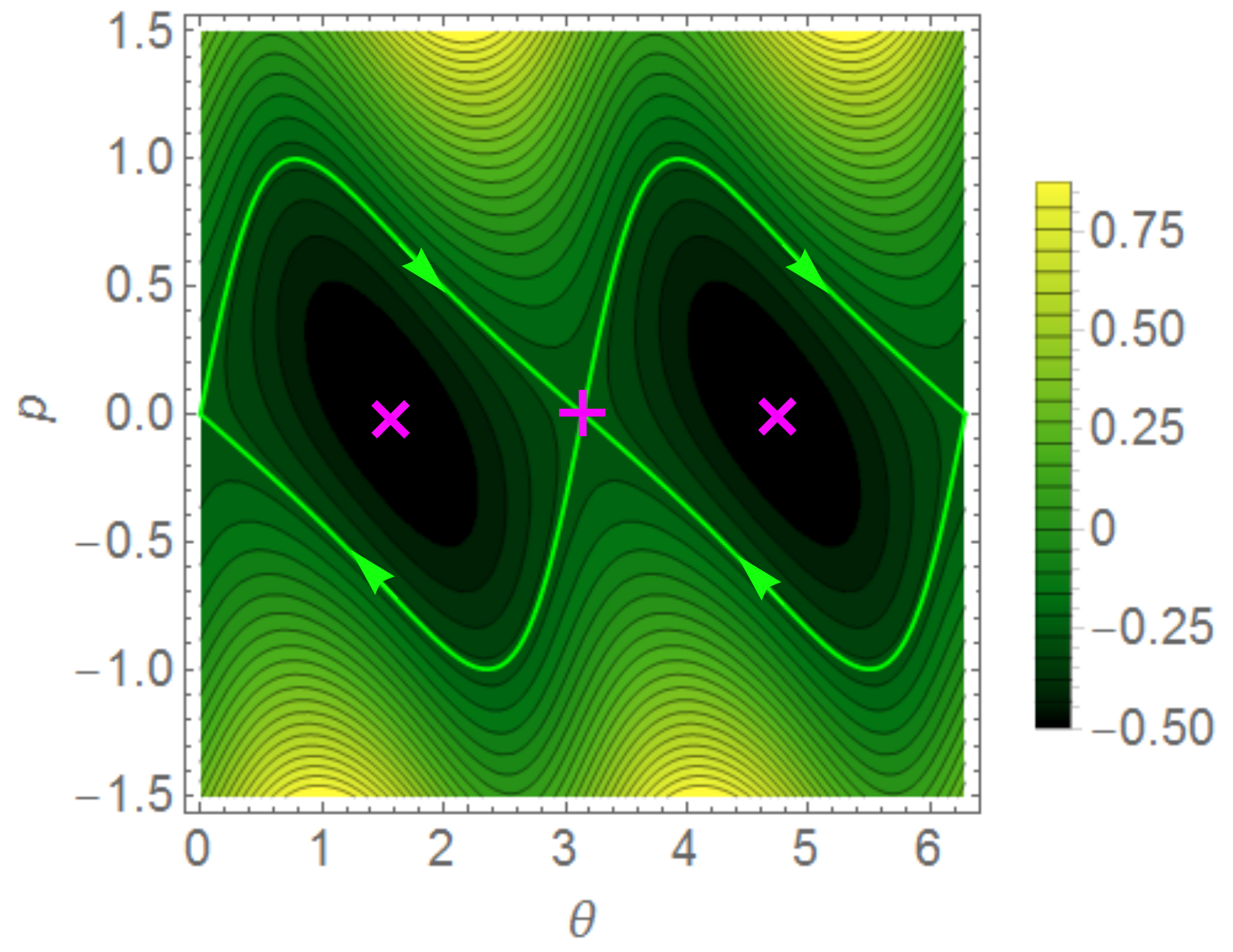}} \put(155,70){\begin{minipage}{0.05\textwidth} $E$ \\ (MHz) \end{minipage}} \put(50,85){(c)} \end{picture} \\ & & \\ & & \\ & & \\
\begin{picture}(100,100) \put(-100,-45){\includegraphics[width=0.32\textwidth]{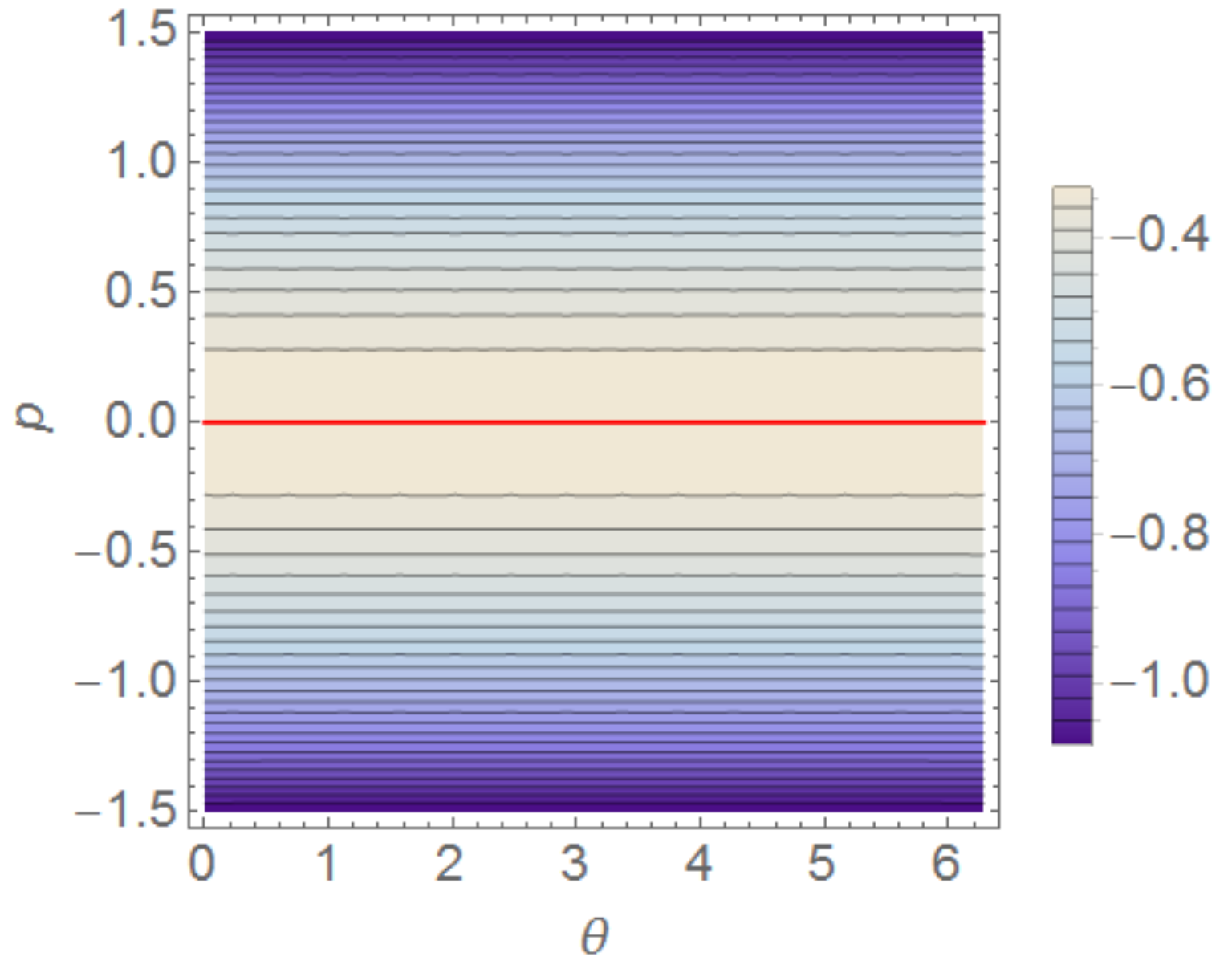}} \put(35,70){\begin{minipage}{0.05\textwidth} $\dot{S}$ \\ (MHz) \end{minipage}} \put(-70,85){(d)}\end{picture} & 
\begin{picture}(100,100) \put(-40,-45){\includegraphics[width=0.32\textwidth]{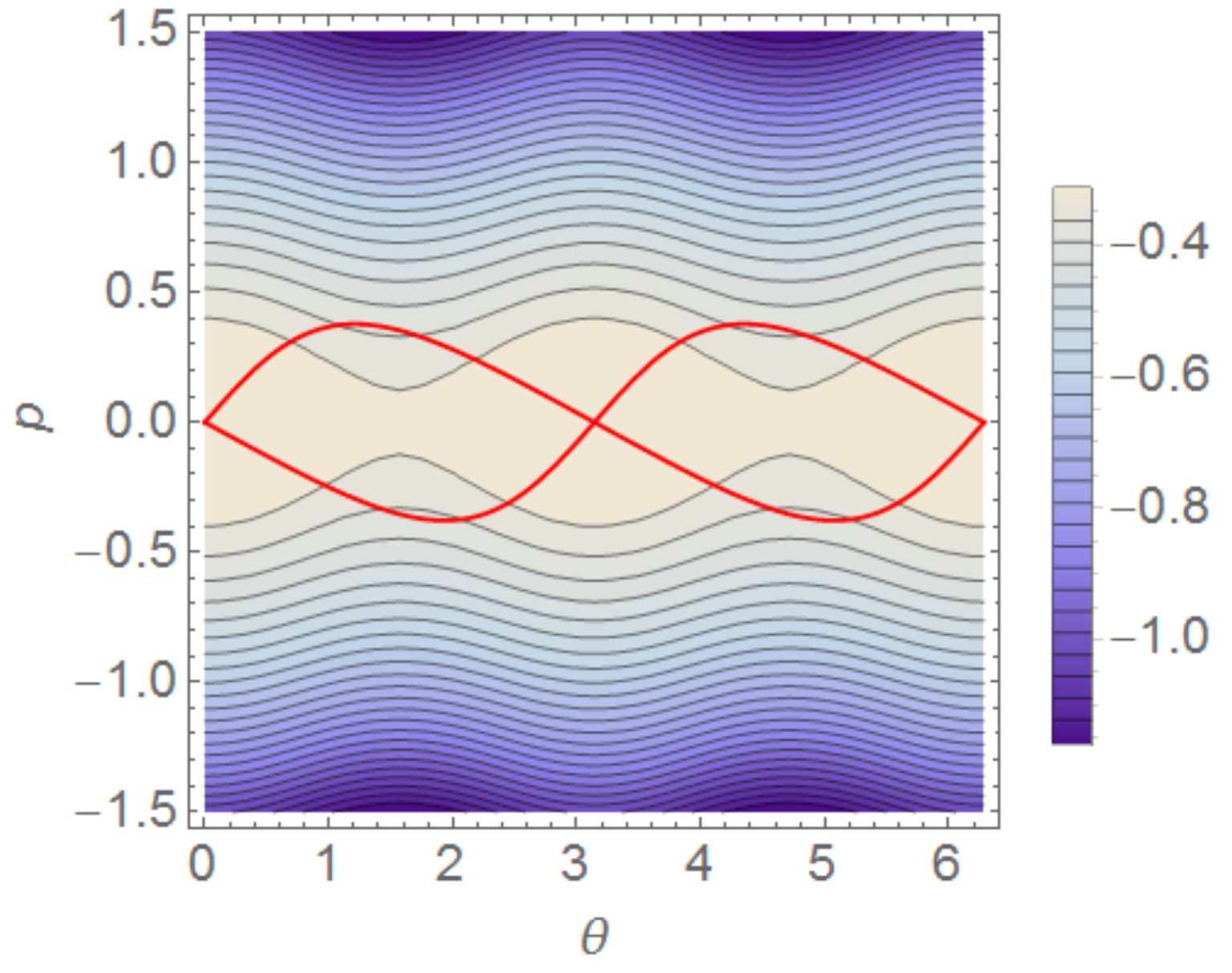}} \put(95,70){\begin{minipage}{0.05\textwidth} $\dot{S}$ \\ (MHz) \end{minipage}} \put(-10,85){(e)}\end{picture} &  
\begin{picture}(100,100) \put(20,-43){\includegraphics[width=0.32\textwidth]{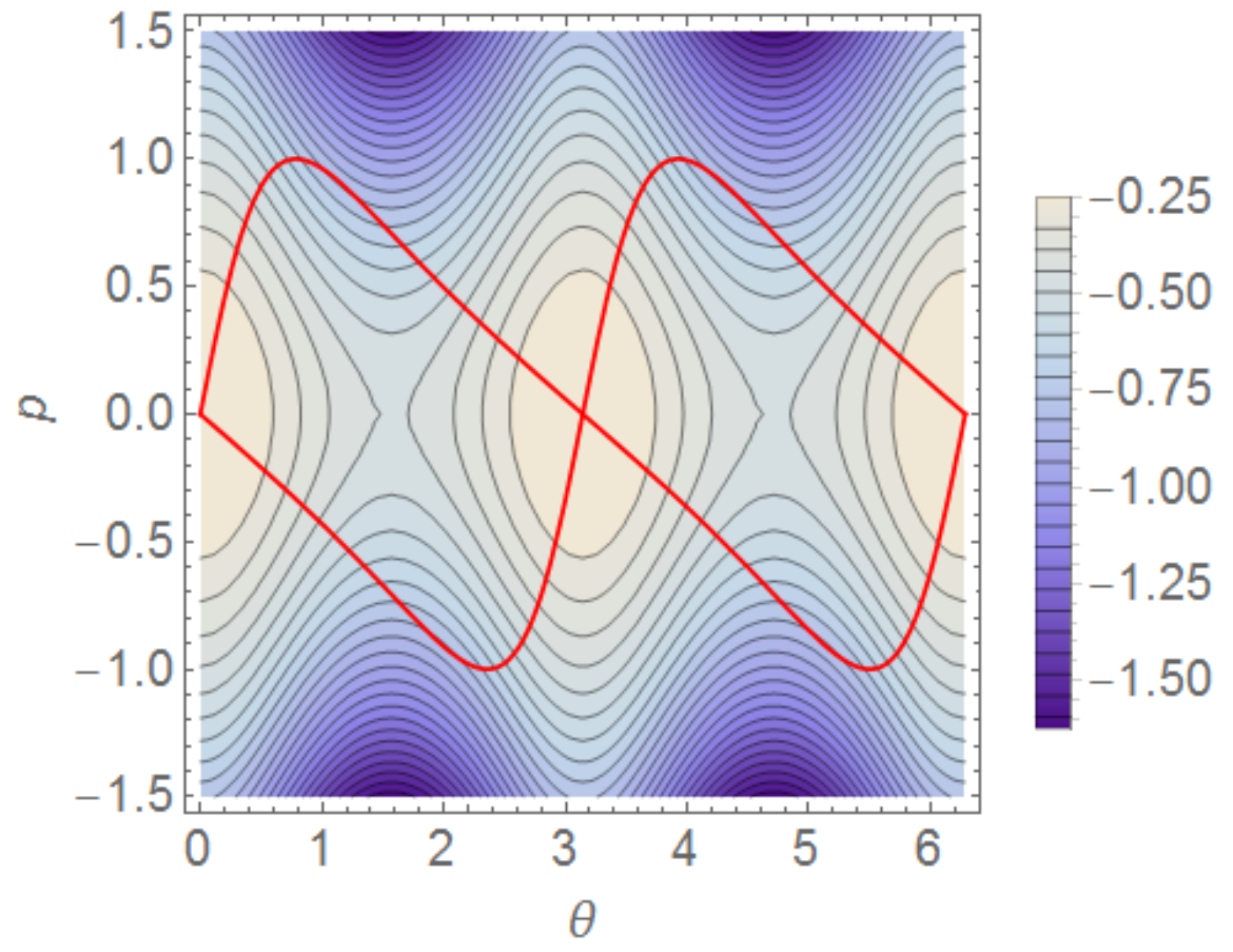}} \put(155,70){\begin{minipage}{0.05\textwidth} $\dot{S}$ \\ (MHz) \end{minipage}} \put(50,85){(f)} \end{picture} \\ & & \\ & & \\ & & \\ & & \\ \end{tabular}
\caption{The phase portrait corresponding to the simultaneous $x$ and $z$ measurement scheme with stochastic Hamiltonian \eqref{hr} is pictured in (a,b,c). $\dot{S}$, given in \eqref{xz_sdot}, is shown in (d,e,f). We have $\tau_z = 1.5 \mu s = \tau_x$ in (a,d), $\tau_z = 1.4 \mu s$ and $\tau_x = 1.6 \mu s$ in (b,e), and $\tau_z = 1 \mu s$ and $\tau_x = 2 \mu s$ in (c,f). $E$ and $\dot{S}$ are in units inverse to those of the $\tau$ values, and are therefore in MHz for times in $\mu s$. (We choose $\mu s$ here and in future figures because that would be typical for an experiment performed with  superconducting transmon devices \cite{Leigh2016,Mahdi2016}). Contour color denotes stochastic energy in (a,b,c), or the value of $\dot{S}$ in (d,e,f), and the green (a,b,c) or red (d,e,f) curve is the separatrix/critical line with energy $E_c$ (see Table 1). Lines of constant stochastic energy in (a,b,c) are the trajectories $p(\theta,E)$. Arrows indicate the direction of Hamiltonian flow. Pink markings in (b) and (c) are fixed points; those marked with $\times$ are elliptic, and those marked with $+$ are hyperbolic/saddle points. The elliptic points $\times$ sit at the eigenstate of the weaker measurement operator, and the hyperbolic points $+$ sit at the eigenstate of the stronger measurement operator. Note that both sets of fixed points sit along the $p=0$ line. For times appreciably longer than the stronger (shorter) $\tau$ in the system, the most-likely overall outcome is collapse toward the nearest $+$ point along the critical line, where $\dot{S}$ is the largest.
}
\label{fig-xz_ps}
\end{figure*}

\section{Simultaneous Continuous Measurement of Two Non-Commuting Observables}
We now proceed to specific examples of the behaviors described above. We first consider a qubit subject to simultaneous continuous weak measurement along $\sigma_x$ and $\sigma_z$ \cite{Jordan2005}, which has been implemented experimentally using a superconducting transmon qubit {\color{black} by the Siddiqi group at UC Berkeley \cite{Leigh2016}}. Two non-commuting measurements tend to compete, since they push a qubit towards different eigenstates. When the measurements are approximately equally strong, the competition between measurements prevents collapse to either set of eigenstates, and results instead in persistent diffusion \cite{Leigh2016}. Below we show that this is consistent with the OP picture, and that detailed insight into the probabilities and dynamics of collapse toward either set of eigenstates may be obtained using \eqref{sg} when measurement strengths along different observables are unequal. We then demonstrate that the OP picture predicts the presence of multipaths in this system.

\subsection*{Stochastic Hamiltonian and Dynamics}
\par We construct the stochastic Hamiltonian from the SPI, used to compute OPs. A Bayesian approach \cite{Korotkov2011,Korotkov1999,Korotkov2001,Korotkov2016} is used to derive the terms $\mathcal{F}$ and $\mathcal{G}$ in the stochastic Hamiltonian for the OPs, as described in detail in appendix \emph{A} and Refs. \cite{Chantasri2013,Chantasri2015}. The main results are the cost function
\be \label{G_xzmeas}
\mathcal{G} =  -\frac{r_x^2-2r_x x +1}{2 \tau_x} - \frac{r_z^2 - 2 r_z z + 1}{2\tau_z},
\ee
and the equations of motion
\be\label{strato_xz} \begin{split}
 \dot{x} &= f_1(\mathbf{q},r_x,r_z) =  \frac{\left(1 - x^2 \right) r_x}{\tau _x} -\frac{x z r_z }{\tau _z},\\
\dot{y} &= f_2(\mathbf{q},r_x,r_z) = -y \left(\frac{z r_z }{\tau _z}+\frac{x r_x}{\tau _x}\right), \\
\dot{z} &= f_3(\mathbf{q},r_x,r_z) = \frac{\left(1-z^2\right) r_z}{\tau _z}-\frac{x z r_x}{\tau _x}.
\end{split} \ee
The variables $x$, $y$, and $z$ are qubit coordinates in the Bloch sphere. The readout from the $x$ and $z$ measurements, respectively, are given by $r_x$ and $r_z$, and $\tau_x$ and $\tau_z$ are the characteristic measurement time for each measurement (a larger characteristic time specifies a weaker measurement, which takes longer to distinguish between measurement eigenstates). The stochastic Hamiltonian for the OPs can then be constructed according to
$H = \mathbf{p}\cdot \mathcal{F}[x,y,z,r_x,r_z] + \mathcal{G}[x,y,z,r_x,r_z], $
where $\mathcal{F}$ is now the vector of $(f_1, f_2, f_3)$ given in \eqref{strato_xz}. 
The optimal readout must satisfy the system of equations $\partial_{r_i} H = 0$, which provides the constraints $r_x^\star = x + p_x(1-x^2) - xyp_y-xzp_z$ and $r_z^\star = z +p_z(1-z^2) - xzp_x - yzp_y$. For the choices $y = 0$ and $p_y = 0$, we find that $\dot{y}=0$ and $\dot{p}_y = 0$; therefore we may choose initial conditions in the $xz$-plane of the Bloch sphere, and work entirely within that plane. If we further restrict our initial states to be pure states (on the edge of the sphere) and assume perfect measurement efficiency, our OPs will be constrained to stay on the great circle of the Bloch sphere in the $xz$-plane. They can be parameterized entirely by the polar angle $\theta$, and momentum $p$ conjugate to $\theta$, by applying the substitution $z = \cos \theta$, $x = \sin \theta$, and $ p = p_x \cos \theta - p_z \sin \theta$. This results in the stochastic Hamiltonian
\begin{equation}\label{hr}\begin{split} H = &p\left( \frac{r_x}{\tau_x} \cos\theta - \frac{r_z}{\tau_z} \sin\theta \right) - \frac{r_z^2 - 2 r_z \cos\theta + 1}{2\tau_z} \\& -\frac{r_x^2-2r_x \sin\theta+1}{2 \tau_x}. \end{split}\end{equation}
After substituting in $r_x^\star = \sin\theta + p \cos\theta$ and $r_z^\star = \cos\theta - p \sin\theta$ (or, equivalently, integrating them both out), we obtain the form \eqref{hgen3} $H = a(\theta) (p^2-1) + b(\theta)$ with
\begin{equation}\label{xz_ab} a \equiv \frac{\sin^2\theta}{2\tau_z}+ \frac{\cos^2\theta}{2\tau_x} ,\quad b \equiv \sin\theta \cos\theta \left( \frac{1}{\tau_x} - \frac{1}{\tau_z} \right).  \end{equation}
As discussed in section II, the log probability density term $g(\theta,\mathbf{r}^\star) = - (p^2+1) a(\theta)$ (\eqref{G_xzmeas}, or the last two terms in \eqref{hr}) reads
\be\begin{split} \label{xz_sdot}
\dot{S}
= H-p\dot{\theta}= - \frac{(p^2+1)(\tau_z \cos^2\theta + \tau_x \sin^2\theta)}{2\tau_x \tau_z},
\end{split}\ee
which expresses the probability cost function in terms of $\theta$, $p$, and characteristic measurement times.
Both the phase portrait and contour plot of $\dot{S}$ are shown in Fig.~\ref{fig-xz_ps}. 

\subsubsection*{Special Case: Equal-Strength Measurements}
\par In the case where $\tau_z = \tau = \tau_x$, the dynamics of the OPs are that of a simple rotor $H = E = (p^2-1)/2\tau$, where higher values of $E$ or $|p|$ result in a faster rotation around the Bloch sphere (see Fig.~\ref{fig-xz_ps}(a,d)). {\color{black} We show that this is consistent with the experimental observation by Hacohen-Gourgy, Martin, et al. \cite{Leigh2016}, of persistent diffusion.} A fixed point in a dynamical system is defined by a pair $(\bar{\theta},\bar{p})$ satisfying $\dot{\theta}(\bar{\theta},\bar{p}) = 0$ and $\dot{p} (\bar{\theta},\bar{p})=0$. The fixed points in the equations of motion ($\dot{p} = -\partial_\theta H = 0$ and $\dot{\theta} = \partial_p H = p/\tau$) in this $\tau_x = \tau_z$ case are the entire $p=0$ line. We may combine this finding with \eqref{sg} to understand the dominant system dynamics purely by reading Fig.~\ref{fig-xz_ps}(a,d). The most-likely final state $\theta_T$ has a path with $p_T=0$, but in this case the paths which satisfy that condition have $p_i =0$ as well, and stay still; the most-likely final state is the same as the initial state. From the plot of $\dot{S}$ (Fig.~\ref{fig-xz_ps}(d)) we see that the likelihood of different events after some elapsed time corresponds directly to the initial $p$ or $E$; probability densities fall off as $|p|$ grows. It is straightforward to translate Fig.~\ref{fig-xz_ps}(a,d) into an equivalent probability distribution $P(\theta|\theta_i)$ of stochastic trajectories / quantum states. We begin with a $\delta$-function around $\theta=\theta_i$, which spreads into a Gaussian $P \sim e^{S} = e^{\int \dot{S} dt}$ for $\dot{S} = -(1+p^2)/2\tau$ over time (the integral of $S$ must be done along a path in the phase space). The peak of the distribution never moves, smaller values of $|p|$ correspond to trajectories near the peak, and larger values of $|p|$ correspond to the low-probability events forming the tails of the distribution $P(\theta|\theta_i)$. It is trivial to integrate the equations of motion since $p$ is conserved, obtaining $\theta(t) - \theta_i = p t /\tau$ or $P(\theta_T|\theta_i) \sim \exp[-(\theta_T - \theta_i)^2 \tau / 2 T]$. If we restrict $\theta$ to the physical domain $[0,2\pi)$, then we may overlap the distribution onto itself, as in 
\be
P(\theta_T) \sim \sum_{n=-\infty}^{\infty} \exp \left[-\frac{(\theta_T-\theta_i+2\pi n)^2 \tau }{2 T} \right], 
\ee
which in this case matches the distribution from simulation exactly. (The fact that $\theta$ and $\theta+2\pi$ correspond to the same state is at the heart of the ``winding number'' multipaths we consider later.) 
In the long-time limit this distribution approaches a uniform distribution of states on the Bloch sphere.

\subsubsection*{Unequal Measurements and Properties of Periodic Islands}
\par In contrast with the previous case, unequal measurement strengths $\tau_x \neq \tau_z$ create elliptic islands along the $p=0$ line in phase-space. These islands are demarcated by a separatrix or critical line of energy $E_c$, drawn in green or red in Fig.~\ref{fig-xz_ps}. There is a minimum energy $E_m$ present in the phase-portrait, which is the energy at the fixed point in the middle of a periodic island (marked $\times$). Another set of fixed points exist at the crossing point on the separatrix (marked $+$). We determine the values of $E_c$ and $E_m$ below, and show that the periods of OPs in the island with $E \in [E_m,E_c]$ decrease monotonically from infinite period at $E \rightarrow E_c$ towards a finite value at $E \rightarrow E_m$. 

\begin{figure}
\includegraphics[width=.95\columnwidth]{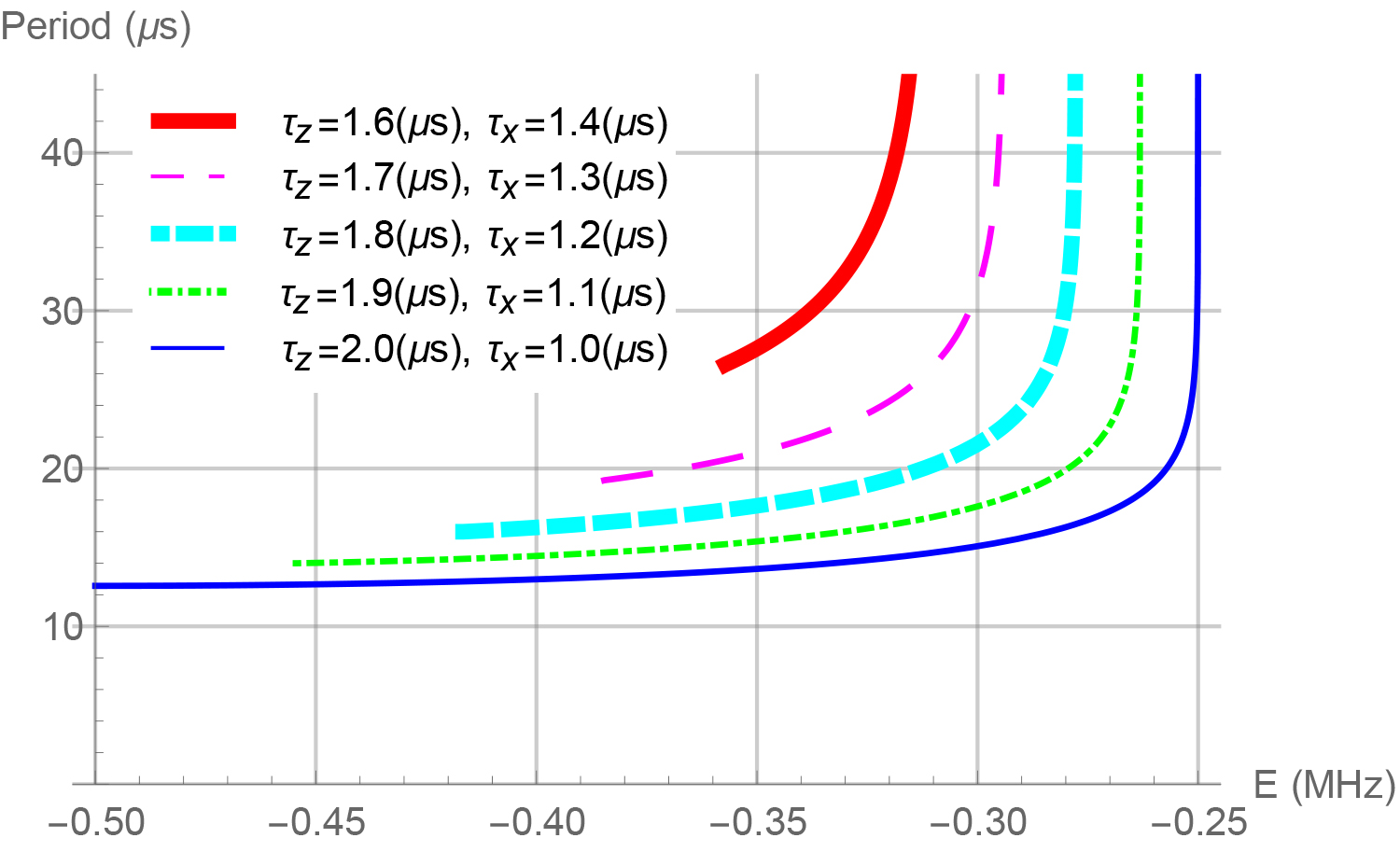}
\caption{(Color online) The period of island OPs is plotted as a function of stochastic energy $\tilde{T}(E)$ for different values of $\tau_z$ and $\tau_x$. 
The smallest (most negative) $E$ for each curve is $E_m$, which sits at the center of the island; the largest $E$ for each curve (at the asymptotes) $E_c$, along the separatrix. The period always tends to infinity on the separatrix because of the fixed points (a path which stops requires infinite time to complete a cycle), and decreases monotonically toward the island's center.}
\label{fig-xz_periods}
\end{figure}
The structure of \eqref{pqE} is closely related to the allowed energy range inside the island. Specifically the contents of the square root
\be\label{pqEsqrt}
\sqrt{1+ \frac{E}{a(\theta)} + \frac{b^2(\theta)}{4 a^2(\theta)}}
\ee
(where $a$ and $b$ are given in \eqref{xz_ab}), are key to determining which $\theta$ are allowed at a specific energy. From Fig.~\ref{fig-xz_ps}, it is obvious that any $E$ outside the stable island has a path which covers all $\theta$, whereas the periodic paths inside the island have a limited range of $\theta$. 
We determine the energy of the separatrix $E_c$ by defining it to be the smallest energy for which no value of $\theta$ makes the expression \eqref{pqEsqrt} imaginary. {\color{black} Correspondingly, the minimum energy $E_m$ is the smallest energy for which \eqref{pqEsqrt} is real for any $\theta$ at all.} We show $E_c$ and $E_m$ in Table 1.
\begin{minipage}{\columnwidth}\vspace{0.2cm}  \begin{center}
Table 1 \\ \begin{tabular}{c|cc} \hline
& $E_{c}$ & $E_{m}$ \\\hline
for $\:$ $\tau_z > \tau_x$ & $-1/2\tau_z$ & $-1/2\tau_x$ \\
for $\:$ $\tau_x > \tau_z$ & $-1/2\tau_x$ & $-1/2\tau_z$
\end{tabular}\end{center}\end{minipage}\vspace{0.2cm} 
We can calculate the minimum ($-$) and maximum ($+$) $\theta$ for a periodic OP as a function of $E$:
\be \label{thminmax}
\theta_M^\pm = \arctan\left( \pm i \frac{\tau_z \sqrt{1+2 E\tau_x}}{\tau_x \sqrt{1+2E\tau_z}} \right).
\ee
The period of such a path is then given by
\be \label{period_integral}
 \tilde{T}= \int_{\theta_{M}^-}^{\theta_{M}^+} \partl{p_+}{E}{} d\theta +  \int_{\theta_{M}^+}^{\theta_{M}^-} \partl{p_-}{E}{} d\theta = 2  \int_{\theta_{M}^-}^{\theta_{M}^+} \partl{p_+}{E}{} d\theta,
\ee
equivalent to \eqref{time2}, using \eqref{thminmax} as the boundary conditions.
It is possible to simplify to doubling the half period in \eqref{period_integral}, because the $\dot{\theta}$ are time reversal symmetric, as shown in \eqref{dotq_form}. Numerical integration of \eqref{period_integral} yields Fig.~\ref{fig-xz_periods}.
\par We note from Table 1 that the $E=0$ trajectories are necessarily always outside the island. There is a reason to expect this, on the basis of the result \eqref{zeroE} ($E=0$ OPs travel between $\theta_i$ and $\theta_f$ in an optimal time): Paths within the island \emph{cannot} necessarily be a valid solution for arbitrary $\theta_i$ and $\theta_f$, because they do not reach all $\theta$. We apply the discussion surrounding \eqref{dotq_form}, to note that regions outside islands in this system cannot have caustics, and that the result \eqref{zeroE} therefore always holds. The qualitative observation that higher energies result in faster rotation speeds also holds in the case of unequal measurement times (see above, Fig.~\ref{fig-xz_ps}(e,f), and section II).

\par In contrast to the case of equal-strength measurements, the fixed points only appear at specific values of $\theta$. As shown in Fig.~\ref{fig-xz_ps}(b,c), fixed points are now found at $(\bar{\theta},\bar{p})= (k\pi/2,0)$ for integer $k$. We recall that $\theta = 0$ is the $+z$-state and $\theta = \pi/2$ is the $+x$-state; the fixed points are now at the eigenstates of the two measurement operators. OPs at these fixed points express trajectories which are pinned to an eigenstate of a measurement operator. These may furthermore correspond to the highest probability-density events at a given time, since they sit along $p=0$ \eqref{sg}. We find that the most-likely $\theta_T$ are the eigenstates of the stronger measurement (ESM, $+$, shorter $\tau$), and the most likely state after times longer than either $\tau$ is dictated by collapse to one of those points along the separatrix. Collapse to $+$ is more likely than to the eigenstate of the weaker measurement (EWM, $\times$, longer $\tau$), since $S(T) = ET$ at a fixed point, and $S$ is less negative at the ESM using Table 1. Alternately, we may note that $\dot{S}$ has its overall maximum value at ESM ($+$), so there is a higher probability cost associated with staying at the EWM ($\times$). We thus have the intuitive result that when two unmatched measurements compete, the stronger one ``wins'' by attracting a higher proportion of stochastic trajectories towards its eigenstates. This behavior is consistent with simulations of stochastic trajectories \cite{JustinCode}. Paths over short times (much shorter than either $\tau$), and/or those leading to less likely final states need not necessarily conform to this prevailing behavior. 
\par We have used the result \eqref{sg}, (which we recall applies when there is no post-selection of the state), to discuss the most-probable dynamics of state collapse in this two-measurement scheme. We can also consider the meaning of the elliptic paths in phase-space over long enough times that they orbit the elliptic fixed point at the EWM. It is apparent from Fig.~\ref{fig-xz_ps}(e,f) that there is generally a higher probability cost per unit time associated with movement outside a stable island compared with inside it. This means that there is a non-negligible probability (fairly high) associated with paths oscillating back and forth between the ESM, across the EWM. Furthermore, the system prefers these island OPs for times approximately larger than either $\tau$, over those outside {\color{black} (necessarily MLPs)}, where MLPs pass over the ESM rather than turning around without going through the ESM.
\par Post-selecting on a $\theta_f$ in the opposite island as $\theta_i$ forces the system to use an OP outside the islands, which necessarily has a lower probability density associated with it, after some elapsed time, than an in-island path. Even with a choice of $\theta_i$ and $\theta_f$ within the same island, the $E=0$ MLPs outside the island are still able to optimize the probability to move between a set $\theta_i$ and $\theta_f$ in un-fixed time, however, by making the trip faster than paths within the islands; the $E=0$ paths have the optimal balance between their travel speed and distance traveled through regions of high probability cost per unit time (very negative $\dot{S}$). Altogether, we see that a phase portrait and $\dot{S}$ plot from the SPI/OP formalism, combined with \eqref{sg} and a simple analysis of fixed points, is sufficient to infer the dominant long-term dynamics of a quantum system subject to a non-trivial measurement scheme. 

\subsection*{Multipaths}
We now search for multipaths in the OPs from the phase space shown in Fig.~\ref{fig-xz_ps}. We identify two mathematically distinct types of multipaths in this phase space: We first discuss those arising from paths which orbit the Bloch sphere a different number of times (paths with different ``winding numbers''), drawn from regions outside of the stable islands in phase-space. Second, we discuss paths within the periodic islands. Winding number multipaths form without a diverging VVD \eqref{vvd1}, while island multipaths necessarily form from a caustic with diverging VVD. Experimentally, we might distinguish between these types of multipaths by noting that those from a caustic may be very close together in phase space. (In fact, they \emph{must} be nearly indistinguishable for post-selections close to the catastrophe where new multipaths are forming.) Winding number paths, however, will have to travel far apart in the Bloch sphere at some point in their evolution, because one of the OPs must contain more windings or oscillations than the other.
\par The relative probabilities between the individual paths meeting a set of boundary conditions to form a multipath group are important. Suppose we collect a finite, but statistically representative, set of stochastic trajectories starting from the same initial state $\theta_i$ over some time $T$. For a multipath meeting the boundary conditions $\theta_i \rightarrow \theta_f$ at $T$ to be experimentally visible, we require that the paths forming it all have probability densities that are not too low compared with 
(a) the highest probability density path (the one leading to the most likely $\theta_T$), and (b) each other. If the overall probability densities (a) of the constituent paths leading to $\theta_f$ are too small, then a prohibitively large data set would be required to get a statistically significant sub-ensemble meeting the desired boundary conditions. If the relative probability densities (b) are highly unbalanced, then the same problem arises within the sub-ensemble; {\color{black} few stochastic trajectories corresponding to low probability density MLPs appear}, and are consequently both difficult observe and less relevant to the dynamics. An example is shown in section IV, in Fig.~\ref{fig-wn}.
\par Combining these types of paths across the different regions of phase space, we demonstrate below that it is possible to obtain multipaths between any two states (anywhere in our phase space), and that there exist abundant MLPs which have high enough probabilities, and form in short enough times, to be realistically observable in experiment. We are able to compute the caustic onset time, using a relationship between that timescale, and the periods of the paths in the island discussed in Fig.~\ref{fig-xz_periods}.

\begin{figure*}
\begin{tabular}{ccc}
\begin{picture}(100,100)\put(-103,-10){\includegraphics[width=0.33\textwidth]{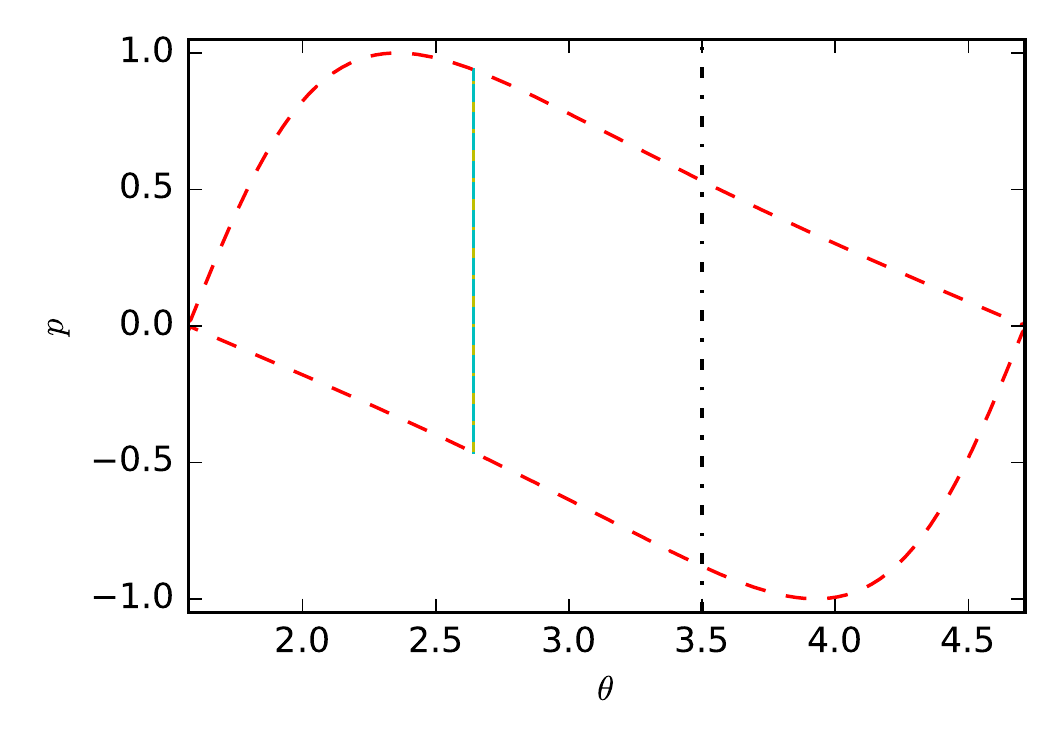}}\put(-71,90){(a)} \end{picture} &
\begin{picture}(100,100)\put(-43,-10){\includegraphics[width=0.33\textwidth]{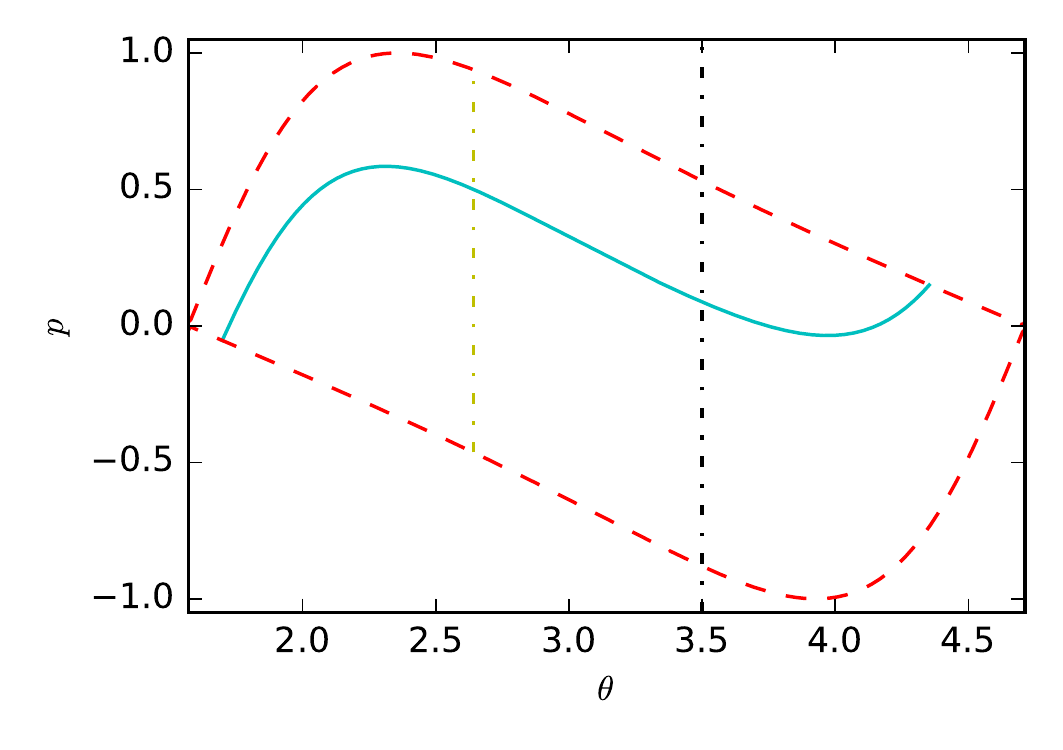}}\put(-11,90){(b)} \end{picture} &
\begin{picture}(100,100)\put(17,-10){\includegraphics[width=0.33\textwidth]{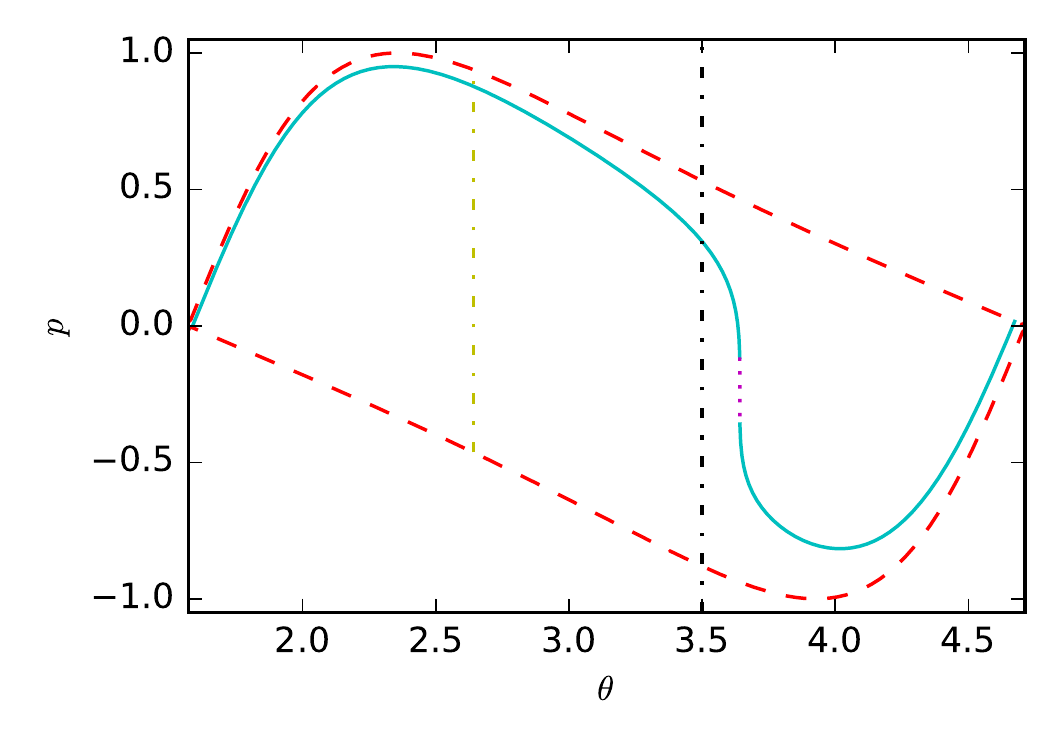}}\put(49,90){(c)} \end{picture} \\
\begin{picture}(100,100)\put(-103,-17){\includegraphics[width=0.33\textwidth]{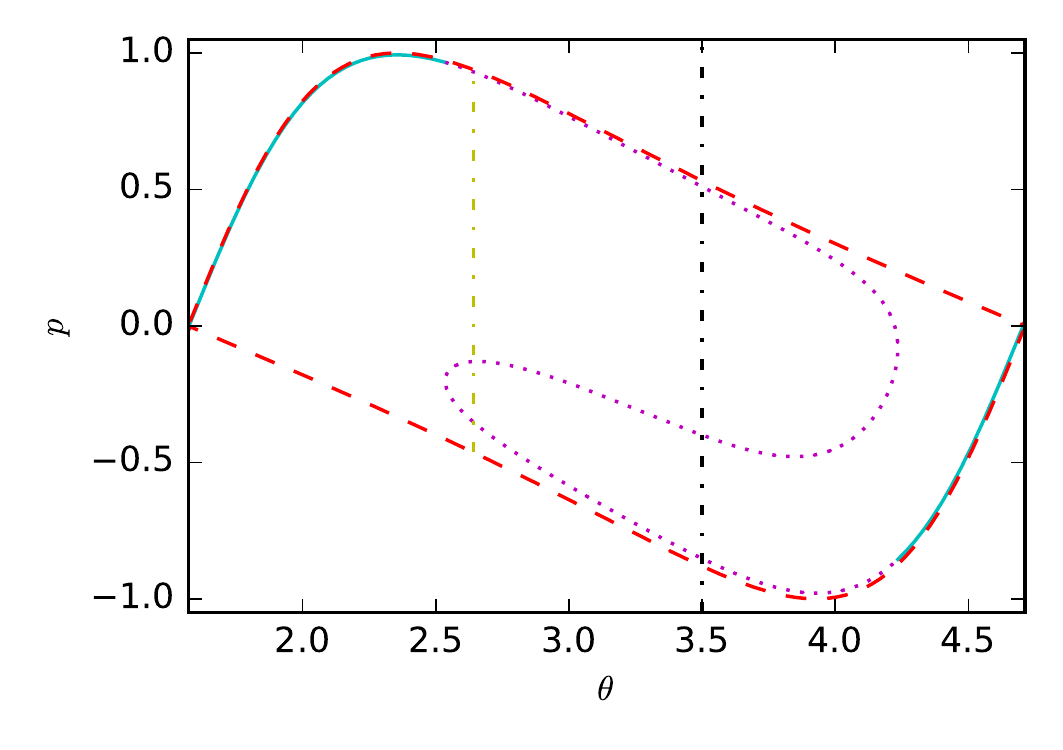}}\put(-71,83){(d)} \end{picture} &
\begin{picture}(100,100)\put(-43,-17){\includegraphics[width=0.33\textwidth]{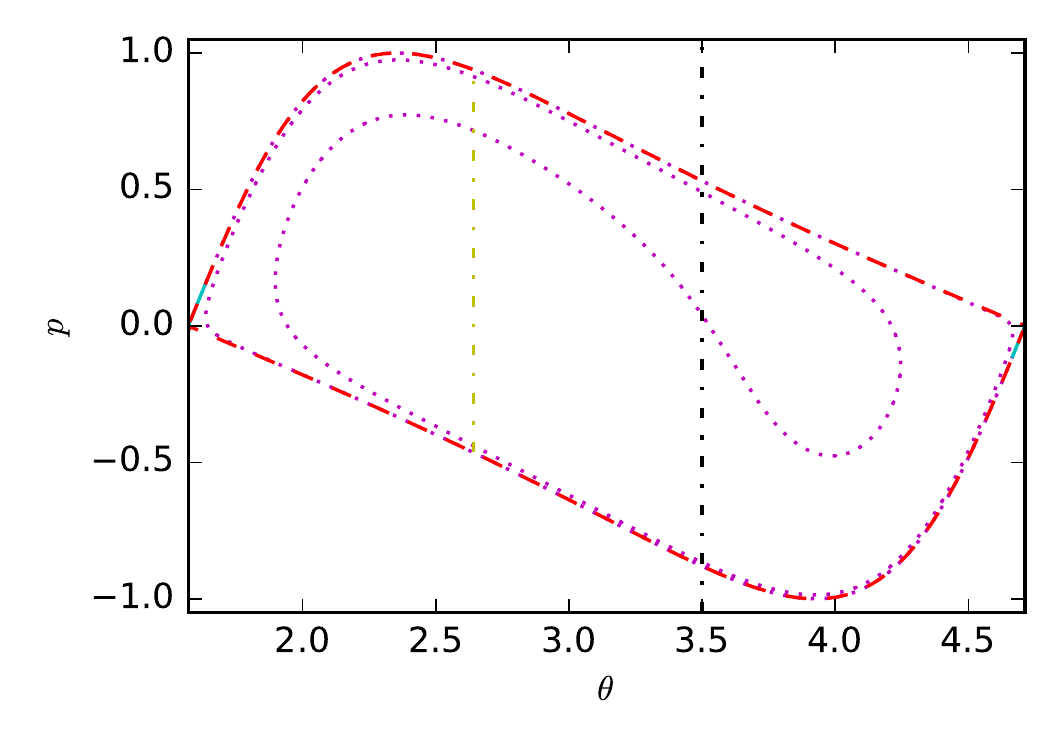}}\put(-11,83){(e)} \end{picture} &
\begin{picture}(100,100)\put(17,-17){\includegraphics[width=0.33\textwidth]{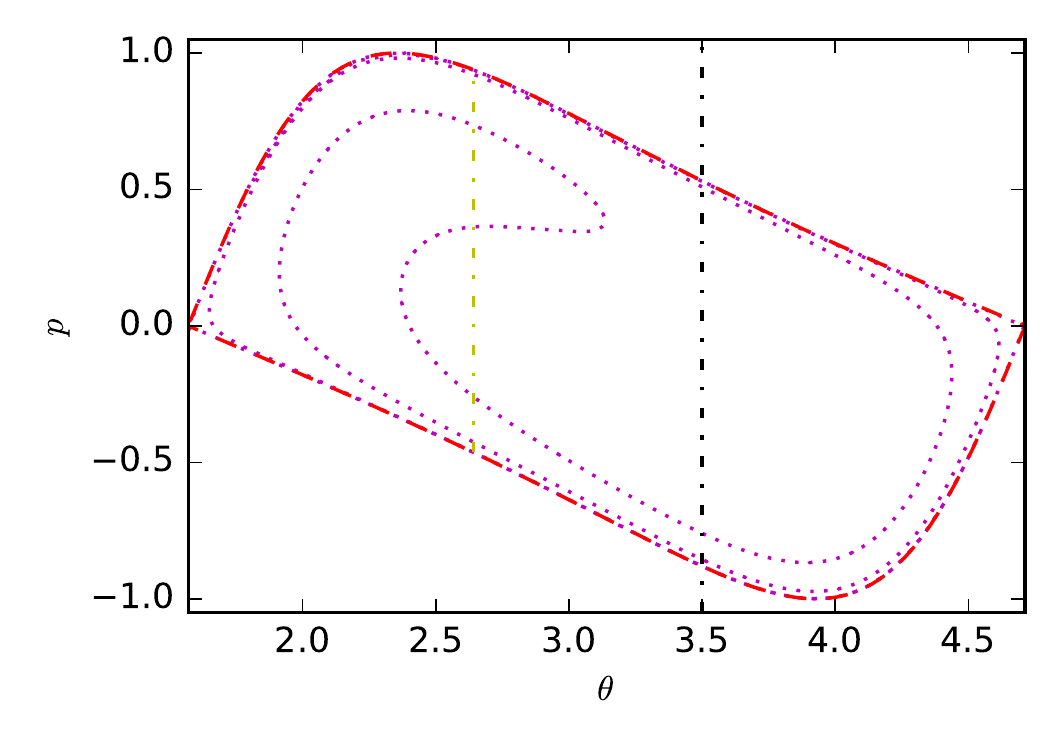}}\put(49,83){(f)} \end{picture}\\
\begin{picture}(100,100)\put(-100,-24){\includegraphics[width=0.33\textwidth]{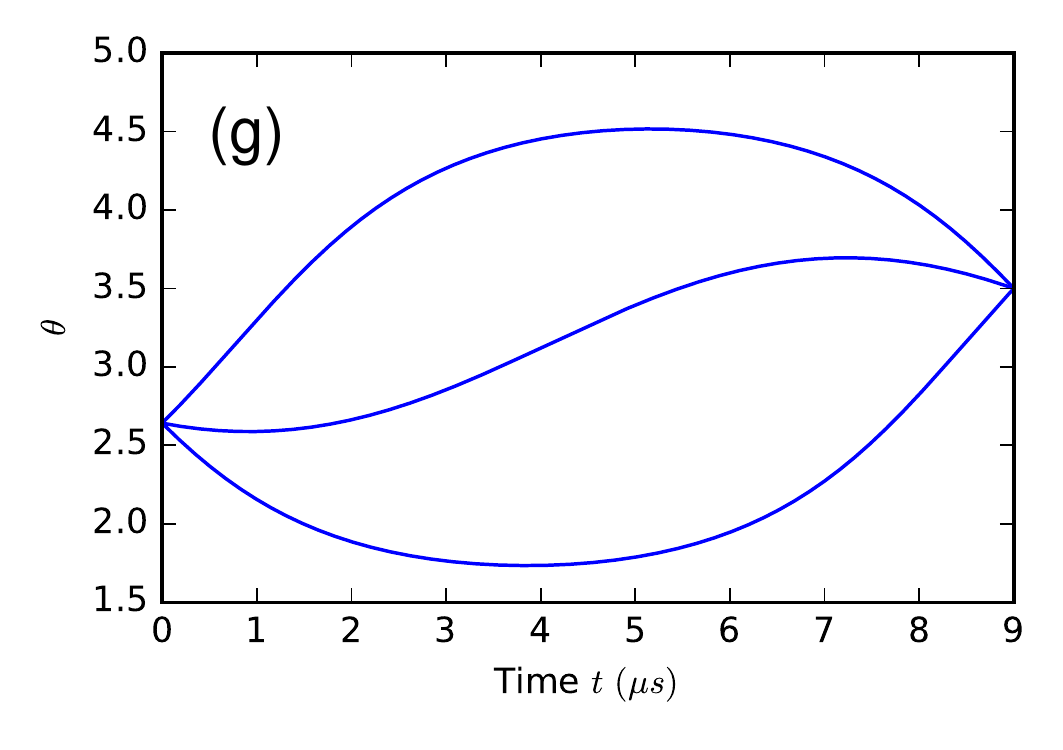}}\put(-71,74){(g)} \end{picture} &
\begin{picture}(100,100)\put(-40,-24){\includegraphics[width=0.33\textwidth]{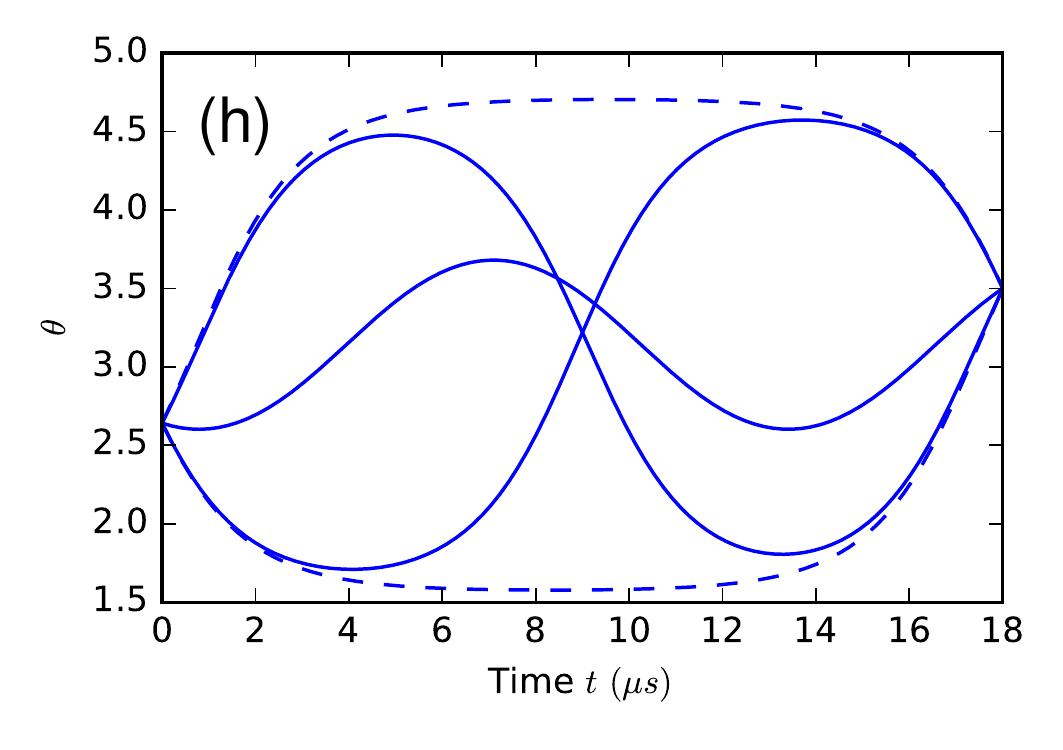}}\put(-11,74){(h)} \end{picture}&
\begin{picture}(100,100)\put(20,-24){\includegraphics[width=0.33\textwidth]{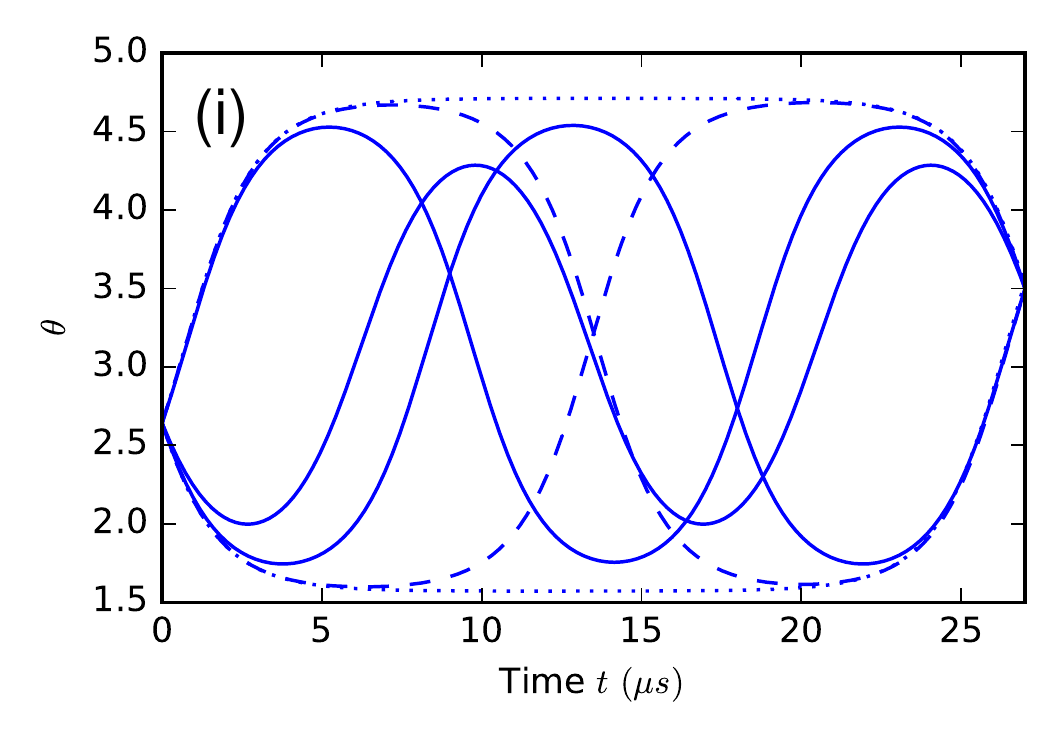}}\put(49,74){(i)} \end{picture} \\ & & \\
\end{tabular}
\caption{We plot the Lagrange manifold within the periodic island in the system with stochastic Hamiltonian \eqref{hr}, and some of the multipaths we find with it. All plots above are for $\tau_z = 2 \mu s$ and $\tau_x = 1 \mu s$. The dynamical variables $\theta$ and $p$ are dimensionless. Plots (a-f) show the Lagrange manifold inside islands at (a) $t=0 \mu s$, (b) $t=3.15\mu s$, (c) $t=6.32 \mu s$, (d) $t=9\mu s$, (e) $t=18\mu s$, and (f) $t=27\mu s$. The dashed red lines in (a-f) are the separatrix marking the edge of the island, the solid teal line in (a-f) denotes the Lagrange manifold at values of $\theta$ where no multipaths are possible, and the dotted magenta line in (c-f) denotes the Lagrange manifold at values of $\theta$ inside a caustic, where multipaths exist. The manifold starts at $\theta_i = \pi-1/2$ for all plots, marked with a yellow dash-dotted line in (a-f). The first caustic allowing multipaths within the island forms at $t = 6.32 \mu s$, as shown in (c). The manifold can then fold over an increasing number of times, as shown in (d-f). Plots (g-i) show the multipaths as functions $\theta(t)$, up to the evolution times shown in (d-f), respectively, for the final position $\theta_T = 3.5$. Each intersection of the Lagrange manifold with the dash-dotted black line at $\theta_T$ in (d-f) corresponds to an optimal path in (g-i). Solid, dashed, and dotted line types in (g-i) show the different groupings of paths which emerge together each time the manifold has gained another fold/catastrophe.}
\label{fig-LMonset}\vspace{-1.1cm}
\end{figure*}
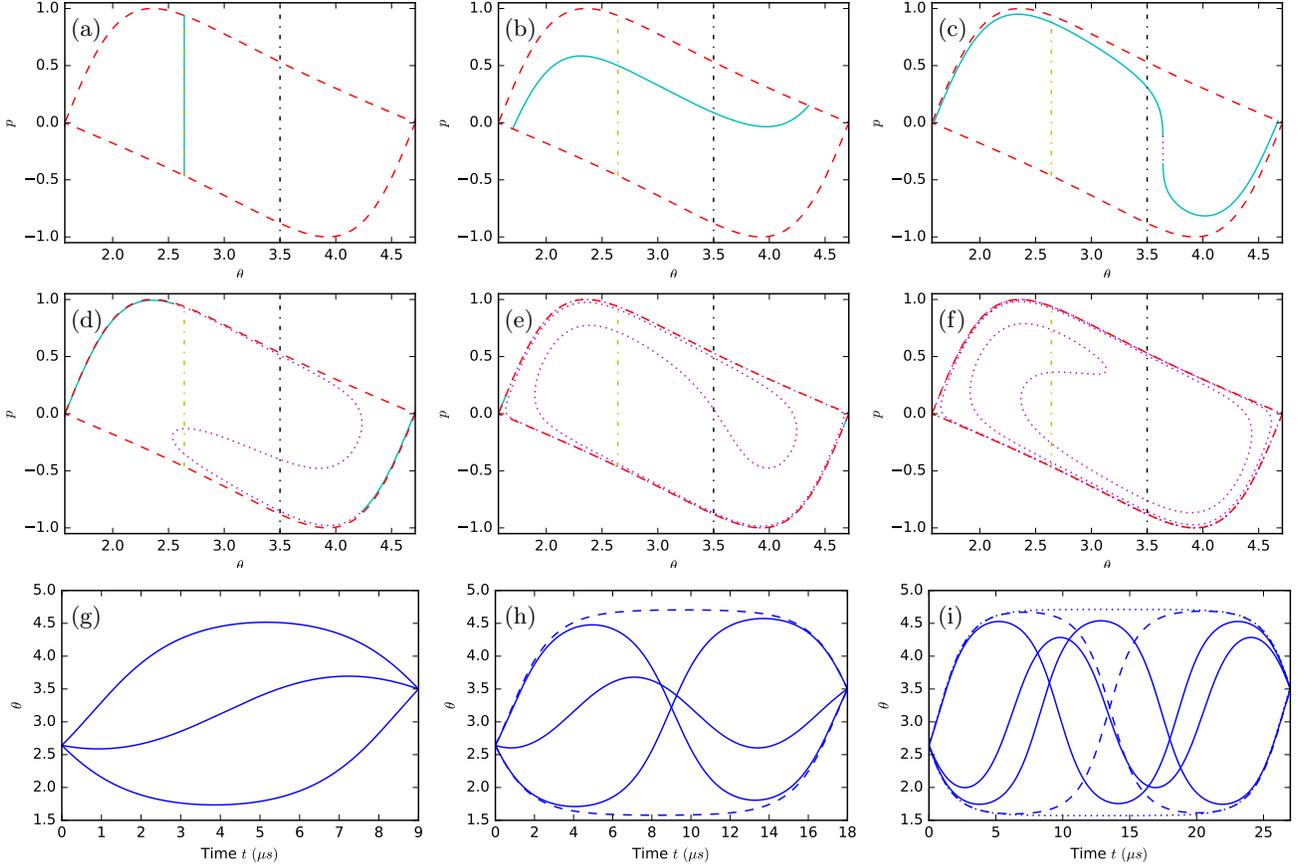

\begin{figure*}
\begin{tabular}{cc}
\begin{minipage}{.55\textwidth}
\begin{tabular}{cc} \includegraphics[width=.08\textwidth,trim={0 0 22pt 0},clip]{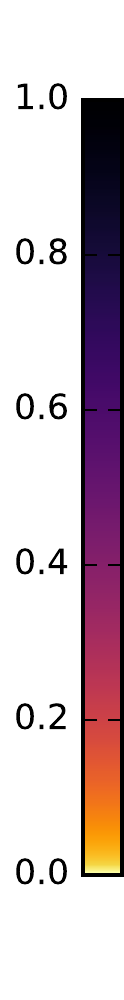} & \includegraphics[width=.9\textwidth, trim={44pt 0 44pt 0},clip]{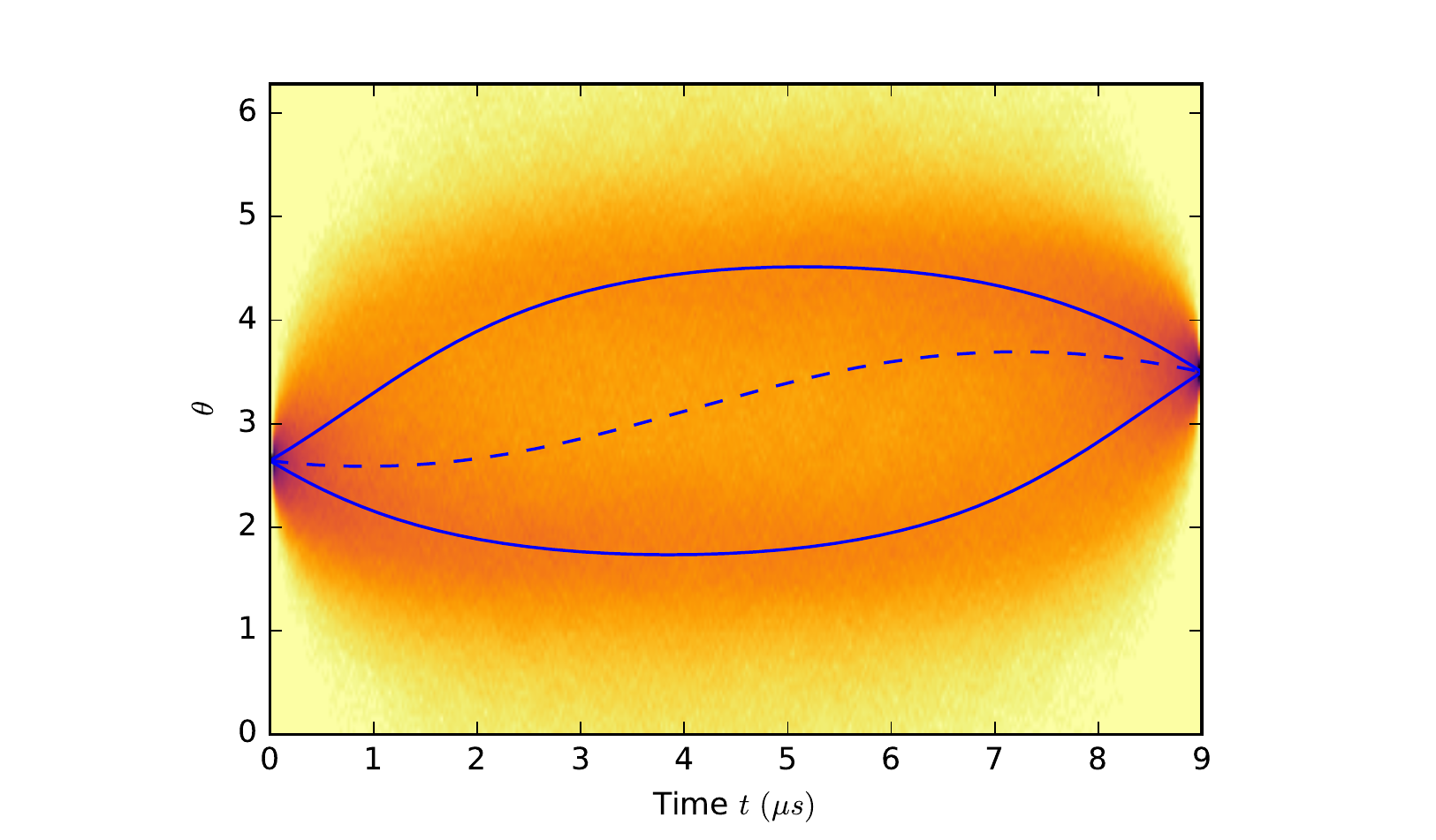} \end{tabular}
\end{minipage} & \begin{minipage}{.42\textwidth}  
\caption{Optimal paths from Fig.~\ref{fig-LMonset}(g) are shown superposed over density plots of simulated stochastic trajectories \cite{JustinCode}, post-selected on $\theta_T = 3.5$ at $T = 9.0 \mu s$. {\color{black} Density is shown as a histogram with equal-area bins (pixels). The colorbar is normalized relative to the highest bin count between boundary conditions, such that 1 is the highest path density, and 0 means no paths at all.} We see that there is good agreement between the two MLPs (solid) with visible peaks in the density distribution. The dashed path, which does not correspond to a peak in the trajectory density, is a LLP, rather than a MLP. {\color{black} See appendix \emph{C}, and Fig.~\ref{fig-5p7p_dens} therein, for further details.}}\label{fig-dens}
\end{minipage}\end{tabular}
\end{figure*}

\subsubsection*{Multipaths I: OPs with different Winding Numbers}
We discuss OPs with differing winding numbers, in the regions where $E > E_c$ {\color{black} (where all OPs are MLPs)}. We define the winding number of an OP to be the number of rotations it makes about the Bloch sphere. Multipaths occur between a fast and slow path traveling in the same direction when $\theta_T^{fast} = \theta_T^{slow} + 2 \pi N$, where $N$ is the difference in winding numbers (and necessarily an integer, so that both $\theta_T$ correspond to the same state). 
In other words: pairs of paths which travel around the Bloch sphere at different speeds form a multipath when one is exactly an integer number of laps ahead of the other (in this case the Lagrange manifold overlaps itself $mod(2\pi)$). This can also happen between paths traveling in opposite directions around the Bloch sphere when a condition $\theta_T^{CW} = \theta_T^{CCW} + 2 \pi N$ is met, assuming the clockwise-rotating paths accrue positive winding counts, and counter-clockwise rotating paths accrue negative winding counts.
\par In practice, observation of such multipaths in laboratory situation should be easier for a smaller difference in winding number, and longer elapsed time. Similar energies result in more similar actions (or probability densities) for each path, and paths one wind apart will have closer energies than those two or more winds apart, for the same elapsed time. Lower overall stochastic energy also corresponds to overall higher probability densities after the same elapsed time, as discussed above. By choosing paths with high stochastic energies, winding number multipaths can be created after an arbitrarily short time evolution, but these faster paths also have a vanishingly small probability density to actually occur. Notice that due to the quadratic dependence of $H$ on $p$ (see \eqref{hgen3}), every energy greater than $E_c$ appears twice in the system \eqref{hr}, once with a clockwise-moving MLP and once with a counterclockwise-moving MLP. It should be readily apparent that such a pair of paths can always form a multipath by meeting at some final $\theta_f$ approximately opposite their $\theta_i$ on the sphere. This is the easiest way to meet the probability conditions described above, leading to multipaths which have the highest physical impact and are the easiest to find in experiment.

\subsubsection*{Multipaths II: OPs within a Periodic Island}  

To locate multipaths within a periodic island, we initialize a Lagrange manifold at some $\theta_i$, and consider the segment of the manifold with $E$ satisfying $E_m \leq E \leq E_c$ (\emph{i.e.} the part in the island). The manifold can stretch in length, but remains continuous for all time. It is forced to spiral as shown in Fig.~\ref{fig-LMonset}(a-f), because OPs on the edges of the island stop at the unstable fixed point (the ends of this section of the manifold collapse to the ESM), while paths inside continue to rotate along elliptic curves with finite period. 

\par We comment on the relative probabilities of OPs comprising these island multipaths. The fixed points at opposite sides of the separatrix ensure that an odd number of optimal paths meet any particular boundary conditions. The first fold in the manifold changes a single path into a triple path, and subsequent foldings add pairs to the group, creating quintuple paths, septuple paths, etc., as shown in Fig.~\ref{fig-LMonset}(d-i). 
{\color{black} We use the second variation in the action around optimal paths to test whether we have a MLP, LLP, or SP; the method is detailed in appendix \emph{C}. For our triple-path example in Fig.~\ref{fig-LMonset}(g), we find that we have two MLPs towards the outside of the island (higher energy), and one LLP more towards its center (lower energy). In Fig.~\ref{fig-dens} we show that these paths are consistent with peaks in the density of simulated stochastic trajectories meeting the desired post-selection. The quintuple and septuple optimal path cases, from Fig.~\ref{fig-LMonset}(h,i) are also shown to have one LLP, with the four or six remaining optimal paths, respectively, being MLPs. Paths with a higher energy (closer to the separatrix), will have a higher probability density associated with them for a given elapsed time $t \gtrsim \tau$, because they sit near the ESM, the least-costly spot in phase-space, for a longer proportion of their evolution. Overall, these results suggest that for this two-measurement system, multipaths containing an arbitrary number of solutions may be obtained from within a periodic island, provided we wait long enough. However, from within the paths meeting the multipath boundary conditions, the pair with the largest stochastic energies will tend to dominate the relative probabilities. Specific examples can be found in Fig.~\ref{fig-5p7p_dens}.
}
\subsubsection*{Caustic Onset Time for Island Multipaths}
\par There is a minimum onset time required to form a caustic within a periodic island, which we calculate exactly. The Lagrange manifold, initialized at some $\theta_i$, will contain OPs with a variety of periods $\tilde{T}$. The fastest period $\tilde{T}_f$ in the manifold belongs to the ellipse of lowest $E$ in the phase portrait which touches the initial manifold at the chosen $\theta_i$ (this is the innermost ellipse in the island which is tangent the initial manifold). Using geometrical arguments, the symmetry of the islands, and the monotonic decrease of $\tilde{T}(E)$ \eqref{period_integral} from Fig.~\ref{fig-xz_periods}, one can show that $\tilde{T}_f /2$ defines the onset time for a caustic. The first failure of the vertical line test will occur along that innermost path with the fastest period on the manifold, exactly opposite the point where the manifold was first tangent to that ellipse at $t=0$. Initial states close to the EWM will generate caustics faster than initial states close to the ESM.
\par Winding number multipaths, by contrast, may appear arbitrarily fast by utilizing a fast-rotating path in regions above or below the islands. The decreasing probability densities associated with fast-rotating path still place a limit on the possibility of observing winding number paths quickly in a finite data set, even though no fundamental onset time exists for them. 

\begin{figure}
\begin{tabular}{c} \begin{picture}(100,100) \put(-50,-80){\includegraphics[width=0.81\columnwidth]{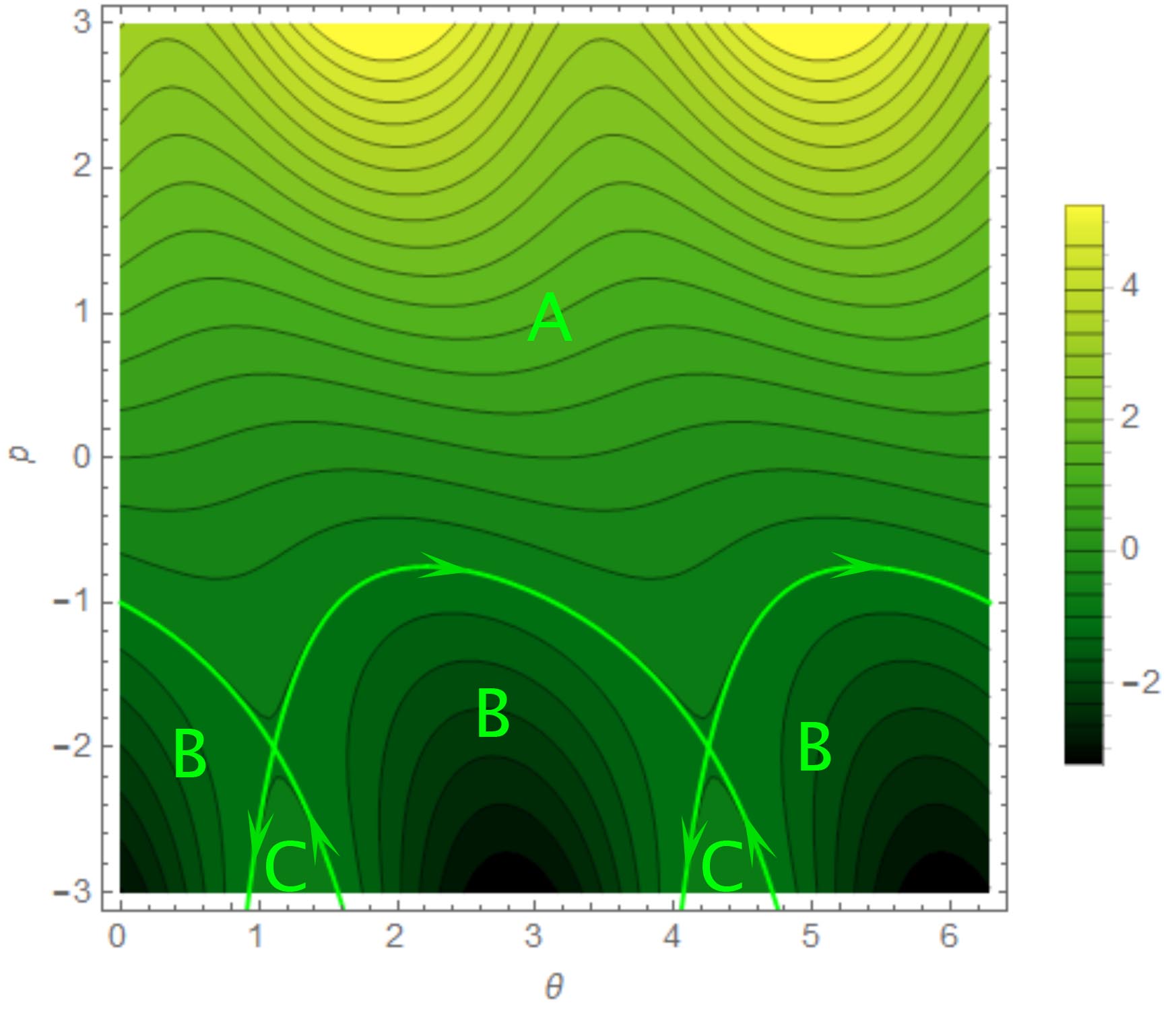}} \put(-55,80){(a)}  \put(125,70){\begin{minipage}{0.05\textwidth} $E$ \\ (MHz) \end{minipage}}\end{picture} \\ \\ \\ \\ \\ \\ \\ \\ \\
\begin{picture}(100,100) \put(-50,-50){\includegraphics[width=0.81\columnwidth]{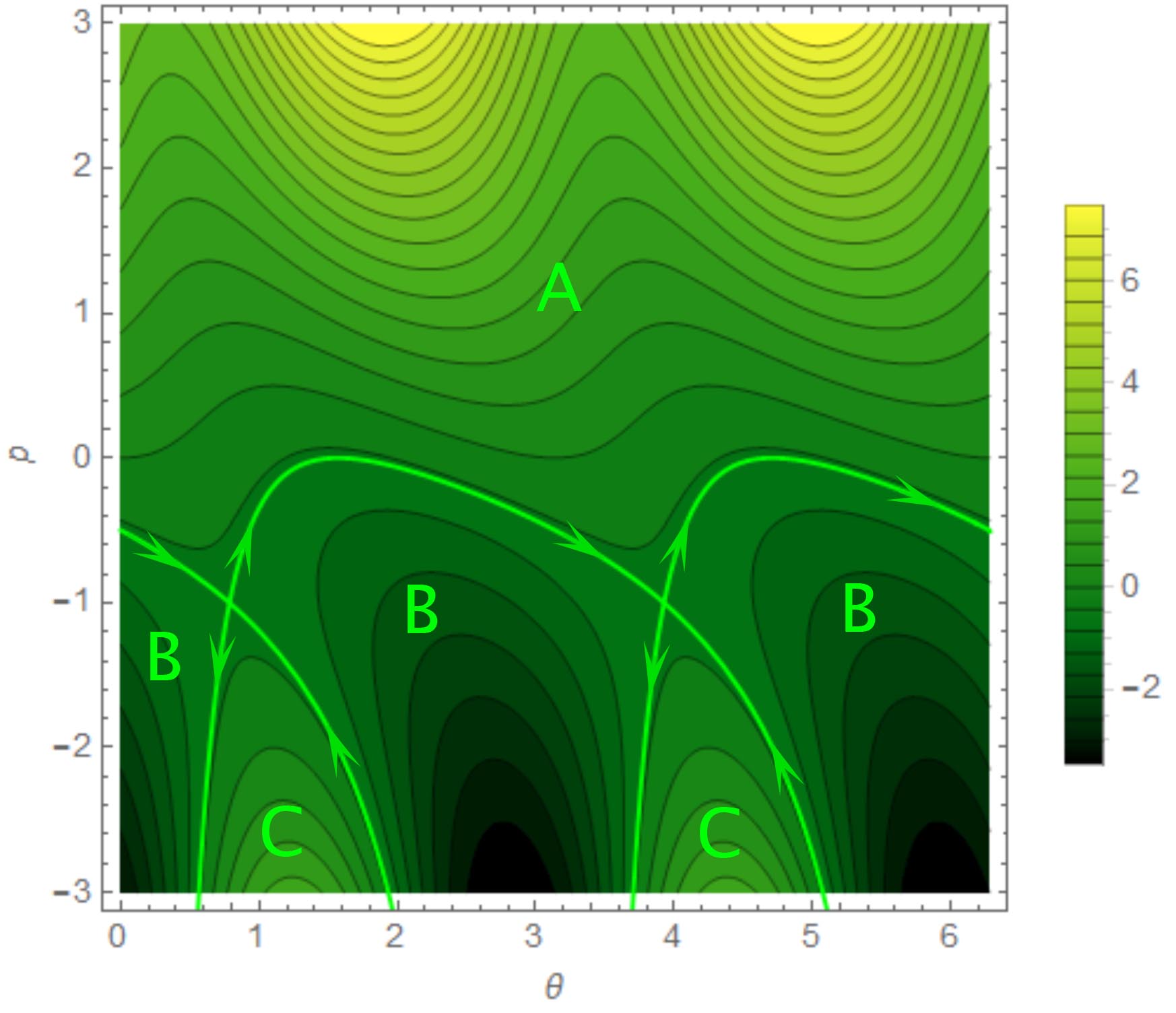}} \put(-55,110){(b)} \put(125,100){\begin{minipage}{0.05\textwidth} $E$ \\ (MHz) \end{minipage}} \end{picture}\\ \\ \\ \\ \\ \\
\begin{picture}(100,100) \put(-50,-55){\includegraphics[width=0.81\columnwidth]{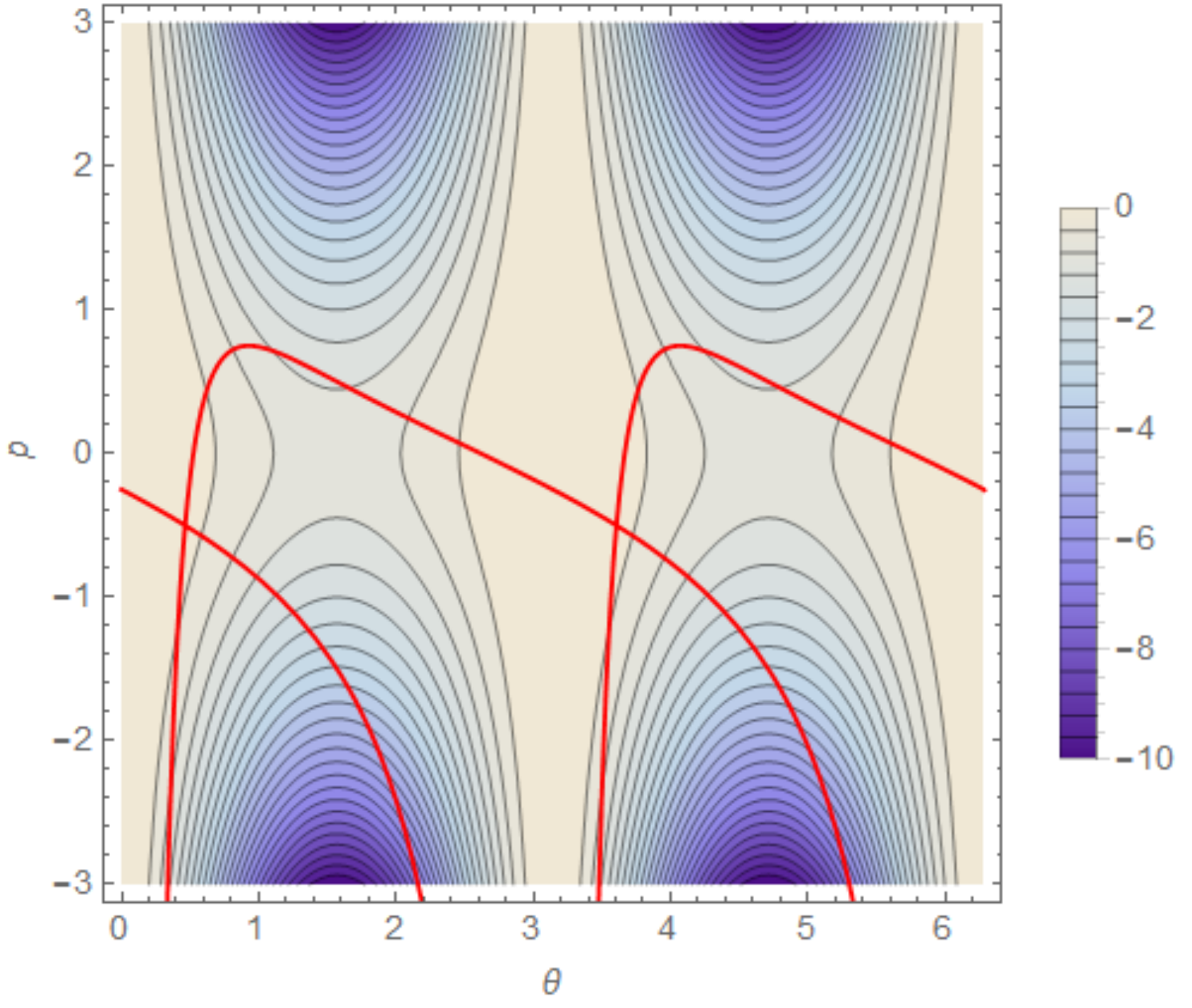}} \put(-55,105){(c)} \put(125,95){\begin{minipage}{0.05\textwidth} $\dot{S}$ \\ (MHz) \end{minipage}}\end{picture} \\ \\ \\ \vspace{0.25cm}
\end{tabular}
\caption{The phase portrait for the $z$ measurement with Rabi drive \eqref{zdriveH} is pictured in (a) and (b). $\dot{S}$, given in \eqref{z_sdot}, is shown in (c). We have $\Delta = 1$MHz and $\tau = 2 \mu s$ in (a), $\Delta = 1$MHz and $\tau = 1 \mu s$ in (b), and $\Delta = 1$MHz and $\tau = 0.5\mu s$ in (c). We choose $\mu s$ and MHz for simplicity, but any units such that $[E] = [\dot{S}] = [\Delta] = [\tau]^{-1}$ would leave these plots unchanged. Note that $\dot{S}$ does not depend on $\Delta$, and that $\tau$ does not change its shape, but only its magnitude; the contour lines of $\dot{S}$ have the same geometry for all $\Delta$ and $\tau$. Contour color denotes stochastic energy in (a) and (b), or the value of $\dot{S}$ in (c), and the green (a,b) or red (c) curve is the separatrix/critical line with energy $E_\star$. Lines of constant stochastic energy in (a), (b), are the trajectories $p(\theta,E)$. 
} \label{fig-z_ps} \end{figure}
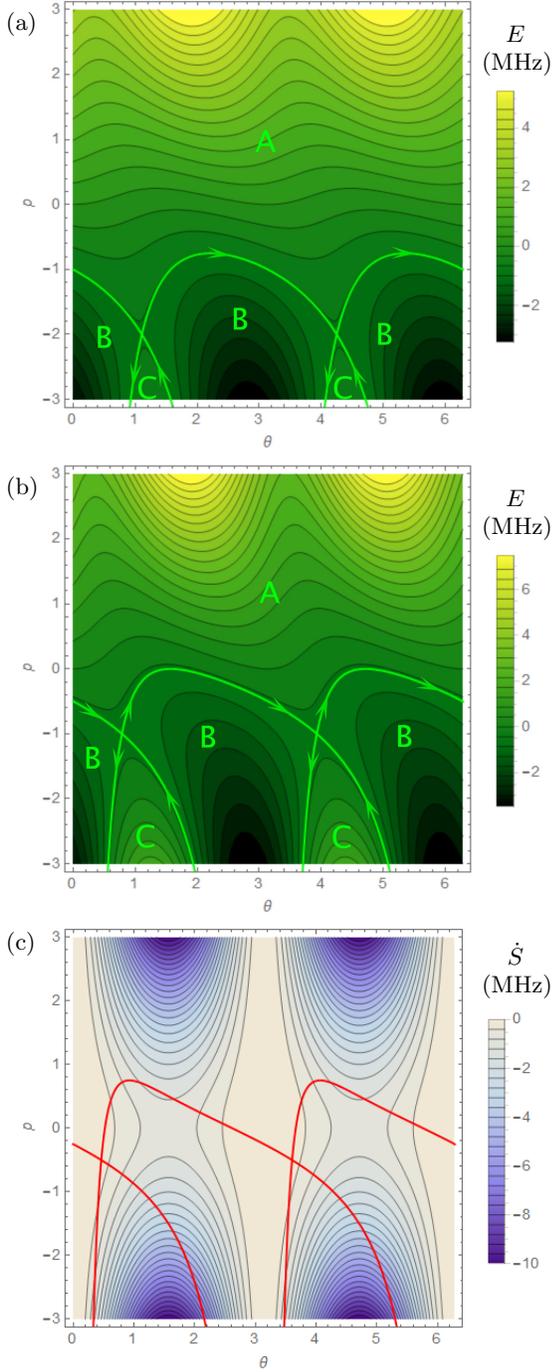
\begin{figure*}
\begin{tabular}{cc}
\begin{tabular}{c} 
\includegraphics[width=0.45\textwidth, trim={12pt 0 40pt 0},clip]{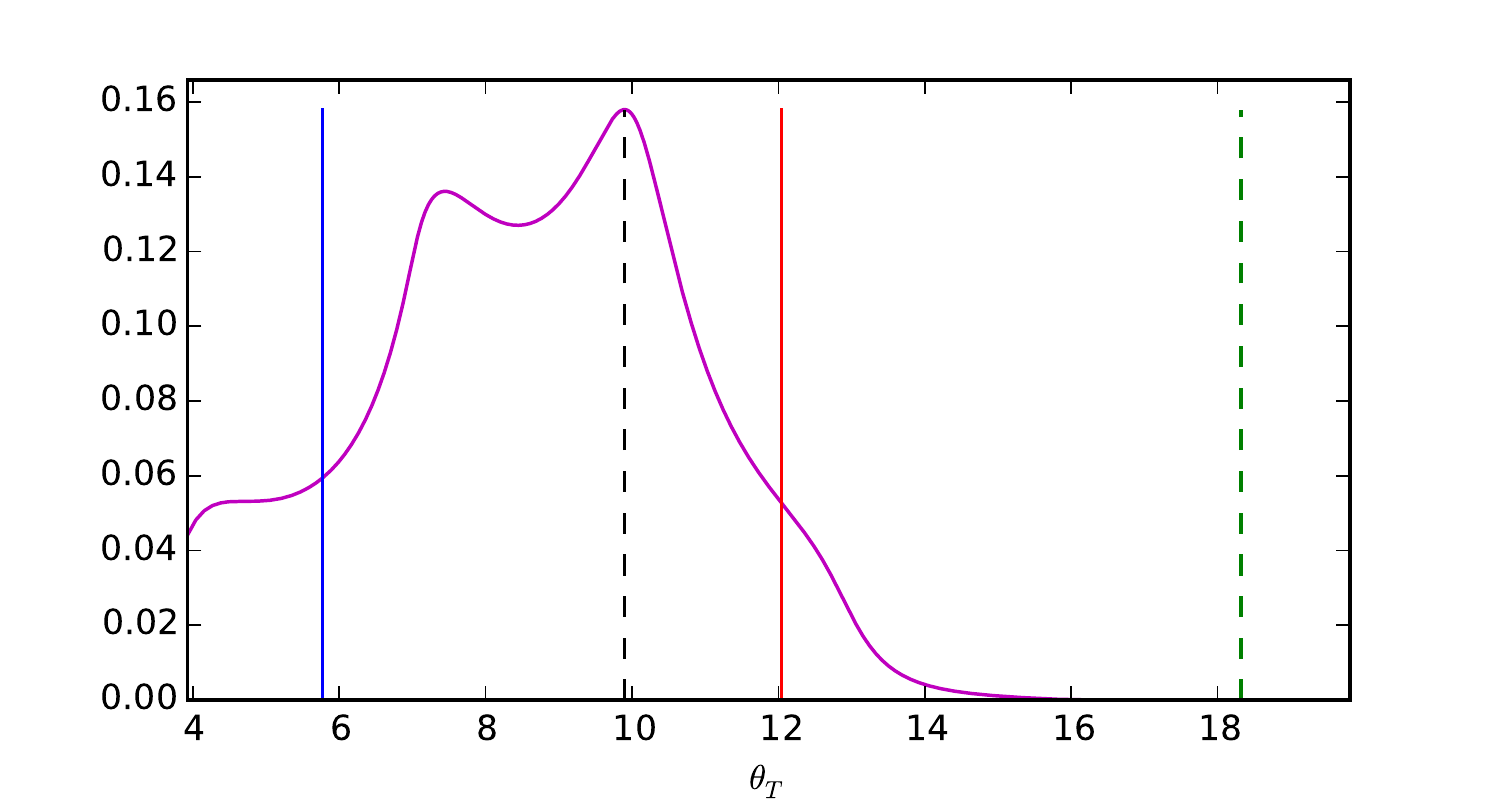} \\
\includegraphics[width=0.45\textwidth, trim={12pt 0 40pt 0},clip]{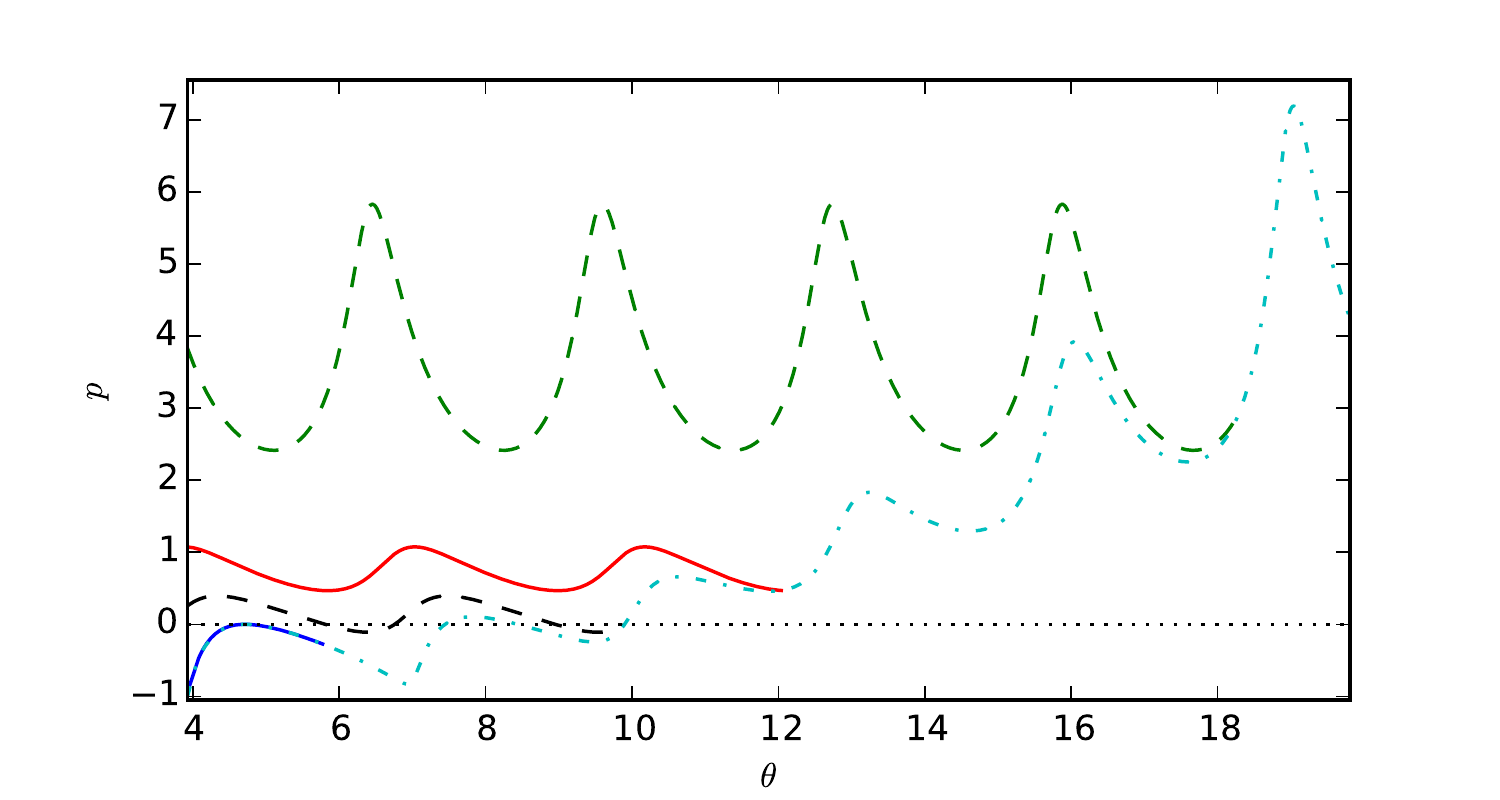}
\end{tabular} & \begin{tabular}{c}
\includegraphics[width=0.45\textwidth, trim={12pt 0 40pt 0},clip]{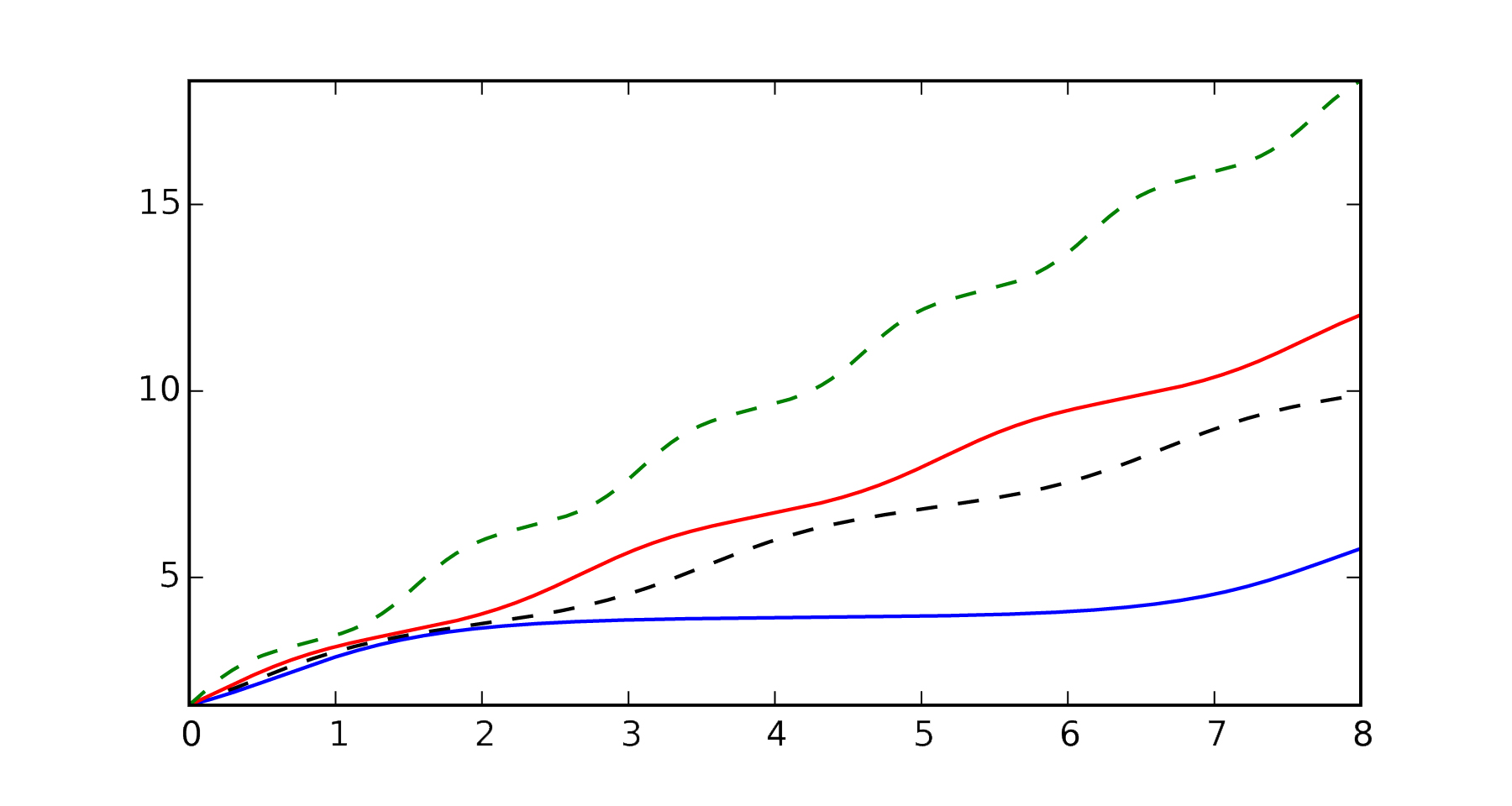} \\  \vspace{0.7cm} \\
\includegraphics[width=.45\textwidth, trim={12pt 0 40pt 0},clip]{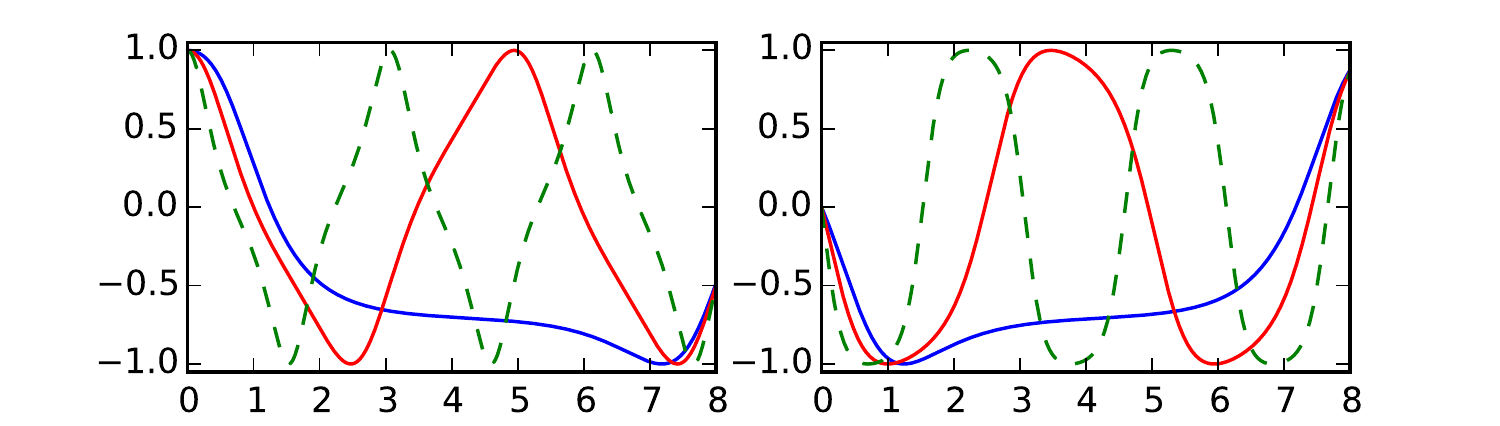} \\ \vspace{0.7cm}
\end{tabular} \end{tabular}
\begin{picture}(1,1) \put(-480,75){$e^S$} \put(-230,75){$\theta$} \put(-120,3){$t$ $(\mu s)$} \put(-220,-30){$y$} \put(-110,-30){$z$} \put(-177,-115){$t$ $(\mu s)$} \put(-67,-115) {$t$ $(\mu s)$} \put(-480,120){(a)} \put(-470,-25){(b)} \put(-230,120){(c)} \put(-172,-30){(d)} \put(-61,-30){(e)} \put(-389,27){$\theta_T^{(0)}$} \put(-361,27){\color{red}$\theta_T^{(2)}$} \put(-442,27){\color{blue}$\theta_T^{(1)}$} \put(-282,27){\color{dgreen}$\theta_T^{(3)}$} \put(-25,63){$\theta_T^{(0)}$} \put(-25,90){\color{red}$\theta_T^{(2)}$} \put(-25,33){\color{blue}$\theta_T^{(1)}$} \put(-25,109){\color{dgreen}$\theta_T^{(3)}$}
\put(-406,108.7){{\color{magenta} $\star$}}
\put(-394,103.3){{\color{magenta} $\star$}}
\put(-375,123){{\color{magenta} $\star$}}
\put(-406,-108.5){{\color{magenta} $\star$}}
\put(-394,-108.5){{\color{magenta} $\star$}}
\put(-375,-108.5){{\color{magenta} $\star$}}
\end{picture}
\caption{A winding-number multipath for the system \eqref{zdriveH} is constructed between the blue, red, and dashed green MLPs, starting at $\theta_i = \pi/2$ with an elapsed time $T = 8 \mu s$. Recall that $\theta$, $p$, $S$, $y$, $z$, and any probabilities and probability densities are dimensionless. The blue MLP reaches $\theta_T^{(1)}$, the red $\theta_T^{(2)} = \theta_T^{(1)} +2\pi$, and the dashed green goes to $\theta_T^{(3)} = \theta_T^{(1)}+4\pi$, such that they all arrive at the same quantum state and form a multipath. The blue and red MLPs are chosen to have relatively high \emph{and} approximately equal probability densities, by choosing final coordinates on either side of the highest probability path (dashed black). The highest probability path at $T = 8\mu s$ is \emph{not} part of the multipath, but we use its action as a reference for the other probability densities at play. In (a) we plot the exponential of the action as a function of $\theta_T$ at $T=8\mu s$ in purple, assuming $S=0$ at $t=0$. The final $\theta_T$ for different paths are marked with vertical lines. The dashed black MLP leads to the the most-likely final coordinate $\theta_T^{(0)} = 9.90$, at one of the three extrema in the action corresponding to a root of the Lagrange Manifold (where $p_T =0$; see the magenta stars {\color{magenta} $\star$}). The manifold itself is shown with all of the aforementioned MLPs in dash-dotten cyan in (b). These MLPs are shown in phase space (b), as plots $\theta(t)$ in (c), and in their projections onto Cartesian $y$ and $z$ in (d) and (e). A path traveling fast enough to go one full winding number further in the specified time (dashed green), is however highly improbable compared with the others. The actions of these MLPs are $S_0 = -1.85$ (dashed black, maximum $S$ at $T=8 \mu s$ for our $\theta_i$, and used for reference only), $S_1 = -2.82$ (blue), $S_2 = -2.94$ (red), and $S_3 = -15.68$ (dashed green). These actions can be used to estimate the numbers $N_1$ and $N_2$ of measured noisy quantum trajectories contributing to each MLP, where $N = N_1+N_2$ is the total number of trajectories meeting the post-selection shared by the blue and red MLPs. Consider $N_1 \approx N e^{S_1} / (e^{S_1} + e^{S_2}) \approx 0.53 N$ and $N_2 \approx N e^{S_2}/(e^{S_1}+e^{S_2}) \approx 0.47 N$; these are quite well-balanced paths. We can neglect contributions from the dashed green path and all higher winding numbers on the basis of the estimate $N_3 \approx N e^{S_3}/(e^{S_1}+e^{S_2}+e^{S_3}) \approx 1.38 \times 10^{-6} N$.
}\label{fig-wn}
\end{figure*}

\section{Measurement of one Observable with Rabi Drive}

We proceed to our second sample system, in which the $x$ measurement from above is turned off, so that continuous monitoring is only performed along $z$, but a Rabi drive is added to the system. The unitary driving term can be described by applying a Hamiltonian $\hat{h} = - \Delta \sigma_x / 2$ to the qubit. Following the same process as in section III (again with details in Appendix \emph{A}), where the angle $\theta$ is now the polar angle in the $y-z$ great circle of the Bloch sphere, we can write the stochastic Hamiltonian \cite{Chantasri2013}
\be \label{zdriveH}
H = p (\Delta - r \sin \theta /\tau) - (r^2 - 2 r \cos \theta + 1 )/2\tau,
\ee
where the optimal readout along $z$ is $r^\star = \cos\theta - p \sin\theta$. Substituting this in once again gives \eqref{hgen3} $H = (p^2-1)a + p b$, for \be \label{zab} a=\sin^2 \theta / 2\tau \quad\text{and}\quad  b=\Delta - \sin \theta \cos \theta / \tau. \ee The corresponding expression for $\dot{S}$ is
\be \label{z_sdot}
\dot{S} = - \frac{(1+p^2)\sin^2\theta}{2\tau}.
\ee
This system contains only one fixed point in a $\pi$-long region of $\theta$ in phase space, and contains no periodic OPs. The fixed point is located at $(\bar{\theta},\bar{p}) = (\arctan(\tau \Delta),-\tau \Delta)$, and marks the point in the phase-space where the measurement backaction and drive exactly cancel out each other's effects. Several different regions defined by the separatrix of energy $E_\star = -\tau \Delta^2/2$ exist in the phase space. We label these regions A, B, and C, in the phase portrait and $\dot{S}$ diagrams shown in Fig.~\ref{fig-z_ps}. The divergence of the separatrix towards $p\rightarrow -\infty$ sends trajectories in regions B and C through regions where $\dot{S}$ tends towards $-\infty$. This biases the highest probability paths, for long evolution times, to sit in region A, by forcing paths over longer evolution times ($t \gtrsim \Delta^{-1}$) in regions B and C towards a probability density of zero much faster than paths in region A. This reflects the asymmetry of the Rabi drive, which pushes paths towards the positive $\theta$ direction; probable paths rotate with the drive rather than against it. In appendix \emph{B} we derive explicit expressions for actions in regions A and C, in the diffusive Rabi limit $\Delta \tau \gg 1$ (in which the drive overwhelms measurement dynamics). The expressions derived there confirm our findings in the discussion above, showing that $S$ in region C ($S_C$) diverges toward $-\infty$ as $\theta_f$ approaches a measurement eigenstate along an OP moving against the Rabi drive.
\begin{figure}
\begin{tabular}{c}
\begin{picture}(100,100) \put(-30,-12){\includegraphics[width=.7\columnwidth]{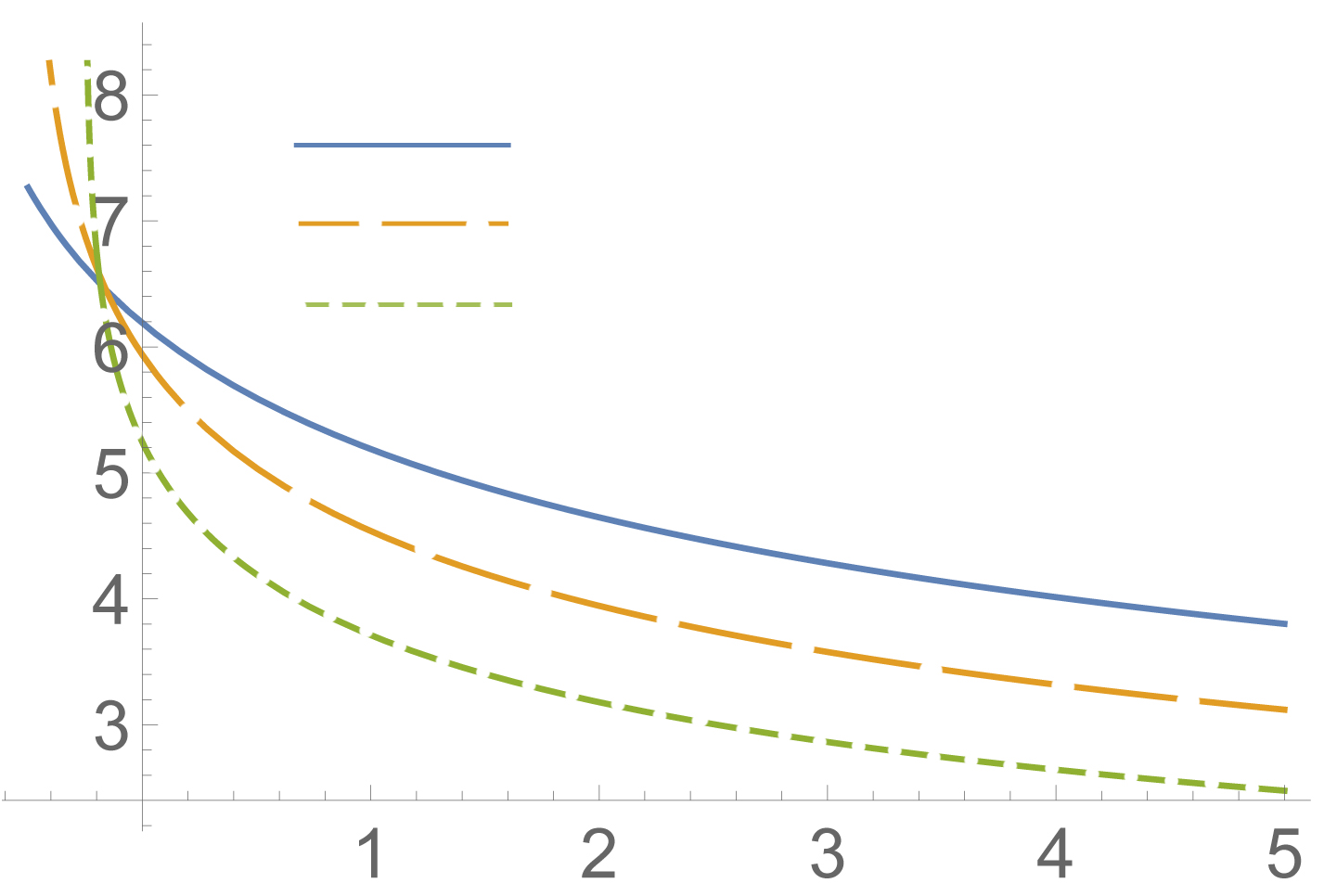}} \put(-60,38){$T_{2\pi} (\mu s)$} \put(40,-22){$E$ (MHz)} \put(42,85){$\Delta = 1$(MHz), $\tau = 2 (\mu s)$} \put(42,73){$\Delta = 1$(MHz), $\tau = 1 (\mu s)$} \put(42,61){$\Delta = 1$(MHz), $\tau = 0.5 (\mu s)$} \put(-50,90){(a)}\end{picture} \\ \\
\begin{picture}(100,100) \put(-30,-32){\includegraphics[width=.7\columnwidth]{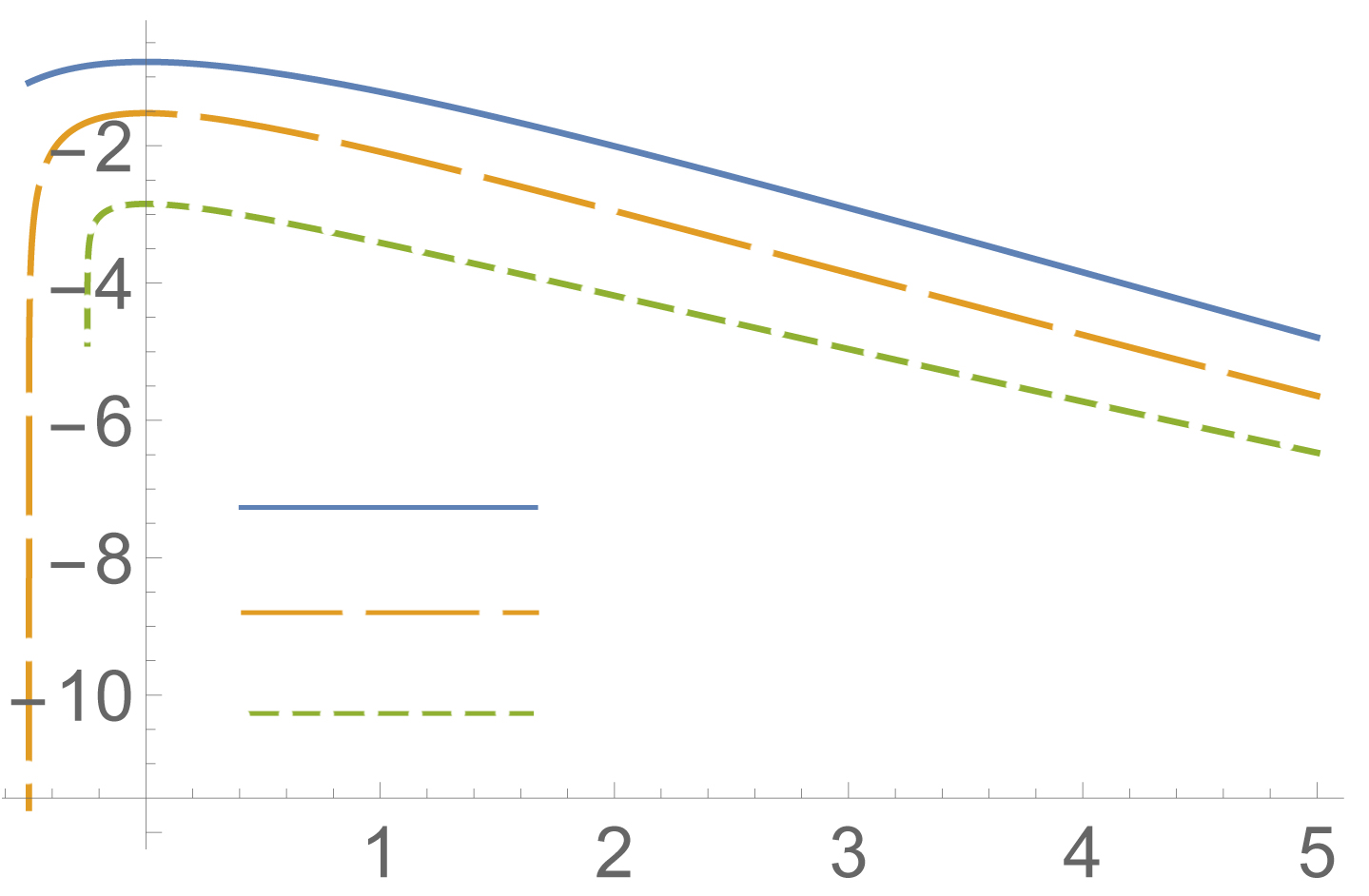}} \put(-50,18){$S$} \put(40,-42){$E$ (MHz)} \put(42,13){$\Delta = 1$(MHz), $\tau = 2 (\mu s)$} \put(42,1){$\Delta = 1$(MHz), $\tau = 1 (\mu s)$} \put(42,-11){$\Delta = 1$(MHz), $\tau = 0.5 (\mu s)$} 
\put(-50,65){(b)}
\end{picture} \\ \\ \\ \\
\end{tabular}
\caption{We plot (a) the time $T_{2\pi}$ required to make one complete rotation about the Bloch sphere in region A, as function of stochastic energy $E$, and (b) the action associated with those same MLPs, moving from $\theta_i = 0$ to $\theta_f = 2\pi$ over the time $T_{2\pi}$. 
Once again $E$ is in units of inverse time, and $S$ is dimensionless.
Notice that the peak at $E=0$ in all curves in (b) is an example of the result \eqref{zeroE}, which states that time taken by the $E=0$ path maximizes the probability to arrive at a chosen $\theta_f$, given $\theta_i$, and is thereby the optimal time in which to move between those two states. 
}\label{fig-act}
\end{figure}

\par The discussion of multipaths here can be relatively short, since we see many of the same features discussed in section III. {\color{black} For initial conditions to the left side of region B, a small caustic will form on the opposite side of the region after a short time; multipaths therein are short-lived due to the loss in probability density experienced by these OPs as they move towards $p\rightarrow -\infty$ after moving through across the top of region B. 
Multipaths from this caustic are thereby restricted to a relatively narrow range of final $T$ and $\theta_f$. A similar example may be found in the supplementary materials of \cite{Mahdi2016}. Aside from this,} we need only consider multipaths due to different winding numbers, which are now restricted to be in region A. All paths there travel in the same direction, and $\dot{\theta}$ changes monotonically with $E$, so no caustics can form within this region (using the logic surrounding \eqref{dotq_form}). 
Getting two paths of the same energy in an experimentally visible multipath is now impossible, because the equal energy pair travel in opposite directions, and we have excluded those in region C. We explore an explicit example of a winding number multipath within region A for the case $\theta_i = \pi/2$ in Fig.~\ref{fig-wn}, where $\tau =1 = \Delta^{-1}$, so that time is in units of $\tau$ and $E$ is in units of $\Delta$. 
We want to find a pair of $\theta_T$ one winding number apart, which have probability densities which are approximately equal \emph{and} reasonably large compared to the most-likely $\theta_T$ at $T$. We do this by finding the $p_T =0$ path in the manifold at $T=8 \tau$, which leads to $\theta_T^{(0)}$, and then post-select on states one winding number apart to either side of the most-likely final coordinate to form our multipath (see Fig.~\ref{fig-wn}(a)). 
We additionally show that the probability density to obtain a path which rotates faster, generating a winding number difference of two (or more), is negligibly small, such that we expect to only observe two paths contributing to the multipath in experiment. 
The expected numbers of quantum trajectories contributing to each OP in a multipath may be estimated using the actions for each path. If $N$ paths meet the multipath post-selection condition, then approximately $N_i = N e^{S_i}/\sum_j e^{S_j}$ trajectories are expected to contribute to the $i^{th}$ path, where $N = \sum_i N_i$, and $j$ indexes all paths meeting the pre- and post-selection with non-negligible probability density $\mathcal{P}_j \sim e^{S_j}$. 
The ratio of any two $N_i$ may be estimated by $N_a / N_b = \mathcal{P}_a/\mathcal{P}_b \approx e^{S_a - S_b}$. We also note that the correspondence of the three roots of the Lagrange manifold in Fig.~\ref{fig-wn}(b) with the local extrema of the action in Fig.~\ref{fig-wn}(a) is an example of the optimization result \eqref{sg} choosing $p_T=0$ that we have used throughout our work above. 
{\color{black} (When there are several possible roots $p_T = 0$, there are several local extrema in the probability density for different $\theta_T$; $\theta_T^{(0)}$ above is chosen as the highest maximum from among those options.)}
\par In Fig.~\ref{fig-act} we explore the relative probability densities for higher winding number differences more generally. We compare the time $T_{2\pi}$ required for trajectories in region A to complete one full rotation about the Bloch sphere, against the action associated with doing so. We see that as energy increases, $T_{2\pi}$ levels off to approach zero asymptotically, whereas $S$ becomes more negative in an approximately linear fashion. We learn that in this system, winding number paths are best found among relatively small $E$, because at larger $E$ an enormous difference in probability density will exist between paths separated by one winding count over an experimentally viable duration of time. Note that the presence of the maximum of the curves in Fig.~\ref{fig-act} at $E=0$ serves as an explicit example of the result \eqref{zeroE} which we have referenced throughout this paper.

\section{Discussion \& Conclusions}
We have used OPs in continuously monitored quantum systems, obtained from extremizing a stochastic path integral \cite{Chantasri2013,Chantasri2015}, to understand these systems' dominant physical behaviors. We did this by optimizing among OPs after relaxing a boundary condition in the SPI formalism, using methods from Hamiltonian mechanics. We may easily obtain the path describing the most-likely travel time between two given states by setting the stochastic energy to zero, and obtain a path leading to the most-likely final state after a chosen elapsed time by choosing a path with zero final momentum. We investigated the dynamics of a qubit system in which two non-commuting observables are measured simultaneously \cite{Jordan2005,Leigh2016}, and found that the relative probability densities for collapse to eigenstates of each observable are determined by the relative strengths of the measurements; measurements of equal strength discourage wave-function collapse altogether, instead favoring persistent diffusion of the quantum state. 
This is an example of how the SPI formalism allows for an insightful, but computationally simple, analysis of the long-term dynamics arising in diffusive quantum trajectories.
\par We also used our SPI/OP formalism to define ``multipaths'', an instability in the stochastic dynamics wherein several distinct routes through Hilbert space dominate the evolution of the system between given boundary conditions. The onset of multipaths may be understood from the folding of a Lagrangian manifold in the OP phase space, in direct analogy with the formation of caustics in ray optics. We study this phenomenon both in the two-observable scheme, and in a qubit subject to one measurement and drive \cite{Chantasri2013,Weber2014}. Our work predicts the presence of experimentally visible multipaths in both systems. Experimental confirmation of multipaths was recently found in quantum trajectories from a weakly-monitored resonance fluorescence system \cite{Mahdi2016}. We consequently expect that the phenomenon is quite common in diffusive quantum trajectories generally. The optimizations of final states and traversal times described above, and the existence of multipaths, are connected through their common mathematical origins in the ``momenta'' conjugate to coordinates parameterizing the quantum state in the SPI/OP formalism. We restricted our analysis by only considering pure states in one-dimensional Hamiltonian phase spaces, but this restriction may be relaxed in future work.
\par The existence of multipath instabilities in qubit dynamics may have consequences for quantum information processing, quantum feedback control problems, and quantum error correction. Taken together, our results open the door for further studies of dynamical instabilities in diffusive quantum trajectories, and the study of chaos and long-term (un)predictability of continuously-measured quantum systems.

\begin{acknowledgments}
We are grateful to Miguel Alonso for introducing us to the concept of the Lagrange Manifold in connection with multipaths. We thank Mahdi Naghiloo and Kater Murch for helpful discussions, and their collaboration \cite{Mahdi2016} in searching for multipaths experimentally. We appreciate Justin Dressel's guidance in using his stochastric trajectory simulation codes. We also acknowledge helpful discussions with Irfan Siddiqi and members of his group, particularly Leigh Martin and Shay Hacohen-Gourgy, and with Ashok Das, Mark Dykman, Rodrigo Gutierrez-Cuevas, Evan Ranken, and Carlos Stroud.
\par Numerical integration of ODEs for Lagrangian manifolds was performed in Python 2.7, using methods based on those in \cite{NumRecC}.
\par This work is supported by NSF grant DMR-1506081 and US  Army  Research  Office Grant No.  W911NF-15-1-0496.
\par This research was supported in part by Perimeter Institute for Theoretical Physics. Research at Perimeter Institute is supported by the Government of Canada through Industry Canada and by the Province of Ontario through the Ministry of Economic Development \& Innovation.
\end{acknowledgments}


\appendix
\section{Bayesian Derivations of $\mathcal{F}$ \& $\mathcal{G}$, and Comparison to SME}
The aim of this appendix is to motivate the forms of $\mathcal{F}$ and $\mathcal{G}$, as defined in \eqref{spif_h}, that we subsequently use in sections II, III, and IV. We will first review how to obtain equations of motion from a measurement operator, and then review the construction of the specific short-time measurement operators for a weak quantum measurement that we will be interested in, as in \cite{Korotkov2011,Korotkov1999,Korotkov2001,Korotkov2016,Chantasri2013}. This will allow us to compare the computation of the term $\mathcal{F}$ using a Bayesian approach \cite{Korotkov2011,Korotkov1999,Korotkov2001,Korotkov2016} to a Stochastic Master Equation (SME) approach \cite{BookBarchielli,BookWiseman,Jacobs2006,Murch2013,Jordan2013rev,BookCarmichael}. 

\subsection{Bayesian Approach}

\subsubsection*{From Measurement Operators to Dynamics}
Suppose we construct some evolution operator $\mathcal{U}$ containing both the unitary dynamics for some short time evolution, and measurement dynamics from some arbitrary number of generalized measurements. Then the density matrix $\rho$ evolves according to \cite{BookNielsen}
\be \label{rhodynamics}
\rho(t+dt) = \frac{\mathcal{U}\rho(t) \mathcal{U}^\dag}{tr(\mathcal{U}\rho(t) \mathcal{U}^\dag)}.
\ee 
We suppose that $\mathcal{U}$ may be decomposed as a product of operators $\mathcal{U} = \mathcal{A} \mathcal{B} \mathcal{C} ...$ where each of $\mathcal{A}$, $\mathcal{B}$, etc. describe some particular measurement or dynamics applied to the system of interest whose state is specified by $\rho$. To first order in $dt$, we may write these operators as
\be
\mathcal{A} = \left( \mathtt{I} + \hat{A} dt\right), \quad \mathcal{B} = \left( \mathtt{I} + \hat{B} dt \right),
\ee
and so on, where $\mathtt{I}$ is the identity matrix. We have stripped the operators of any overall constants they may be carrying, as these will necessarily cancel between the numerator and denominator of \eqref{rhodynamics} anyway. Then 
\be \begin{split}
\mathcal{U} &=  \left( \mathtt{I} + \hat{A} dt\right)\left( \mathtt{I} + \hat{B} dt \right) ... \\
&\approx  \mathtt{I} + dt \left( \hat{A} + \hat{B} + ... \right) 
\end{split}\ee
to first order in $dt$. This is important, because by neglecting terms of order $dt^2$ and up, we also eliminate the need to worry about the commutators between $\hat{A}$, $\hat{B}$, etc.; we are assuming that $dt$ is small enough that the ordering of the operators $\mathcal{A} \mathcal{B} ...$ in $\mathcal{U}$ is irrelevant. 
\par We continue by finding an expression accurate to first order in $dt$ for \eqref{rhodynamics}. We may write the numerator
\be \begin{split}
\mathcal{U}&\rho(t) \mathcal{U}^\dag \\ &\approx \left( \mathtt{I} + dt \left( \hat{A} + \hat{B} + ... \right)  \right) \rho(t) \left( \mathtt{I} + dt \left( \hat{A}^\dag + \hat{B}^\dag + ... \right)  \right)
\\ & \approx \rho(t) + dt \left( (\hat{A} + \hat{B}+...)\rho(t) + \rho(t) (\hat{A}^\dag+\hat{B}^\dag+...) \right),
\end{split} \ee
and define $\zeta \equiv (\hat{A} + \hat{B}+...)\rho(t) + \rho(t) (\hat{A}^\dag+\hat{B}^\dag+...)$ for ease of notation. When the operators $\hat{A}$, $\hat{B}$ etc. are Hermitian, we may simplify to $\hat{\zeta}=\lbrace \hat{A} + \hat{B} + ..., \rho \rbrace$ where $\lbrace , \rbrace$ denotes the anti-commutator (this generally applies to operators describing a measurement). Likewise, if the operators are anti-Hermitian, we may write $\hat{\zeta} = [\hat{A} + \hat{B} +..., \rho]$ where $[,]$ is the commutator (and this generally applies to operators arising from some unitary evolution process). The denominator of \eqref{rhodynamics} is the trace of the numerator, \emph{i.e.}
\be \begin{split}
\left(tr(\mathcal{U}\rho(t) \mathcal{U}^\dag) \right)^{-1} &\approx \left( 1 +  tr \left( \hat{\zeta} \right)dt \right)^{-1}
\\ & \approx 1 -  tr \left(  \hat{\zeta} \right)dt.
\end{split} \ee
Taking these together, we have
\be \begin{split}
\rho(t+dt) &= \mathcal{U}\rho(t) \mathcal{U}^\dag \left(tr(\mathcal{U}\rho(t) \mathcal{U}^\dag) \right)^{-1}
\\ & \approx \rho(t) + dt \left(\hat{\zeta}  - \rho(t) tr\left( \hat{\zeta} \right) \right)
\end{split} \ee
or alternately
\be \label{o1_evolution}
\dot{\rho} \approx \frac{\rho(t+dt)-\rho(t)}{dt} \approx \hat{\zeta}  - \rho(t) tr\left( \hat{\zeta} \right).
\ee

\subsubsection*{Measurement Operator and $\mathcal{G}$}
We will consider a measurement along the $z$-axis of a qubit, as described in \cite{Korotkov2011,Korotkov1999,Chantasri2013}. The eigenstates of $\sigma_z$ are either $\ket{+1}$ or $\ket{-1}$. We may write a Gaussian probability density associated with each of those outcomes to a $z$ measurement {\color{black} (for $z=+1$ as the excited state, and $z=-1$ as the ground state) 
\be
P_\mp = \sqrt{\frac{dt}{2\pi \tau_z}}\exp \left[-\frac{(r_z\mp1)^2 dt}{2 \tau_z}\right],
\ee} 
where $r$ is the readout and $\tau_z$ is a characteristic measurement time as used in sections III and IV. A weak measurement is characterized by a $\tau_z$ large enough that there is considerable overlap between $P_+$ and $P_-$, meaning that the measurement does not distinguish strongly between the two states (but does not collapse the quantum state to an eigenstate either). By Bayes' rule, we may write
\be \label{bayes}
P(\pm1| r)P(r) = P(r|\pm1)P(\pm1).
\ee
We will represent our qubit state as a density matrix, which can be related to the Bloch sphere coordinates $x$, $y$, and $z$ according to the usual decomposition
\be \label{densmat}
\rho = \left( \begin{array}{cc} \rho_{11} & \rho_{12} \\ \rho_{12}^\ast & \rho_{22} \end{array} \right) = \frac{1}{2} \left( \begin{array}{cc} 1+z & x-i y \\ x+ iy & 1-z \end{array} \right).
\ee
We invoke \eqref{bayes} to express the evolution of the density matrix when a weak $z$ measurement is applied between times $t$ and $t+dt$. We may write:
\be\label{rho11}\begin{split}
\rho_{11}(t+dt) &= P(-1|r) = \frac{P(r|-1)P(-1)}{P(r)} \\&= \frac{P_- \rho_{11}(t)}{P_- \rho_{11}(t) + P_+ \rho_{22}(t)},
\end{split} \ee and \be\label{rho22}\begin{split}
\rho_{22}(t+dt) &= P(+1|r) = \frac{P(r|+1)P(+1)}{P(r)} \\&= \frac{P_+ \rho_{22}(t)}{P_- \rho_{11}(t) + P_+ \rho_{22}(t)}.
\end{split} \ee
A similar relation for the off-diagonal terms in $\rho$ can also be written \cite{Korotkov2011,Korotkov1999,Korotkov2001,Korotkov2016}, where for pure states $\rho_{12}/\sqrt{\rho_{11} \rho_{22}} = constant$, leading to:
\be\label{rho12}
\rho_{12}(t+dt) = \frac{\rho_{12}(t) \sqrt{P_- P_+}}{P_- \rho_{11}(t) + P_+ \rho_{22}(t)}.
\ee
\par It is now easy to verify that the choice of measurement operator
\be
\mathcal{Z} = \left( \begin{array}{cc} \sqrt{P_-} & 0 \\ 0 & \sqrt{P_+} \end{array} \right)
\ee
reproduces \eqref{rho11}, \eqref{rho22}, and \eqref{rho12} under application of \eqref{rhodynamics} for $\mathcal{U} \rightarrow \mathcal{Z}$. 
Expanding $\mathcal{Z}$ to first order and dropping a constant $(dt/2\pi \tau_z)^\frac{1}{4}$, we may write
\be
\mathcal{Z} \approx \mathtt{I} - dt\frac{(r_z-\sigma_z)^2}{4\tau_z} = \mathtt{I} + \hat{Z} dt
\ee
for $\hat{Z} = -(r_z -\sigma_z)^2/(4\tau_z)$.
In the language of section II, we may make the association 
\be P(r|\rho) = tr(\mathcal{Z} \rho(t) \mathcal{Z}^\dag) = P_- \frac{1+z}{2} + P_+ \frac{1-z}{2} \ee
as in \cite{Chantasri2013}. Note that $\rho$ and the coordinates $\mathbf{q} = (x,y,z)$ contain exactly the same information. The term $\mathcal{G}$ or $g(q,r)$ describing the probability in section II is derived by expanding $\ln(P(r|\rho)) = g(z,r)dt + C + \mathcal{O}(dt^2)$, where $C$ is a constant that does not impact the dynamics, to obtain the form used in sections II, III, and IV (see equations \eqref{assum_iii}, \eqref{G_xzmeas}, and \eqref{zdriveH} specifically):
\be \label{gz}
g(z,r_z)dt = - \frac{dt}{2\tau_z} (r_z^2 - 2 r_z z +1).
\ee
By similar arguments to those applied above, we may find that a measurement along $x$ can be expressed by an operator
\be
\mathcal{X} \approx  \mathtt{I} - dt\frac{(r_x-\sigma_x)^2}{4\tau_x} = \mathtt{I} + \hat{X} dt
\ee
where $\hat{X} =  -(r_x - \sigma_x)^2/(4\tau_x)$ and
\be \label{gx}
g(x,r_x)dt = - \frac{dt}{2\tau_x} (r_x^2 - 2 r_x x +1).
\ee
The two $g$ terms may simply be added together when both measurements are present, as is done in the construction of \eqref{G_xzmeas} and \eqref{hr} used in section III.
\par We note that it is possible to simplify $\hat{X}$ and $\hat{Z}$ further to first order in $dt$, for the purposes of applying \eqref{o1_evolution}. Expanding the square in $\hat{X}$, for instance, leads to a form
\be \begin{split}
\mathcal{X} &= \mathtt{I}\left(1 - dt \frac{r_x^2+1}{4\tau_x} \right) + dt \frac{\sigma_x r_x}{2\tau_x}
\\ & = \mathtt{I} (1 + \lambda dt) + \mu \sigma_x dt
\end{split} \ee
and similarly for $\mathcal{Z}$. In \eqref{o1_evolution} the terms $\lambda$ always cancel out of the dynamics completely. Therefore we may simply keep the $\mu$ terms, taking
\be \label{final_xz_measop}
\hat{x} = \frac{r_x}{2 \tau_x} \sigma_x, \quad\text{and}\quad \hat{z} = \frac{r_z}{2\tau_z} \sigma_z
\ee
as our measurement operators when computing the equations of motion.

\subsubsection*{Dynamics $\mathcal{F}$, for $\mathcal{U} \rightarrow \mathcal{X} \mathcal{Z}$ \& $\mathcal{U} \rightarrow \mathcal{R} \mathcal{Z}$}
In the section above, we have justified the use of the terms \eqref{gz} and \eqref{gx} as the expressions for $\mathcal{G}$ in our stochastic Hamiltonians. We must now obtain expressions for $\mathcal{F}$. The only unitary dynamics we will have for our qubits are a Rabi drive, described by a Hamiltonian $\hat{h} = - \Delta \sigma_x /2$, where $\Delta$ is the Rabi frequency. The associated time evolution operator is $\mathcal{R} = e^{-i \hat{h} dt}$. 
To first order in $dt$ we may write
\be
\mathcal{R} \approx \mathtt{I} - i \frac{\Delta dt}{2} \sigma_x = \mathtt{I} + \hat{R} dt
\ee
for 
 $
\hat{R} =  - i \Delta \sigma_x / 2.
$
We have now assembled all of the tools required to use \eqref{o1_evolution}.
\par The system treated in section III has $\mathcal{U} = \mathcal{X} \mathcal{Z}$ representing simultaneous monitoring along the $x$ and $z$ axes of the Bloch sphere. We use
$ \hat{\zeta} = \lbrace \hat{x} + \hat{z},\rho \rbrace $
to compute $\dot{\rho}$ via \eqref{o1_evolution}. Given $\dot{\rho}$, the equations of motion \eqref{strato_xz} are then obtained via $\dot{q} = Tr(\dot{\rho} \sigma_q)$ where $q = x,y,z$. Similarly, the system with a single measurement and Rabi drive treated in section IV has $\mathcal{U} = \mathcal{R} \mathcal{Z}$. We obtain $\dot{\rho}$ by applying \eqref{o1_evolution} with $\hat{\zeta} = \lbrace \hat{z}, \rho \rbrace + [\hat{R},\rho] = \lbrace \hat{z} ,\rho \rbrace - i [\hat{h},\rho]$. The application of  $\dot{q} = Tr(\dot{\rho} \sigma_q)$ then yields the equations \cite{Chantasri2013}
\be\label{RZequs}\begin{split}
\dot{x} &= - \frac{x z r_z}{\tau_z}, \\
\dot{y} &= \Delta z - \frac{y z r_z}{\tau_z}, \\
\dot{z} &= - \Delta y + \frac{(1-z^2) r_z}{\tau_z}, 
\end{split}\ee
used to construct \eqref{zdriveH}.

\subsection{SME-based Computation of $\mathcal{F}$}
We may compare the above equations of motion \eqref{strato_xz} and \eqref{RZequs} with those obtained from a stochastic master equation (SME), commonly used in the literature \cite{BookBarchielli,BookWiseman,Jacobs2006}. We write it here in units $\hbar\rightarrow 1$, and assume perfect measurement efficiency, such that:
\be \label{sme} 
d \rho = i[\rho,\hat{h}]+ \sum_{i} \mathcal{L}[L_i, \rho] dt + \mathcal{M}[L_i,\rho]dW_i.
\ee
The index $i$ is over different measurement operators applied to the system, and $\hat{h}$ is a Hamiltonian describing unitary dynamics.  The Lindblad superoperator term is $\mathcal{L}[L_i,\rho] = L_i \rho L_i^\dag - (L_i^\dag L_i \rho + \rho L_i^\dag L_i)/2$, and the measurement superoperator is $\mathcal{M}[L_i,\rho] = L_i\rho + \rho L_i^\dag - Tr[\rho(L_i + L_i^\dag)]\rho$. The dimensionless readout in each channel goes as $r_i = \sqrt{\tau_i}dW_i/dt + q_i$. 
The $dW_i$ are Gaussian white noise (Wiener process) in each output channel.
\par For the $x$ and $z$ measurements of section III: We only have measurement dynamics, and the Hamiltonian describing unitary evolution in the system is $\hat{h} = 0$. We have $\rho$ representing the qubit density matrix, and take $L_i$ to be the measurement operators along $x$ or $z$ such that $L_x = \sigma_x/2\sqrt{\tau_x}$ and $L_z = \sigma_z/2\sqrt{\tau_z}$. The times $\tau_x$ and $\tau_z$ are the characteristic measurement times along each quadrature, defined in the same manner as above. The $\sigma_i$ are Pauli matrices.
\par For the $z$ measurement with drive of section IV: We take $\hat{h} = - \Delta \sigma_x /2$, and have only $i = z$ with $L_z = \sigma_z/2\sqrt{\tau_z}$. 
\par One may ask whether or not the Stochastic Differential Equations (SDEs) derived from \eqref{sme} agree with \eqref{strato_xz} and \eqref{RZequs}, respectively, after the readout(s) $r_i = \sqrt{\tau_i}dW_i/dt + q_i$ are substituted in. The immediate expressions which come out of the SME will be a set of equations
\be \label{ito}
dq = \omega_q dt + \xi_{qi} dW_i.
\ee
Eliminating $dW_i$ in favor of $r_i$ in \eqref{ito} will \emph{not} give us the equations obtained from the Bayesian approach above. We do however find that the equations
\be \label{strato} \begin{split}
dq &= \left( \omega_q - \frac{1}{2} \sum_{ij} \xi_{ji} \partial_j \xi_{qi} \right)dt + \xi_{qi} dW_i \\&= \bar{\omega}_q dt + \xi_{qi} dW_i
\end{split} \ee
match those of the Bayesian approach exactly. When the form \eqref{strato} is subjected to a Stratonovich integral, it gives the same solutions as when \eqref{ito} is integrated with an It\^{o} integral \cite{BookGardiner2}; the form \eqref{strato} is standard for converting between It\^{o} and Stratonovich SDEs.

\section{Diffusive Rabi Oscillations}

This section considers a special case of the Rabi-driven system described in section IV; specifically, we examine the diffusive oscillation limit $\Delta \gg 1/\tau$. In this limit we are able to evaluate many of the claims we have just made about multipaths more rigorously. In our paper \cite{Chantasri2013}, we worked out analytic results for the quantum jump limit, wherein the measurement dynamics overwhelm the Rabi drive. We are now working in the opposite limit, where the drive overwhelms the measurement backaction. 
\par We begin with the time taken to move between two states $\theta_i$ and $\theta_f$. We examine the zero energy lines for simplicity, noting from Fig.~\ref{fig-z_ps} that these OPs are qualitatively representative of all of those in regions A and C. The traversal time between two states is given by the zero-energy simplification of \eqref{time2} with \eqref{zab} 
(we let $\theta_f > \theta_i$; we may reverse the sign in the opposite case to get a positive $T$ in either case)
\be
T = \tau \int_{\theta_i}^{\theta_f} \frac{d \theta}{\sqrt{(\tau \Delta- \sin \theta \cos \theta)^2 + \sin^4 \theta}}.
\ee
We now make an expansion in $\Delta \tau$ as a large parameter, keeping the order $\Delta^2$ and order $\Delta$ terms in the denominator, pulling out a factor of $\Delta$ from this, and expanding the square root and denominator to leading order to find,
\be
T \approx \frac{1}{\Delta} \int_{\theta_i}^{\theta_f}  d\theta \left(1 + \frac{\sin \theta \cos \theta}{\Delta \tau} \right).
\ee
Carrying out the integral leads to
\be
T \approx \frac{\theta_f - \theta_i}{\Delta} - \frac{\cos 2 \theta_f - \cos 2 \theta_i}{4 \Delta^2 \tau}.
\ee
The first term in the above result is from the Rabi drive, where the subtended angle is the Rabi rate times the elapsed time, and the second term is a correction to it from the diffusive measurement dynamics. The time should be the same for either the trajectory moving in positive $\theta$ (region A), and in negative $\theta$. We noted above however, from Fig.~\ref{fig-z_ps}(c), that the probability associated with the path in region A should be much larger than in region C. 

We consider both the clockwise and counter clockwise action.  First we consider the trajectory in the positive $\theta$ direction, and again expand using $\Delta \tau$ as a large parameter.  When $p_+$ is expanded, the leading order term proportional to $\Delta \tau$ vanishes; the constant order terms also cancels, leaving the inverse order term as the leading one,
\be
p_+ \approx \frac{\sin^2 \theta}{2 \Delta \tau}.
\ee
Integrating this function gives the action for this trajectory,
\be
S_A \approx -\int_{\theta_i}^{\theta_f} d\theta p_+ = -\frac{\theta_f - \theta_i}{4 \Delta \tau} + \frac{\sin 2\theta_f - \sin 2\theta_i}{8\Delta \tau},
\ee
where we use the subscript A in association with the region these paths inhabit, as labeled in Fig.~\ref{fig-z_ps}. This action gives the probability of reaching these angles on the zero energy line.
We note that as $\Delta \tau$ becomes very large, the action vanishes, recovering deterministic dynamics.

We can now ask about the dynamics in the opposite direction.  In this case, the terms proportional to $\Delta \tau$ do not cancel, giving a leading order momentum of $p_- = - 2 \Delta \tau/\sin^2 \theta$.  Integrating this gives (for the $E=0$ path in region C)
\be
S_C = 2 \Delta \tau (\cot \theta_i - \cot \theta_f).
\ee
Note that when $\theta$ approaches integer multiples of $\pi$ (including $\theta=0$), the action diverges.  This indicates the measurement diffusion cannot cause the state to move backwards beyond the poles of the Bloch sphere.  This makes sense because the measurement causes no diffusion at the poles, while the Rabi drive takes the state in the positive $\theta$ direction. This is entirely consistent with arguments in section IV based on Fig.~\ref{fig-z_ps} regarding the vanishingly small probabilities for long time evolutions in regions B and C. This also lends quantitative backing to our assertion that equal energy winding-number multipaths are not likely to be experimentally viable in this system. Paths in region C would take the same amount of time to traverse $\theta_f \rightarrow \theta_i$ as those in region A do to traverse $\theta_i\rightarrow\theta_f$, if they could squeeze past the asymptotes at the poles. The paths in region C get stuck at the poles however, an must wait for the path in region A to catch up, while their probability grows vanishingly small.

\section{MLP/LLP/SP}

\begin{figure*}
\begin{tabular}{cc}
\begin{minipage}{.63\textwidth}
\begin{picture}(350,227)
\put(-14,-10){\includegraphics[width=310pt,trim={20pt 20pt 20pt 20pt},clip]{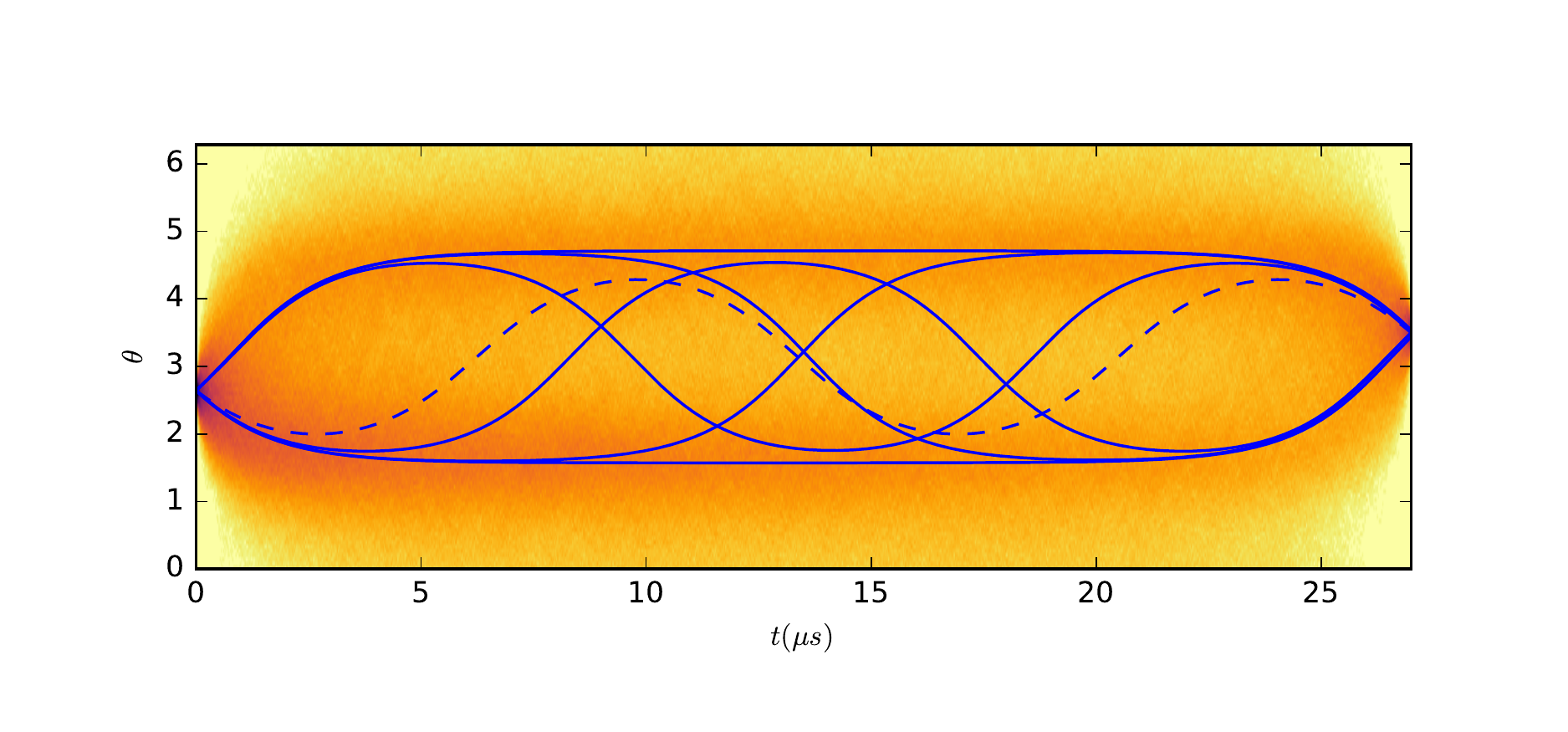}}
\put(-14,105){\includegraphics[width=310pt,trim={20pt 20pt 20pt 20pt},clip]{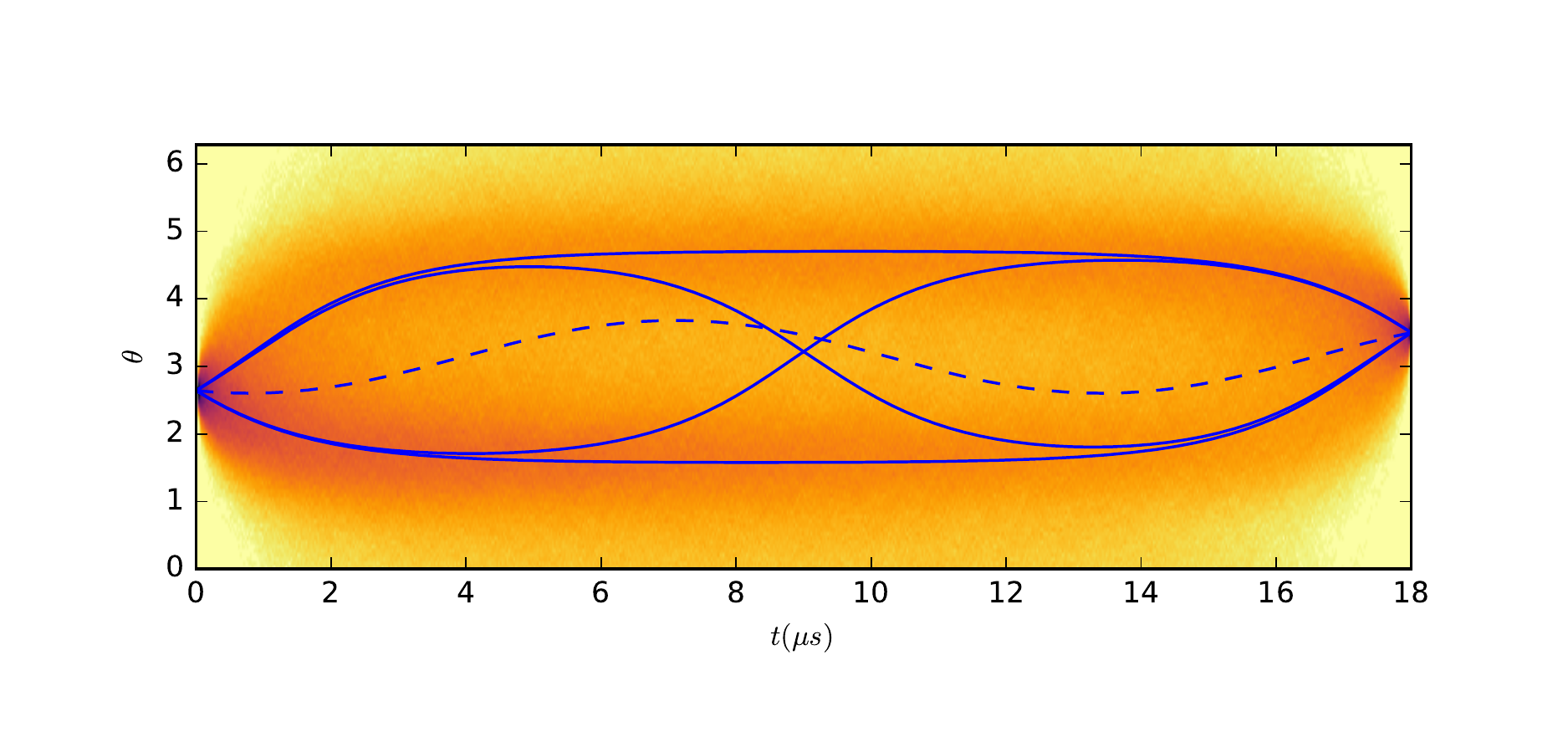}}
\put(287,7){\includegraphics[width=.08\textwidth,trim={0 0 22pt 0},clip]{dens_cbar.pdf}}
\put(19,207){(a)}
\put(19,92){(b)}
\put(110,143){{\footnotesize\it 1}}
\put(100,157){{\footnotesize\it 2}}
\put(80,167){{\footnotesize\it 3}}
\put(88,184){{\footnotesize\it 4}}
\put(110,197){{\footnotesize\it 5}}
\put(125,30){{\footnotesize\it a}}
\put(105,38){{\footnotesize\it b}}
\put(82,41){{\footnotesize\it c}}
\put(61,51){{\footnotesize\it d}}
\put(116,51){{\footnotesize\it e}}
\put(140,68){{\footnotesize\it f}}
\put(157,83){{\footnotesize\it g}}
\end{picture}
\end{minipage} & \begin{minipage}{.35\textwidth}  
\begin{center} 
\begin{tabular}{c|c|c|c|c}\hline 
 {\footnotesize\#} &{\footnotesize $p_i$} & {\footnotesize OP Type} & {\footnotesize Path \% 1} & {\footnotesize Path \% 2} \\ \hline      
&-0.44247943 & MLP & 41.4 & 54.8 \\   
&0.10592792 & LLP & 24.4 & -- \\
&0.90882604 & MLP & 34.2 & 45.2\\ \hline\hline       
{\footnotesize\it 1}& -0.46340433 & MLP & 44.2 & 45.4 \\       
{\footnotesize\it 2}& -0.44816125 & MLP & 12.9 & 13.3 \\  
{\footnotesize\it 3}& 0.12631625 & LLP & 2.6 & -- \\  
{\footnotesize\it 4}& 0.89614502 & MLP & 4.0 & 4.1 \\  
{\footnotesize\it 5}& 0.93875909 & MLP & 36.3 & 37.2\\ \hline\hline
{\footnotesize\it a}&-0.46343702 & MLP & 44.6 & 44.8 \\                 
{\footnotesize\it b}&-0.46284337 & MLP & 12.3 & 12.3 \\  
{\footnotesize\it c}&-0.43987448 & MLP & 1.3 & 1.4 \\ 
{\footnotesize\it d}&-0.33363249 & LLP & 0.5 & -- \\  
{\footnotesize\it e}&0.91225871 & MLP & 1.1 & 1.1  \\
{\footnotesize\it f}&0.93735002 & MLP & 3.7 & 3.7 \\  
{\footnotesize\it g}&0.93880374 & MLP & 36.5 & 36.7 \\ \hline\hline
\end{tabular}\end{center}
\end{minipage}\end{tabular}
\caption{We plot the example with five OPs from Fig.~\ref{fig-LMonset}(e,h) in (a), and the example with seven OPs from Fig.~\ref{fig-LMonset}(f,i) in (b), superposed over density plots of simulated stochastic trajectories \cite{JustinCode}, as in Fig.~\ref{fig-dens}. The sub-ensembles of trajectories whose densities are shown are those meeting the post-selection $\theta_f = 3.5$ at $T = 18 \mu s$ (a), and $T = 27 \mu s$ (b). Density is shown as a histogram with equal-area bins (pixels), and the colorbar is normalized relative to the highest bin count between boundary conditions, such that 1 is the highest trajectory density, and 0 means no trajectories at all. We find that all but one optimal path is a MLP in each case (solid lines), and the remaining optimal path is a LLP (dashed). In the table on the right we list extra information for each of the sets of paths from the triple path in Fig.~\ref{fig-dens}, the quintuple path in (a) and the septuple path in (b). Labels in the \# column match those in the plots (a) and (b). Additional columns give the initial momenta $p_i$ generating the OP, its type (MLP/LLP/SP), and the percentage of trajectories expected to belong to each path based on the paths' actions. ``Path \% 1'' gives the percentage of post-selected paths $ 100 \cdot e^{S_i}/\sum_j e^{S_j}$ including the LLPs in the normalization sum, whereas ``Path \% 2'' excludes the LLPs.}\label{fig-5p7p_dens}
\end{figure*}

We establish some expressions for the second variation in the stochastic action, and explain its use in distinguishing between MLPs, LLPs, and SPs.

\par Consider some generic action $S[q,\dot{q}] = -\int dt \mathcal{L}(q,\dot{q})$ written in terms of a Lagrangian $\mathcal{L}$. (The minus sign here is used to make $\mathcal{L} = \dot{q}p-H = - \dot{S}$, such that the definition of the conjugate momentum $p\equiv \partial_{\dot{q}}\mathcal{L}$ is preserved. The sign could be be put on the definition of momentum instead.) We let $Q$ and $\dot{Q}$ describe an OP obtained from $\delta S = 0$, where
\be\begin{split} \label{eleq}
\delta S &= \int_0^T dt \left( \eta(t) \partl{\mathcal{L}}{q}{} + \dot{\eta}(t) \partl{\mathcal{L}}{\dot{q}}{} \right) \\&= \int_0^T dt  \left( \partl{\mathcal{L}}{q}{} - \frac{d}{dt} \partl{\mathcal{L}}{\dot{q}}{}  \right) \eta + \eta \partl{\mathcal{L}}{\dot{q}}{} \bigg|_0^T.
\end{split}\ee
Solutions $Q$ and $\dot{Q}$ to the Euler-Lagrange equation $\partial_Q \mathcal{L} = d_t \partial_{\dot{Q}} \mathcal{L}$ for $\mathcal{L} = - \dot{S} = (p^2+1)a(Q)$ (OPs, and using notation conventions from section II), are equivalent to solutions of Hamilton's equations used everywhere else in this work, where $\theta\leftrightarrow Q$, and $p$ and $\dot{Q}$ are related by the transformation
\be
p = \frac{\dot{Q}-b(Q)}{2 a(Q)},
\ee
equivalent to \eqref{dotq_form}. 
Substituting this into $\mathcal{L} = -\dot{S} = (p^2+1)a(Q)$ (recall that this is the form we get after the optimal readout(s) is (are) put in), we find
\be
\mathcal{L} = a(Q) +\frac{\dot{Q}^2 + b(Q)^2 -2 \dot{Q} b(Q)}{4 a(Q)}.
\ee
Note also that the variation functions $\eta(t)$ are required to satisfy $\eta(0) = 0 = \eta(T)$, and be continuous and differentiable everywhere on $t \in [0,T]$. Furthermore, it is apparent from \eqref{eleq} that the validity of the Euler-Lagrange equation depends on the \emph{absence} of zeros in $\eta(t)$ anywhere on the interval $t \in [0,T]$ \emph{except} the endpoints (additional roots could admit trivial solutions to $\delta S = 0$). Any arbitrary variation $\eta$ which meets these conditions is allowed. 

\par Around an OP, we may approximate the action with a series of the form
\be \label{var_series}
S[Q+\epsilon \eta] \approx S_Q + \frac{\epsilon^2}{2}\delta^2 S|_{Q,\dot{Q}} + \mathcal{O}(\epsilon^3),
\ee
where $S_Q$ is the action evaluated along the OP.
The second order term may be written \cite{StroudSC,BookDas}
\be\begin{split}
-\delta^2 S&|_{Q,\dot{Q}} = \\ &\int_{0}^{T} dt\left(\eta^2 \partl{\mathcal{L}}{Q}{2} + 2 \eta \dot{\eta} \frac{\partial^2 \mathcal{L}}{\partial Q \partial \dot{Q}} + \dot{\eta}^2 \partl{\mathcal{L}}{\dot{Q}}{2}  \right).
\end{split}\ee
We compute $\delta^2 S$ numerically, for two different variations of the form $\epsilon \eta = \epsilon \sin(t\pi/T)$, and $\epsilon \eta = \epsilon (t^2-t T)$, respectively. Notice that $\epsilon$ comes out of the second variation altogether, and that the resulting expression in the expansion \eqref{var_series} is symmetrical (parabolic) in $\epsilon$. Therefore the sign of $\delta^2S$ tells us whether we have a max/min/saddle; if $\delta^2 S$ is positive we are at a minimum of the action functional (LLP), if $\delta^2 S$ is negative we are at a maximum of the action functional (MLP), and if $\delta^2 S$ is zero we are at a SP.
\par We find that in the cases with five and seven optimal paths, plotted in Fig.~\ref{fig-LMonset}(e,h) and Fig.~\ref{fig-LMonset}(f,i), respectively, we have one LLP (always in the center-most path, when arranged in order of ascending initial momenta), and the remaining paths are MLPs. These cases are plotted along with the density plots of simulated stochastic trajectories in Fig.~\ref{fig-5p7p_dens}. Recall that a path being a MLP means that any small variation in the path, uniformly to one side of the curve or the other (\emph{i.e.} for valid a $\eta(t)$ as described above), which still meets the same boundary conditions, forces the action to fall.


\bibliography{references}

\end{document}